\documentclass[a4paper,11pt]{article}

\usepackage[a4paper,left=2.73cm,right=2.7cm,top=3cm,bottom=3.5cm]{geometry}

\usepackage{amsfonts,amsmath,amssymb,tabstackengine}
\usepackage[usenames, dvipsnames]{color}
\usepackage{array}
\usepackage{tikz}

\usepackage{color, colortbl}
\definecolor{Gray}{gray}{0.95}

\usepackage[colorlinks=true,linktocpage=true,linkcolor=blue,citecolor=blue]{hyperref}
\usepackage{graphicx}

\addtocontents{toc}{\protect\setcounter{tocdepth}{3}}
\numberwithin{equation}{section}

\def\cN{{\cal N}}

\begin{document}

\begin{titlepage}

\thispagestyle{empty}

\begin{center}

{\LARGE \textbf{S-folds and holographic RG flows on the D3-brane}}

\vspace{40pt}
		
{\large \bf Adolfo Guarino}$\,^{a, b}$\,\,\,\,  \large{and}  \,\,\,\, {\large \bf Colin Sterckx}$\,^{c, a}$ 
		
\vspace{25pt}
		
$^a$\,{\normalsize  
Departamento de F\'isica, Universidad de Oviedo,\\
Avda. Federico Garc\'ia Lorca 18, 33007 Oviedo, Spain.}
\\[7mm]

$^b$\,{\normalsize  
Instituto Universitario de Ciencias y Tecnolog\'ias Espaciales de Asturias (ICTEA) \\
Calle de la Independencia 13, 33004 Oviedo, Spain.}
\\[7mm]

$^c$\,{\normalsize  
Universit\'e Libre de Bruxelles (ULB) and International Solvay Institutes,\\
Service  de Physique Th\'eorique et Math\'ematique, \\
Campus de la Plaine, CP 231, B-1050, Brussels, Belgium.}
\\[10mm]

\texttt{adolfo.guarino@uniovi.es} \,\, , \,\, \texttt{colin.sterckx@ulb.ac.be}

\vspace{20pt}


\vspace{20pt}
				
\abstract{
\noindent 

Type IIB S-folds of the form $\,\textrm{AdS}_{4} \times \textrm{S}^1 \times \textrm{S}^5\,$ are conjectured to correspond to new strongly coupled three-dimensional CFT's on a localised interface of $\,\textrm{SYM}_{4}\,$. In this work we construct holographic RG flows on the D3-brane that generically connect anisotropic deformations of $\,\textrm{SYM}_{4}\,$ in the UV to various S-fold $\textrm{CFT}$'s in the IR with different amounts of supersymmetry and flavour symmetries. Examples of holographic RG flows between \mbox{S-fold} CFT's are also presented.  Lastly a geometric interpretation of axion deformations is provided in terms of monodromies on the internal $\,\textrm{S}^{5}\,$ when moving around the $\,\textrm{S}^{1}$. Special attention is paid to the monodromy-induced patterns of symmetry breaking as classified by the mapping torus $T_{h}(\textrm{S}^5)$.

}

\end{center}

\end{titlepage}

\tableofcontents

\hrulefill
\vspace{10pt}

\section{Motivation and outlook}
\label{sec:intro}

Since its discovery, electromagnetic duality has and continues to provide plenty of intriguing phenomena both in supergravity and quantum field theory. From a string theory perspective, it plays a prominent role amongst the various string/M-theory dualities \cite{Hull:1994ys}. Within the realm of four-dimensional maximal gauged supergravity \cite{deWit:2007mt}, electromagnetic duality was shown to generate new classes of dyonic semi-simple gaugings which hitherto lack a string/M-theory origin. The prototypical example being the SO(8) gauged supergravity which, in its purely electric version \cite{deWit:1982ig}, arises from the dimensional reduction of eleven-dimensional supergravity on a seven-sphere $\,\textrm{S}^{7}\,$ \cite{deWit:1986iy,Nicolai:2011cy}. The SO(8) supergravity was shown to admit a generalisation to a one-parameter family of dyonic theories based on an electromagnetic deformation parameter originally denoted by $\,c\,$ in \cite{Dall'Agata:2012bb}. However, despite the plethora of new AdS$_{4}$ \cite{Dall'Agata:2012bb,Borghese:2012qm,Borghese:2012zs,Borghese:2013dja},  black hole \cite{Anabalon:2013eaa,Lu:2014fpa,Wu:2015ska}, Janus \cite{Karndumri:2020bkc} and domain-wall \cite{Guarino:2013gsa,Tarrio:2013qga} solutions that appear when turning on $\,c\,$, a higher-dimensional interpretation of such parameter (if any) remains elusive and some no-go theorems against such a higher-dimensional geometric origin have been stated \cite{deWit:2013ija,Lee:2015xga}.

After \cite{Dall'Agata:2012bb}, a characterisation and classification of electromagnetic deformations for other gaugings of maximal supergravity was presented in \cite{Dall'Agata:2014ita} and extended to less supersymmetric theories in \cite{Inverso:2015viq}. Amongst the non-semisimple gaugings investigated in \cite{Dall'Agata:2014ita}, the $\,\textrm{ISO}(7)\,$ and $\,[\,\textrm{SO}(1,1) \times \textrm{SO}(6)\,] \ltimes \mathbb{R}^{12}\,$ gauged maximal supergravities have received most of the attention due to their connection with (massive) type IIA \cite{Guarino:2015jca} and type IIB \cite{Inverso:2016eet} string theory, respectively.

\subsubsection*{Type IIA}

The dyonic $\,\textrm{ISO}(7)\,$ supergravity \cite{Guarino:2015qaa} has been shown to contain a rich structure of AdS$_{4}$ vacua \cite{Guarino:2015qaa,Guarino:2019jef,Guarino:2020jwv,Bobev:2020qev}, asymptotically AdS black hole \cite{Guarino:2017eag,Guarino:2017pkw}, Janus \cite{Suh:2018nmp} and domain-wall \cite{Guarino:2016ynd,Guarino:2019snw} solutions. Using the uplift formulae in \cite{Guarino:2015vca} the AdS$_{4}$ vacua have been uplifted to $\,\textrm{AdS}_{4} \times \textrm{S}^{6}\,$ backgrounds of massive type IIA supergravity \cite{Guarino:2015jca,Guarino:2015vca,Varela:2015uca} which further extend to well-defined backgrounds of string theory. This string-theoretic origin permitted the higher-dimensional and holographic understanding of the electromagnetic parameter $\,c\,$ within the massive IIA context \cite{Guarino:2015jca}:  the parameter $\,c\,$ was identified with the Romans mass parameter $\,\hat{F}_{0}\,$ \cite{Romans:1985tz} in ten dimensions, and interpreted holographically as the Chern--Simons (CS) level $\,k\,$ of the dual three-dimensional Chern--Simons-matter theories. Further confirmation came from the successful holographic counting of asymptotically AdS black hole microstates carried out in \cite{Hosseini:2017fjo,Benini:2017oxt}.

\begin{figure}[t!]
\begin{center}
\mbox{
\includegraphics[width=0.50\textwidth]{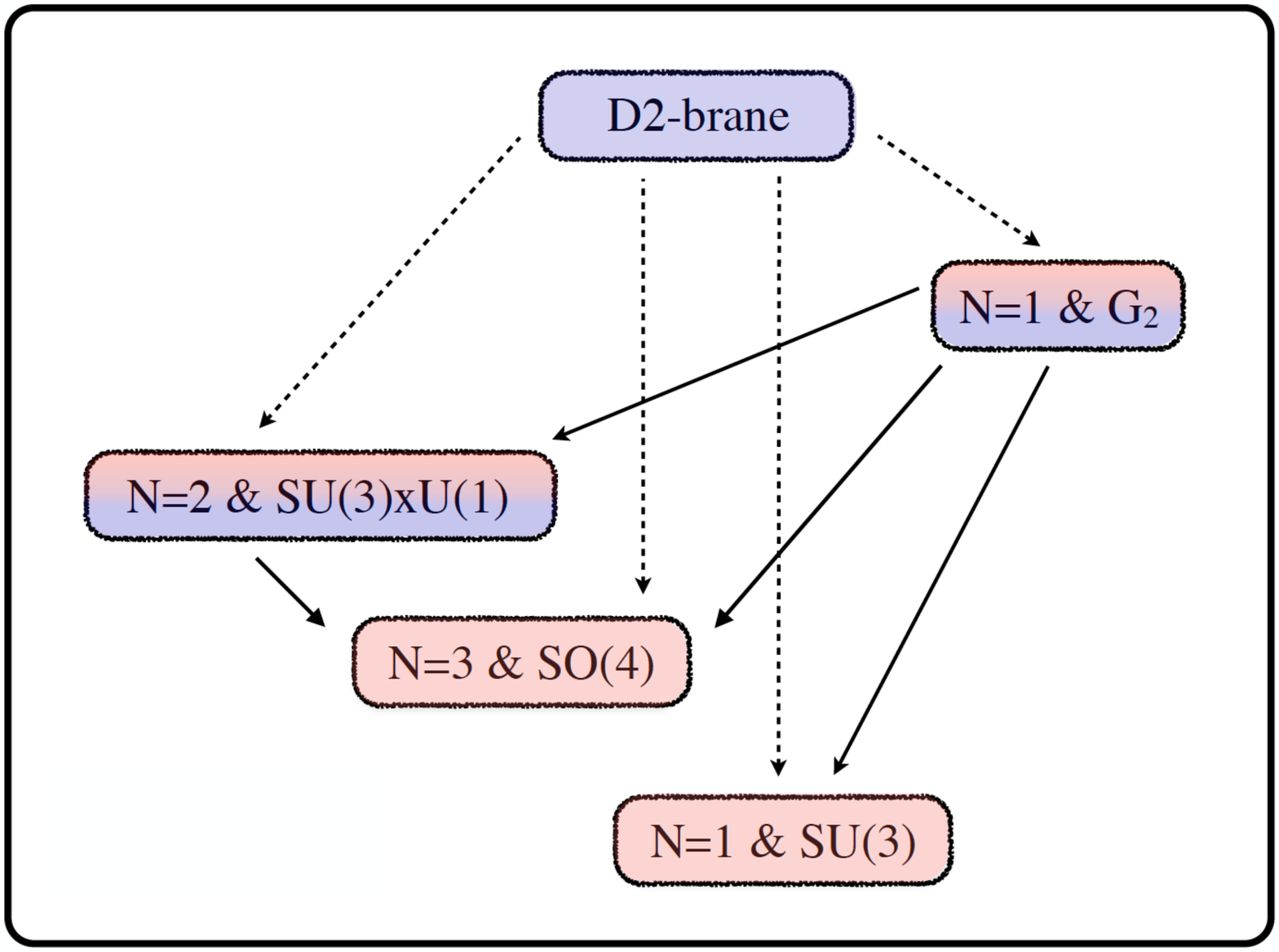}
\includegraphics[width=0.50\textwidth]{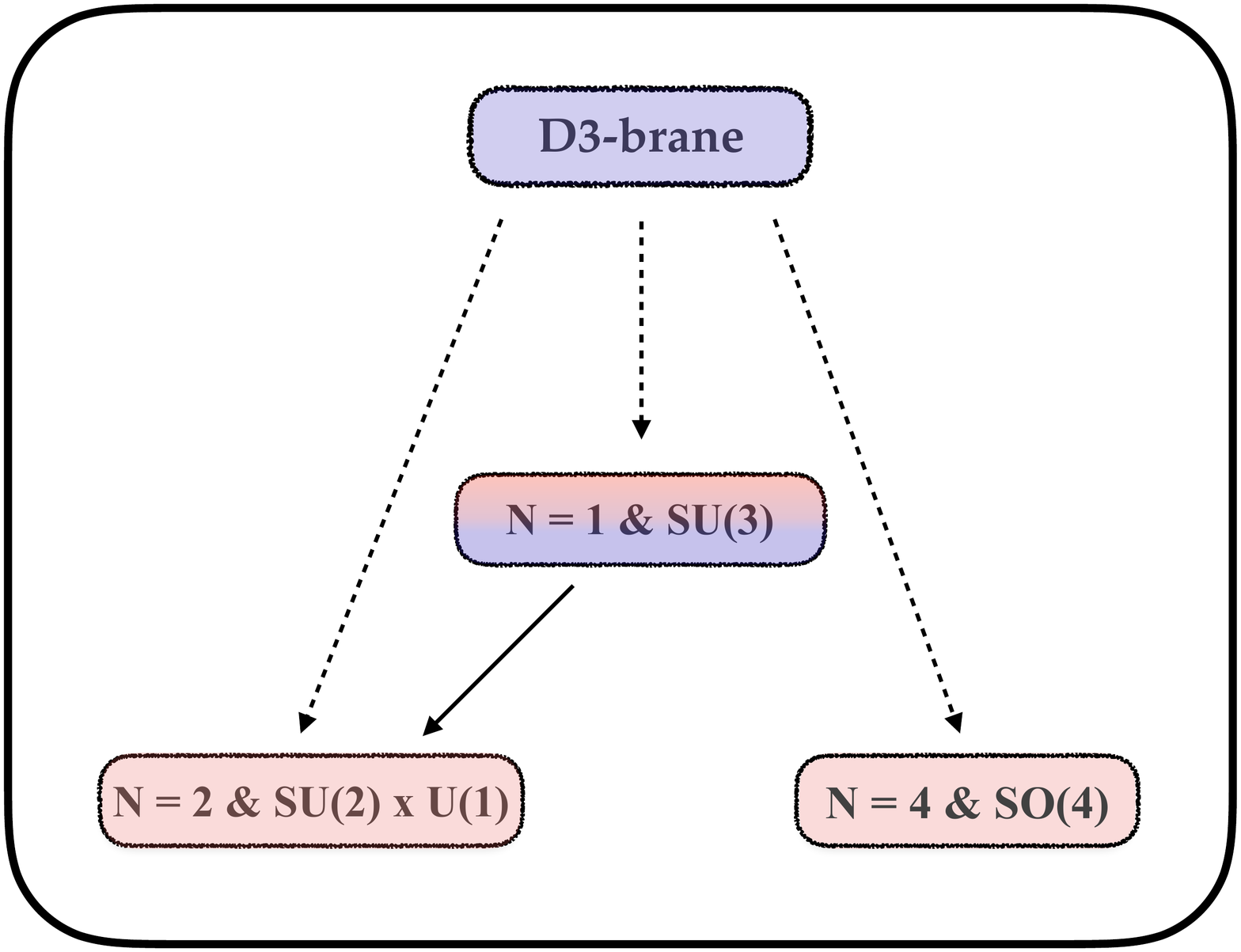}}
\caption{Left: Network of domain-walls connecting the D2-brane behaviour (SYM$_{3}$-CS) and the known supersymmetric $\textrm{AdS}_{4}$ solutions (CFT$_{3}$'s) of the ISO(7) maximal supergravity. Right: Network of domain-walls connecting the D3-brane behaviour (anisotropic SYM$_{4}$) and the known supersymmetric $\textrm{AdS}_{4}$ solutions (J-fold $\,\textrm{CFT}_{3}$'s) of the dyonically-gauged $\, [\,\textrm{SO}(1,1) \times \textrm{SO}(6)\,] \ltimes \mathbb{R}^{12}\,$ maximal supergravity. In both diagrams, domain-walls connecting \textrm{AdS}$_{4}$ solutions are denoted by solid lines.}
\label{fig:net_DW}
\end{center}
\end{figure}

Getting closer to the goal of this work, domain-wall solutions describing holographic RG flows involing the $\,\textrm{CFT}_{3}\,$'s dual to the various AdS$_{4}$ vacua of the $\,\textrm{ISO}(7)\,$ maximal supergravity were obtained in \cite{Guarino:2016ynd,Guarino:2019snw}. Two types of supersymmetric holographic RG flows were constructed numerically: 
\begin{itemize}

\item[$i)$] $\,\textrm{SYM}_{3}\,$ to $\,\textrm{CFT}_3\,$ flows connecting the non-conformal behaviour of the D2-brane in the UV (dual to super-Yang--Mills theory deformed by the CS term) to the various AdS$_{4}$ vacua (dual to CFT$_{3}$'s) in the IR.

\item[$ii)$] $\,\textrm{CFT}_3\,$ to $\,\textrm{CFT}_3\,$ flows connecting two AdS$_{4}$ vacua of the theory.

\end{itemize}

\noindent A summary of such supersymmetric domain-walls is displayed in Figure~\ref{fig:net_DW} (left diagram) where we have denoted by $\,\mathcal{N}\,\&\,\tilde{\textrm{G}}_{0}\,$ an AdS$_{4}$ vacuum preserving $\,\mathcal{N}\,$ supersymmetries and a residual gauge symmetry group $\,\tilde{\textrm{G}}_{0}\,$. The flavour symmetry group $\,\textrm{G}\,$ of the dual CFT$_{3}$ is then extracted from $\,\tilde{\textrm{G}}_{0}=\textrm{G} \times \textrm{SO}(\mathcal{N})\,$ where $\,\textrm{SO}(\mathcal{N})\,$ is the corresponding R-symmetry group.

\subsubsection*{Type IIB}

The dyonic $\,{[\textrm{SO}(1,1) \times \textrm{SO}(6)] \ltimes \mathbb{R}^{12}} \,$ maximal supergravity has also been shown to contain a rich structure of AdS$_{4}$ vacua \cite{Inverso:2016eet,Guarino:2019oct,Guarino:2020gfe}. Unlike for their type IIA counterparts, these AdS$_{4}$ vacua organise into multi-parametric families depending on a set of constant axions~$\chi$'s \cite{Guarino:2020gfe}. Each of these families preserves a fixed amount of supersymmetry and has a fixed AdS$_{4}$ radius. Therefore, the axions~$\chi$'s are interpreted as exactly marginal deformations in the dual CFT$_{3}$'s. Within a family of AdS$_{4}$ vacua, different values of the axions cause a different breaking of the residual gauge symmetry. More concretely, they induce a breaking of flavour symmetries in the dual CFT$_{3}$'s.

The AdS$_{4}$ vacua with vanishing axions feature the largest possible symmetry within their respective families. Focusing on supersymmetric solutions only, three such AdS$_{4}$ vacua have been found preserving various amounts of (super) symmetry: $\,{\mathcal{N}=1\,\&\,\textrm{SU(3)}}\,$ \cite{Guarino:2019oct}, $\,{\mathcal{N}=2\,\&\,\textrm{SU(2)} \times \textrm{U(1)}}\,$ \cite{Guarino:2020gfe} and $\,\mathcal{N}=4\,\&\,\textrm{SO(4)}\,$ \cite{Gallerati:2014xra}. Upon subtraction of the corresponding R-symmetry factors, these AdS$_{4}$ vacua are dual to CFT$_{3}$'s with $\,\textrm{SU(3)}\,$, $\,\textrm{SU(2)}\,$ and trivial flavour symmetry groups, respectively. These vacua of the $\,[\textrm{SO}(1,1) \times \textrm{SO}(6)] \ltimes \mathbb{R}^{12}\,$ maximal supergravity have been uplifted to S-fold backgrounds of type IIB supergravity on $\,\textrm{AdS}_{4} \times \textrm{S}^{1} \times \textrm{S}^{5}\,$ that involve non-trivial $\,\textrm{SL}(2,\mathbb{Z})_{\textrm{IIB}}\,$ monodromies along $\,\textrm{S}^{1}\,$ in the hyperbolic class \cite{Inverso:2016eet,Guarino:2019oct,Guarino:2020gfe} -- commonly denoted by $\,J \in \textrm{SL}(2,\mathbb{Z})_{\textrm{IIB}}\,$ -- and thus the corresponding CFT$_{3}$ duals have been referred to as J-fold CFT$_{3}$'s in the literature \cite{Assel:2018vtq}. These J-fold CFT$_{3}$'s and their gravity duals have also been investigated using a five-dimensional gauged supergravity approach within the context of holographic interfaces in $\,\mathcal{N}=4\,$ SYM$_{4}$ \cite{Bobev:2019jbi,Bobev:2020fon}\cite{Arav:2021tpk}, and arise as limiting cases of Janus solutions.

In this work we will construct supersymmetric domain-wall (DW$_4$) solutions describing holographic RG flows involving the J-fold $\,\textrm{CFT}_{3}\,$'s dual to the various AdS$_{4}$ vacua of the $\,{[\textrm{SO}(1,1) \times \textrm{SO}(6)] \ltimes \mathbb{R}^{12}} \,$ maximal supergravity. As in the type IIA case previously reviewed, two types of holographic RG flows appear in the type IIB context: 
\begin{itemize}

\item[$i)$] $\,\textrm{SYM}_{4}\,$ to $\,\textrm{CFT}_3\,$ flows connecting a $\,c$-deformation of the D3-brane behaviour in the UV (dual to an anisotropic deformation of $\textrm{SYM}_{4}$) to the various axion-vanishing AdS$_{4}$ vacua (dual to J-fold CFT$_{3}$'s) in the IR.

\item[$ii)$] $\,\textrm{CFT}_3\,$ to $\,\textrm{CFT}_3\,$ flows connecting two AdS$_{4}$ vacua of the theory, the one at the UV generically involving non-zero axions (equivalently marginal deformations).

\end{itemize}

\noindent A summary of these new supersymmetric domain-walls is also displayed in Figure~\ref{fig:net_DW} (right diagram) where the notation $\,\mathcal{N}\,\&\,\tilde{\textrm{G}}_{0}\,$ has been maintained for the various AdS$_{4}$ vacua. The $\,\textrm{CFT}_3\,$ to $\,\textrm{CFT}_3\,$ flow depicted in the diagram corresponds to the case with vanishing axions (no marginal deformations) in the UV, so that the largest possible $\,\textrm{SU}(3)\,$ symmetry within the $\,\mathcal{N}=1\,$ family of AdS$_4$ vacua is realised in the UV. Notwithstanding, $\,\textrm{CFT}_3\,$ to $\,\textrm{CFT}_3\,$ flows also exist that reach $\,\mathcal{N}=1\,$ AdS$_4$ vacua  in the UV with non-vanishing axions so that a smaller $\,\textrm{SU}(2)\times \textrm{U}(1) \subset \textrm{SU}(3)\,$ symmetry is realised in the UV.

As we will show in this work, the RG flows of type $\,i)\,$ will involve a set of sub-leading corrections in the parameter $\,c\,$ of the four-dimensional DW$_4$ description of the D3-brane solution at $\,c=0\,$ dual to (isotropic) $\,\mathcal{N}=4\,$ $\,\textrm{SYM}_{4}\,$. These corrections will induce anisotropy in the UV. Type IIB examples of such an anisotropic behaviour of $\,\textrm{SYM}_{4}\,$ have previously been obtained in terms of a charge density of dissolved D7-branes \cite{Mateos:2011ix} or backreacted  geometries corresponding to the intersection of D3- and (smeared) D5-branes along $\,2+1\,$ dimensions \cite{Conde:2016hbg,Hoyos:2020zeg}. The type IIB background in \cite{Conde:2016hbg,Hoyos:2020zeg} involves non-trivial RR fluxes $\,F_{3}\,$ and $\,\widetilde{F}_{5}\,$ and involves a stack of $\,N_{c}\,$ D3-branes and $\,N_{f}\,$ flavour D5-branes. Here we obtain anisotropy in a purely closed string setup \textit{without sources} -- sourceless Bianchi identities are satisfied in ten dimensions as the effective four-dimensional supergravity enjoys maximal supersymmetry -- by implementing the locally geometric $\,\textrm{SL}(2)_{\textrm{IIB}}\,$ twist that generates the S-fold background in the IR. As a by-product of the mechanism the results preserve $\,\textrm{SL}(2)_{\textrm{IIB}}\,$  covariance.

Another interesting aspect regarding the UV behaviour of the RG flows presented in this work has to do with the role being played by the axions $\,\chi$'s. As we will show explicitly, these axions can be locally reabsorbed in a reparameterisation of the angular coordinates on the five-sphere so that a non-trivial monodromy arises on $\,\textrm{S}^5 \times \textrm{S}^1\,$. We will investigate this issue for the two type IIB backgrounds controlling the UV behaviour of the RG flows: the D3-brane and the $\,\mathcal{N}=1\,$ family of S-folds. We will build upon \cite{1852633}, where this issue has been investigated for the $\,\mathcal{N}=2\,$ family of S-folds, and generalise the results therein to accommodate the two cases of relevance in this work and, as a by-product, also the non-supersymmetric S-folds in \cite{Guarino:2020gfe}. In this manner we will understand the geometric role played by the axions $\,\chi$'s and relate it to the patterns of (flavour) symmetry breaking that these axions induce as classified by the mapping torus $T_{h}(\textrm{S}^5)$.

The paper is organised as follows. In Section~\ref{sec:4D} we present the supergravity model, the BPS flow equations and the four-dimensional AdS$_{4}$ and non-AdS$_{4}$ solutions relevant for the work. In Section~\ref{sec:RG flows} we numerically construct the holographic RG flows, both $\,\textrm{SYM}_{4}\,$ to $\,\textrm{CFT}_3\,$ and $\,\textrm{CFT}_3\,$ to $\,\textrm{CFT}_3\,$ flows. In Section~\ref{sec:10D} we investigate the flows from a ten-dimensional perspective. First we focus on the issue of anisotropy when the UV regime of the flows is controlled by the deformed D3-brane at $\,c \neq 0\,$, and then discuss global \textit{vs} local geometric aspects of the type IIB backgrounds when the axions $\,\chi$'s are activated, as well as the implications they have for symmetry breaking. We conclude and discuss some future lines in Section~\ref{sec:conclus}. Some technicalities regarding group theoretical aspects of the four-dimensional supergravity model and its potential oxidation to five dimensions are collected in the Appendix.

\section{Four dimensions}
\label{sec:4D}

Let us consider the maximal supergravity with gauge group $\,{[\textrm{SO}(1,1) \times \textrm{SO}(6)] \ltimes \mathbb{R}^{12}}\,$ that arises upon dimensional reduction of type IIB supergravity on $\,\textrm{S}^5\times \mathbb{R}\,$ \cite{Inverso:2016eet}. For the sake of tractability, we restrict ourselves to the $\,\mathbb{Z}_{2}^{3}$-invariant sector of the theory constructed in \cite{Guarino:2020gfe}.

\subsection{Supergravity model and flow equations}

This sector describes a minimal $\,\mathcal{N}=1\,$ supergravity coupled to seven chiral multiplets with complex scalars $\,z_{i}\,$ and no vector multiplets. We parameterise the complex scalars of the chiral multiplets as 
\begin{equation}
\label{Phi_def_7}
z_{i} = - \chi_{i} + i \, y_i
\hspace{10mm} \textrm{ with } \hspace{10mm} i=1, \ldots, 7\hspace{8mm} \textrm{   and   }\hspace{8mm} y_i>0 \ .
\end{equation}
The seven complex fields serve as coordinates on a K\"ahler scalar geometry
\begin{equation}
\mathcal{M}_{\textrm{scal}} = [\textrm{SL}(2)/\textrm{SO}(2)]^7 \ ,
\end{equation}
characterised by a K\"ahler potential of the form
\begin{equation}
\label{K_7_chirals}
K = - \displaystyle\sum_{i=1}^7   \log[-i(z_{i}-\bar{z}_{i})] \ .
\end{equation}
The scalar potential for the $\,\mathbb{Z}_{2}^{3}$-invariant sector of the theory follows from an holomorphic superpotential of the form
\begin{equation}
\label{W_7_chiral}
W = 2 \, g \, \big[  \, z_{1} z_{5} z_{6} +  z_{2} z_{4} z_{6} + z_{3} z_{4} z_{5} + (  z_{1} z_{4} +  z_{2} z_{5} + z_{3} z_{6} ) \,  z_{7} \, \big] + 2 \,  g \, c  \, ( 1 - z_{4}  z_{5} z_{6}  z_{7} )  \ ,
\end{equation}
which is originated from the gauging in the maximal theory. Truncating away the fermions the Lagrangian acquires an Einstein-scalar form and reads
\begin{equation}
\label{L_scalars_real}
\begin{array}{llll}
e^{-1} \, \mathcal{L}_{\textrm{E-s}} & = & \tfrac{1}{2} \, R - K_{IJ} \, \partial_{\mu}\Sigma^{I} \, \partial^{\mu}\Sigma^{J} -  V \ , \\[2mm]
& = & \tfrac{1}{2} \, R -\frac{1}{4} \displaystyle\sum_{i=1}^7 \frac{1}{y_i^2} \left[  (\partial y_{i})^2 +  (\partial \chi_{i})^2 \right]  -  V \ ,
\end{array}
\end{equation}
where $\,\Sigma^{I}=\left\lbrace \, y_{1} \, , \, \chi_{1} \, , \, y_{2} \, , \, \chi_{2} \, , \, \ldots \, , \,  y_{7} \, , \, \chi_{7} \, \right\rbrace\,$ $\,(I=1,\ldots , 14)\,$ denotes the real and imaginary components of the complex fields $\,z_{i}\,$. The kinetic matrix for these (real) fields reads
\begin{equation}
K_{IJ} = \textrm{diag} \left( \, \frac{1}{4 y_1^{2}}  \, , \,  \frac{1}{4 y_1^{2}} \, , \, \frac{1}{4 y_2^{2}}  \, , \,  \frac{1}{4 y_2^{2}} \, , \,  \ldots \, , \,  \frac{1}{4 y_7^{2}}  \, , \,  \frac{1}{4 y_7^{2}}\, \right) \ .
\end{equation}
The scalar potential $\,V\,$ can be recovered from the holomorphic superpotential (\ref{W_7_chiral}) using the standard result
\begin{equation}
\label{V_scalar}
V = e^{K} \Big[ K^{z_{i} \bar{z}_{j}}  \, D_{z_{i}} W  \, D_{\bar{z}_{j}} \bar{W}   - 3 \, W \, \bar{W}  \Big] \ ,
\end{equation}
where $\,D_{z_{i}} W \equiv \partial_{z_{i}} W + (\partial_{z_{i}} K) W\,$ is the K\"ahler derivative and $\,K^{z_{i} \bar{z}_{j}} \,$ is the inverse of the K\"ahler metric $\,K_{z_{i} \bar{z}_{j}} \equiv \partial^{2}_{z_{i} , \bar{z}_{j}} K \,$. This is the $\,\mathcal{N}=1\,$ supergravity model we will investigate during the rest of the work.

\subsubsection*{First-order flow equations}

In order to study RG flows holographically, we will investigate flat-sliced domain-wall (DW$_{4}$) solutions whose metric takes the form
\begin{equation}
\label{DW4_metric}
ds_{\textrm{DW}_{4}}^2 = e^{2 A(z)} \, \eta_{\alpha \beta} \, dx^{\alpha} dx^{\beta} + dz^2
\hspace{10mm} \textrm{ with } \hspace{10mm}
\eta_{\alpha \beta} = \textrm{diag}(-1,1,1) \ ,
\end{equation}
where $\,z \in (-\infty , \infty)\,$ is the coordinate transverse to the domain-wall and $\,A(z)\,$ is the scale factor. Asking for the vanishing of the supersymmetry variations of fermions (gravitino and chiralini) in the $\,\mathcal{N}=1\,$ supergravity model, a set of first-order BPS equations consisting of
\begin{equation}
\label{BPS_equations}
\partial_{z} A = \mp  \, |\mathcal{W}|
\hspace{8mm} \textrm{ and } \hspace{8mm}
\partial_{z} \Sigma^{I} = \pm \, K^{IJ} \, \partial_{\Sigma^{J}} |\mathcal{W}| \ ,
\end{equation}
is obtained to which we will refer as \textit{flow equations}. The real superpotential $\, |\mathcal{W}|\,$ is constructed from the (complex) gravitino mass term
\begin{equation}
\label{N1_gravitino_mass}
\mathcal{W} = e^{\frac{K}{2}}\,W =m_{3/2} \ ,
\end{equation}
with $\,W\,$ in (\ref{W_7_chiral}), and fully specifies the flow equations in (\ref{BPS_equations}).

\subsection{AdS$_4$ vacua: modes and dimensions of dual operators}
\label{sec:AdS_vacua}

The simplest solutions to the BPS equations  (\ref{BPS_equations}) are supersymmetric AdS$_4$ vacua. These solutions have constant scalars and thus satisfy (\ref{BPS_equations}) provided 
\begin{equation}
\partial_{\Sigma^{I}} |\mathcal{W}| = 0
\hspace{10mm} \textrm{ and } \hspace{10mm}
A(z)=\mp  \, |\mathcal{W}_{0}| \, z + C \ ,
\end{equation}
where $\, |\mathcal{W}_{0}|^{-2} = L^2=-3/V_{0}\,$ corresponds to values evaluated at the AdS$_{4}$ vacuum and $\,C\,$ is an arbitrary constant that can be reabsorbed by a rescaling of the coordinates $\,x^{\alpha}\,$ in (\ref{DW4_metric}).

The AdS$_4$/CFT$_{3}$ holographic dictionary then states that, at an AdS$_{4}$ vacuum, scalars fields with a normalised mass $\,m^2 L^2 <0\,$ correspond to relevant operators, $\,m^2 L^2 = 0\,$ to marginal operators and $\,m^2 L^2 >0\,$ to irrelevant operators in the dual field theory. Each scalar field comes along with two modes with conformal dimensions $\,\Delta_\pm\,$, where $\,\Delta_{+}\,$ ($\,\Delta_{-}\,$) is the larger (smaller) root of 
\begin{equation}
\label{Deltas_eq}
\Delta (\Delta - 3) = m^2L^2 \ .
\end{equation}
The conformal dimension of the dual operator is then identified with $\,\Delta_{+}\,$. The question about which of the two modes is selected by supersymmetry is answered by the diagonalisation of the matrix
\begin{equation}
\label{Delta_matrix_def}
\Delta^{I}{}_{J}  \equiv  L \,\, K^{IM} \, \partial_{M , J} |\mathcal{W}|  \ .
\end{equation}
From a field theory perspective, $\,\Delta_{-}\,$ is interpreted as a source for the operator whereas $\,\Delta_{+}\,$ is interpreted as a vacuum expectation value (VEV) for the operator. There is an ambiguity with this interpretation whenever the masses lie within the window $\,-\frac{9}{4}< m^2 L^2 < -\frac{5}{4}\,$, for which an alternative quantisation of the scalar field is possible that interchanges the source and the VEV \cite{Klebanov:1999tb}.

In the following we review the three families of $\,\mathcal{N}=1\,$, $\,\mathcal{N}=2\,$ and $\,\mathcal{N}=4\,$ supersymmetric AdS$_4$ vacua presented in ref.~\cite{Guarino:2020gfe}. These vacua feature a hierarchy of vacuum energies given by
\begin{equation}
\label{V_chain}
V^{\mathcal{N}=1} \,\, > \,\, V^{\mathcal{N}=2} \,\,  = \,\,  V^{\mathcal{N}=4} \ .
\end{equation}
We will present these families of solutions as well as the spectrum of scalars around the most symmetric vacuum within each family. A summary of such most symmetric AdS$_{4}$ vacua can also be found in Table \ref{vacua_summary}.

\subsubsection{$\mathcal{N}=1\,$ family: vacuum with $\, \textrm{SU}(3) \,$ symmetry}
\label{sec:N=1_AdS4_vacua}

There is a two-parameter family of $\,\mathcal{N}=1\,$ supersymmetric AdS$_{4}$ solutions that preserves $\,\textrm{U}(1)^2\,$. It is located at
\begin{equation}
\label{VEVs_z_N1}
z_{1}=z_{2}= z_{3} = c \left( -\chi_{1,2,3} +  i \, \frac{\sqrt{5}}{3} \right)
\hspace{6mm} \textrm{and} \hspace{6mm}
z_{4}=z_{5}=z_{6}=z_{7}= \frac{1}{\sqrt{6}} ( 1+ i \, \sqrt{5}) \ ,
\end{equation}
so that $\,|z_{4,5,6,7}|=1\,$, and is subject to the constraint\footnote{Here $\,\chi_{1,2,3}\,$ are understood as constant parameters.}
\begin{equation}
\label{chi_SU3_enhancement}
\sum_{i=1}^{3} \chi_{i} = 0 \ .
\end{equation}
This family of AdS$_{4}$ solutions has a vacuum energy given by
\begin{equation}
V_{0} = -\frac{162}{25 \sqrt{5}} \, \, g^2 \, c^{-1} \ ,
\end{equation}
and a spectrum of $\,\mathbb{Z}_{2}^{3}$-invariant normalised scalar masses of the form
\begin{equation}
\label{spectrum_SU3_scalars}
\begin{array}{lll}
m^2 L^2 &=&
 0 \,\,\, ( \times  2)
 \hspace{3mm} , \hspace{3mm}
 4 \pm \sqrt{6} \,\,\, ( \times 2 )
 \hspace{3mm} , \hspace{3mm}
 -2 \,\,\, ( \times 2 ) \ ,  \\[4mm]
& &
 -\frac{14}{9} + 5 \chi _i^2 \pm \frac{1}{3} \sqrt{4 + 45 \chi _i^2}    \,\,\, ( \times 1 )
 \hspace{8mm} i=1,2,3 \ ,
\end{array}
\end{equation}
where $\,L^2=-3/V_{0}\,$ is the AdS$_{4}$ radius. As discussed in \cite{Guarino:2020gfe}, a generic solution in this family preserves $\,\textrm{U}(1)^2\,$. However, the residual symmetry gets enhanced to $\,\textrm{SU}(2) \times \textrm{U}(1)\,$ when imposing a pairwise identification between the axions $\,\chi_{1,2,3}\,$. Finally there is a symmetry enhancement to $\,\textrm{SU}(3)\,$ when setting $\,\chi_{1,2,3}=0\,$.

This $\,\mathcal{N}=1 \,\, \& \,\, \textrm{SU}(3)\,$ symmetric AdS$_4$ vacuum was uplifted to a family of type IIB S-folds with $\,\mathcal{N}=1\,$ supersymmetry in \cite{Guarino:2019oct}. Setting the moduli $\,\chi_{1,2,3}=0\,$ yields
\begin{equation}  
\label{VEVs_z_N1_2}
z_{1}=z_{2}= z_{3} = i \, c  \, \frac{\sqrt{5}}{3}
\hspace{10mm} \textrm{ and } \hspace{10mm}
z_{4}=z_{5}=z_{6}=z_{7}= \frac{1}{\sqrt{6}} ( 1+ i \, \sqrt{5}) \ ,
\end{equation}
with masses in (\ref{spectrum_SU3_scalars}) given by
\begin{equation}
\label{spectrum_SU3_scalars_chi=0}
\begin{array}{lll}
m^2 L^2 &=&
-\frac{20}{9}    \,\,\, ( \times 3 )
\hspace{3mm} , \hspace{3mm}
-\frac{8}{9}    \,\,\, ( \times 3 )
\hspace{3mm} , \hspace{3mm}
 0 \,\,\, ( \times  2)
 \hspace{3mm} , \hspace{3mm}
 4 \pm \sqrt{6} \,\,\, ( \times 2 )
 \hspace{3mm} , \hspace{3mm}
 -2 \,\,\, ( \times 2 ) \ .
\end{array}
\end{equation}
By virtue of (\ref{Deltas_eq}), the set of normalised scalar masses in (\ref{spectrum_SU3_scalars_chi=0}) implies a set of conformal dimensions $\,\Delta_{\pm}\,$ for the dual operators given by
\begin{equation}
\label{Deltas_SU3_scalars_chi=0}
\resizebox{\hsize}{!}{$
\begin{array}{rrrrrrrrrrrrrr}
m^2 L^2 &=& -\frac{20}{9} \, ( \times 3 ) & , & -2 \, ( \times 2 ) & , & -\frac{8}{9} \, ( \times 3 ) & ; &  0 \, ( \times 2 ) & ; &  4 - \sqrt{6} \, ( \times 2 ) & , &  4 + \sqrt{6}  \, ( \times 2 ) & , \\[3mm]
\Delta_{+} &=& \frac{\textbf{5}}{\textbf{3}} \, ( \times 3 ) & , & \textbf{2} \, ( \times 2 ) & , &   \frac{8}{3} & ; & 3 & ; &   \textbf{1} + \sqrt{\textbf{6}} \, ( \times 2 ) & , & 2 + \sqrt{6} & , \\[3mm]
\Delta_{-} &=& \frac{4}{3} & , &  1 & , &   \frac{\textbf{1}}{\textbf{3}} \, ( \times 3 ) & ; & \textbf{0} \, ( \times 2 )  & ; &   2 - \sqrt{6} & , & \textbf{1} - \sqrt{\textbf{6}} \, ( \times 2 )   & .
\end{array}$}
\end{equation}
The highlighted conformal dimensions in (\ref{Deltas_SU3_scalars_chi=0}) appear as eigenvalues of the matrix \eqref{Delta_matrix_def} and will play a role later on when studying holographic RG flows involving this conformal fixed point.

\subsubsection{$\mathcal{N}=2\,$ family: vacuum with $\,\textrm{SU}(2) \times \textrm{U}(1) \,$ symmetry}
\label{sec:N=2_AdS4_vacua}

There is a one-parameter family of $\,\mathcal{N}=2\,$ supersymmetric AdS$_{4}$ solutions that preserves $\,\textrm{U}(1)^2\,$. It is located at
\begin{equation}
\label{VEVs_z_N2}
z_{1}=-\bar{z}_{3}= c \left( -\chi \, + \,  i \, \frac{1}{\sqrt{2}} \right)
\hspace{3mm} \textrm{,} \hspace{3mm}
z_{2} = i \, c
\hspace{3mm} \textrm{,} \hspace{3mm}
z_{4}=z_{6}= i
\hspace{2mm} \textrm{ and } \hspace{2mm}
z_{5}=z_{7} =  \frac{1}{\sqrt{2}} (1 \, + \, i ) \ ,
\end{equation}
so that $\,|z_{4,6}|=|z_{5,7}|=1\,$. This family of AdS$_{4}$ solutions has a vacuum energy given by
\begin{equation}
V_{0} = -3 \,\, g^2 \, c^{-1}\ ,
\end{equation}
and a spectrum of $\,\mathbb{Z}_{2}^{3}$-invariant normalised scalar masses of the form
\begin{equation}
\label{spectrum_SU2xU1_scalars}
\begin{array}{lll}
m^2 L^2 &=&
0 \,\,\, ( \times 1 )
\hspace{3mm} , \hspace{3mm}
3 \pm \sqrt{17} \,\,\, ( \times 2 )
\hspace{3mm} , \hspace{3mm}
-2 \,\,\, ( \times 1 )
\hspace{3mm} , \hspace{3mm}
2 \,\,\, ( \times 4 )
\hspace{3mm} , \hspace{3mm}
-2 + 4 \chi^2 \,\,\, ( \times 2 )  \\[2mm]
& & -1 + 4 \chi^2 \pm \sqrt{16 \chi^2+1} \,\,\, ( \times 1 )
\ ,
\end{array}
\end{equation}
where $\,L^2=-3/V_{0}\,$ is the AdS$_{4}$ radius. A generic solution in this family preserves $\,\textrm{U}(1)^2\,$, but the residual symmetry gets enhanced to $\,\textrm{SU}(2) \times \textrm{U}(1)\,$ when $\,\chi=0\,$.

This $\,\mathcal{N}=2 \,\, \& \,\, \textrm{SU}(2) \times \textrm{U}(1)\,$ symmetric AdS$_4$ vacuum was uplifted to a family of type IIB S-folds with $\,\mathcal{N}=2\,$ supersymmetry in \cite{Guarino:2020gfe} (see \cite{1852633} for the uplift including the axion $\,\chi\,$). Setting the modulus $\,\chi=0\,$ yields
\begin{equation}  
\label{VEVs_z_N2_2}
z_{1} = z_{3}=  i \, c \, \frac{1}{\sqrt{2}}
\hspace{5mm} \textrm{,} \hspace{5mm}
z_{2} = i \, c
\hspace{5mm} \textrm{,} \hspace{5mm}
z_{4}=z_{6}= i
\hspace{5mm} \textrm{ and } \hspace{5mm}
z_{5}=z_{7} =  \frac{1}{\sqrt{2}} (1 \, + \, i ) \ ,
\end{equation}
with masses in (\ref{spectrum_SU2xU1_scalars}) given by
\begin{equation}
\label{spectrum_SU2xU1_scalars_chi=0}
\begin{array}{lll}
m^2 L^2 &=&
-2 \,\,\, ( \times 4 )
\hspace{3mm} , \hspace{3mm}
0 \,\,\, ( \times 2 )
\hspace{3mm} , \hspace{3mm}
3 \pm \sqrt{17} \,\,\, ( \times 2 )
\hspace{3mm} , \hspace{3mm}
2 \,\,\, ( \times 4 ) \ .
\end{array}
\end{equation}
Upon solving (\ref{Deltas_eq}), the set of normalised scalar masses in (\ref{spectrum_SU2xU1_scalars_chi=0}) implies a set of conformal dimensions $\,\Delta_{\pm}\,$ for the dual operators given by
\begin{equation}
\label{Deltas_SU2xU1_scalars_chi=0}
\resizebox{\hsize}{!}{$
\begin{array}{rrrrrrrrrrrrrr}
m^2 L^2 &=&  -2 \, (\times 4)  &,&  3 - \sqrt{17} \, (\times 2) &;&  0 \, (\times 2) &;&  2 \, (\times 4) &,&  3 + \sqrt{17}  \, (\times 2) & , \\[3mm]
\Delta_{+} &=&  \textbf{2} \, (\times 2)  &,&  \frac{ \textbf{1} }{ \textbf{2} } (\textbf{1}+\sqrt{\textbf{17}})  \, (\times 2) &;&  3 &;&   \frac{ \textbf{1} }{ \textbf{2} } (\textbf{3}+\sqrt{\textbf{17}})  \, (\times 2)  & ,  & \frac{ 1 }{ 2 } ( 5 + \sqrt{17})    & , \\[3mm]
\Delta_{-}  &=&   \textbf{1} \, (\times 2)  &,&    \frac{1}{2} (5-\sqrt{17})  &;& \textbf{0} \, (\times 2)  &;&   \frac{ \textbf{1} }{ \textbf{2} } (\textbf{3} - \sqrt{\textbf{17}})  \, (\times 2)  & ,  & \frac{ \textbf{1} }{ \textbf{2} } (\textbf{1} - \sqrt{\textbf{17}})  \, (\times 2)  & .	
\end{array}$}
\end{equation}
As in the previous case, some of the conformal dimensions in (\ref{Deltas_SU2xU1_scalars_chi=0}) have been highlighted as they will play a role later on when studying holographic RG flows involving this conformal fixed point.

\subsubsection{$\mathcal{N}=4\,$ vacuum with $\,\textrm{SO}(4)\,$ symmetry}
\label{sec:N=4_AdS4_vacua}

There is an $\,\mathcal{N}=4\,$ supersymmetric AdS$_{4}$ solution that preserves $\,\textrm{SO}(4)\,$. It is located at
\begin{equation}
\label{VEVs_z_N4}
z_{1}=z_{2}=z_{3} =  i \, c
\hspace{10mm} \textrm{and} \hspace{10mm}
z_{4}=z_{5}=z_{6}=-\bar{z}_{7}=  \frac{1}{\sqrt{2}} ( 1+ i ) \ ,
\end{equation}
so that $\,|z_{4,5,6}|=|z_{7}|=1\,$. This AdS$_{4}$ solution has a vacuum energy given by
\begin{equation}
V_{0} = -3 \,\, g^2 \, c^{-1} \ ,
\end{equation}
as for the previous solution, and a spectrum of $\,\mathbb{Z}_{2}^{3}$-invariant normalised scalar masses of the form
\begin{equation}
\label{spectrum_SO4_scalars}
m^2 L^2 \,\,=\,\,
-2 \, ( \times 3)
 \hspace{5mm} , \hspace{5mm}
0 \, ( \times 6)
\hspace{5mm} , \hspace{5mm}
 4 \, ( \times 4)
\hspace{5mm} , \hspace{5mm}
10 \, ( \times 1)  \ ,
\end{equation}
where $\,L^2=-3/V_{0}\,$ is the AdS$_{4}$ radius.

This $\,\mathcal{N}=4 \,\, \& \,\, \textrm{SO}(4)\,$ symmetric AdS$_4$ vacuum was first reported in \cite{Gallerati:2014xra}, and then uplifted to a family of type IIB S-folds with $\,\mathcal{N}=4\,$ supersymmetry in \cite{Inverso:2016eet}. Solving (\ref{Deltas_eq}) for the set of normalised scalar masses in (\ref{spectrum_SO4_scalars}) yields a set of conformal dimensions $\,\Delta_{\pm}\,$ for the dual operators given by
\begin{equation}
\label{Deltas_SO4_scalars}
\begin{array}{rrrrrrrrrrrrrr}
m^2 L^2 &=&  -2 \, (\times 3)  &;&  0 \, (\times 6) &;&  4 \, (\times 4) &,&   10 \, (\times 1) & , \\[3mm]
\Delta_{+} &=&  \textbf{2} \, (\times 3)  &;&  \textbf{3}  \, (\times 3) &;&  \textbf{4} \, (\times 1) &,&  5 & , \\[3mm]
\Delta_{-}  &=&  1 &;&   \textbf{0}  \, (\times 3) &;&  -\textbf{1}  \, (\times 3) &,&  -\textbf{2}  \, (\times 1) & .
\end{array}
\end{equation}
As in the previous cases, some of the conformal dimensions in (\ref{Deltas_SO4_scalars}) have been highlighted as they will play a role later on when studying holographic RG flows involving this conformal fixed point.

\begin{table}[t]
\small{
\[
\arraycolsep=2pt\def\arraystretch{1.8}
		\begin{array}{|c|c|c|c|c|}
			\hline \phantom{aaaa} & \,\, \cN = 1\,\, \textrm{vacuum}  \,\,   &  \,\, \cN = 2 \,\, \textrm{vacuum} \,\,  & \,\,  \cN = 4 \,\, \textrm{vacuum}  \,\,  & \,\,  \text{D3-brane at } c=0 \,\, \\
			\hline \textrm{Re}z_1  & \,-\chi_1                 & -\chi                & 0                   & -\chi^{(0)}_1 \\
			       \textrm{Re}z_2  & \, -\chi_2       & 0                   & 0                   & -\chi^{(0)}_2 \\
			       \textrm{Re}z_3  & \,\chi_1+\chi_2                 & \chi               & 0                   & -\chi^{(0)}_3 \\
			\hline \textrm{Re}z_4  & \frac{1}{\sqrt{6}}   & 0                   & \frac{1}{\sqrt{2}} & 0 \\
			       \textrm{Re}z_5  & \frac{1}{\sqrt{6}}   & \frac{1}{\sqrt{2}} & \frac{1}{\sqrt{2}} & 0 \\
			       \textrm{Re}z_6  & \frac{1}{\sqrt{6}}   & 0                   & \frac{1}{\sqrt{2}} & 0 \\
			       \textrm{Re}z_7  & \frac{1}{\sqrt{6}}   & \frac{1}{\sqrt{2}} & - \frac{1}{\sqrt{2}} & 0 \\
			\hline  \textrm{Im}z_1    & c\,\frac{\sqrt{5}}{3} & c \, \frac{1}{\sqrt{2}}  & c                   & \frac{(g \, z )^2}{8} \\
						  \textrm{Im}z_2    & c\,\frac{\sqrt{5}}{3} & c                   & c                   & \frac{(g \, z )^2}{8} \\
						  \textrm{Im}z_3    & c\,\frac{\sqrt{5}}{3} & c \, \frac{1}{\sqrt{2}}  & c                   & \frac{(g \, z )^2}{8}  \\
			\hline  \textrm{Im}z_4    & \sqrt{\frac{5}{6}}    & 1                   & \frac{1}{\sqrt{2}}  & e^{- \frac{1}{2} \Phi_{0}} \\
						  \textrm{Im}z_5    & \sqrt{\frac{5}{6}}    & \frac{1}{\sqrt{2}}  & \frac{1}{\sqrt{2}}  & e^{- \frac{1}{2} \Phi_{0}} \\
						  \textrm{Im}z_6    & \sqrt{\frac{5}{6}}    & 1                   & \frac{1}{\sqrt{2}}  & e^{- \frac{1}{2} \Phi_{0}} \\
						  \textrm{Im}z_7    & \sqrt{\frac{5}{6}}    & \frac{1}{\sqrt{2}}  & \frac{1}{\sqrt{2}}  & e^{- \frac{1}{2} \Phi_{0}} \\
			\hline
							V_0   & -\frac{162}{25 \sqrt{5}} \, \, g^2 \, c^{-1} &-3 \, g^2 c^{-1} & -3 \, g^2 c^{-1} &V(z)= - g^2 \,  \frac{24}{(g z)^2}  \\
							\hline
							\,\, \# \Delta_{J} <0 \,\, & 2 & 4 & 4 & \times \\ 
			\hline
		\end{array}
\]
}
\caption{Summary of the AdS$_{4}$ supersymmetric vacua with the largest possible residual symmetry within their respective families. The VEVs of $\,z_{i}\,$ and the values of the scalar potential at the vacua are provided. In the last line, $\,\# \Delta_{J}<0\,$ denotes the number of dual irrelevant operators at each such vacua.}
\label{vacua_summary}
\end{table}

\subsection{Non-AdS$_4$ solutions and the D3-brane}
\label{sec:Non-AdS_vacua}

In this section we will obtain (semi-)analytic non-AdS$_4$ solutions of the BPS flow equations (\ref{BPS_equations}): first in the purely electric case with $\,c=0\,$, and then turning on the electromagnetic deformation $\,c\,$. 

\subsubsection{Analytic flow at $\,c=0\,$}
 
Let us first focus on the BPS equations (\ref{BPS_equations}) when the gauging in the maximal theory is purely electric, namely, $\,c=0\,$. In this case there is a simple solution of the BPS equations given by
\begin{equation}
\label{analytic_sol_c=0}
z_{1, 2, 3}= - \chi_{1,2,3}^{(0)} + i \, \dfrac{(g \, z )^2}{8}
\hspace{4mm} \textrm{ , } \hspace{4mm}
z_{4}=z_{5}=z_{6}=z_{7}= i  \, e^{- \frac{1}{2} \Phi_{0}}
\hspace{4mm} \textrm{ and } \hspace{4mm}
e^{A}=(g \, z )^3 \ ,
\end{equation}
subject to the constraint\footnote{The axions $\,\chi^{(0)}_{1,2,3}\,$ must be constant by virtue of the BPS equations (\ref{BPS_equations}) when setting $\,c=0\,$.}
\begin{equation}
\label{chi1+chi_2+chi_3=0}
\sum_{i=1}^{3} \textrm{Re}z_{i} \,\, = \,\, - \sum_{i=1}^{3} \chi^{(0)}_{i} \,\,= \,\, 0 \ ,
\end{equation}
and with $\,\Phi_{0}\,$ being an arbitrary constant. The four-dimensional solution (\ref{analytic_sol_c=0}) with arbitrary (constant) values of the axions $\,\chi^{(0)}_{1,2,3}\,$ has an uplift to a ten-dimensional background of type IIB supergravity that is locally equivalent to the D3-brane solution. More concretely, the axions $\,\chi^{(0)}_{1,2,3}\,$ can be locally reabsorbed in a reparameterisation of the angular coordinates $\,\theta_{1,2,3}\,$ along the three commuting translational (shift) isometries on $\,\textrm{S}^{5}\,$. This is explicitly shown in Section~\ref{sec:D3-brane_c=0} and further discussed in Section~\ref{sec:axions_SU3}.

It is worth mentioning that the condition (\ref{chi1+chi_2+chi_3=0}) is required by the BPS equations (\ref{BPS_equations}) but \textit{not} by the (local) second-order equations of motion that follow from the Lagrangian (\ref{L_scalars_real}). The scalar potential (\ref{V_scalar}) evaluated at the solution (\ref{analytic_sol_c=0}) yields
\begin{equation}
\label{analytic_V_c=0}
g^{-2} \, V(z)= - \frac{24}{(g z)^2}   \ ,
\end{equation}
whereas the $\,\mathcal{N}=1\,$ gravitino mass (\ref{N1_gravitino_mass}) reads
\begin{equation}
\label{analytic_m3/2_c=0}
g^{-2} \, m_{3/2}^2 = \dfrac{9}{(g z)^2} + \dfrac{64}{(g z)^6} \left(  \sum_{i=1}^{3} \chi^{(0)}_{i} \right)^2  \ ,
\end{equation}
thus being independent of the arbitrary parameter $\,\Phi_{0}\,$ in (\ref{analytic_sol_c=0}). Lastly, the constraint (\ref{chi1+chi_2+chi_3=0}) further eliminates the dependence of (\ref{analytic_m3/2_c=0}) on the axion fields $\,\chi^{(0)}_{1,2,3}\,$.

\subsubsection*{Axions and supersymmetry}

The amount of four-dimensional supersymmetry preserved by a solution can be assessed by direct evaluation of the eight gravitino masses, namely, the eigenvalues of $\, |A_{1}|^2= A_{1} A^{\dagger}_{1}\,$, where $\,A_{1}(z_{i})=A_{IJ}(z_{i})\,$ is the scalar-dependent gravitino mass matrix in the maximal theory \cite{deWit:2007mt}. Substituting the analytic BPS solution (\ref{analytic_sol_c=0}) into the expression for $\,A_{1}(z_{i})\,$ one finds a set of (normalised) eigenvalues given by
%
%
\begin{equation}
\label{analytic_m3/2_c=0_N8}
g^{-2} \, m_{3/2}^2  \,\, = \,\, g^{-2} \, \textrm{Eigen}\left( |A_{1}|^2 \right) 
\,\, = \,\, \dfrac{9}{(g z)^2} + \dfrac{64}{(g z)^6} \left(  \pm \chi^{(0)}_{1} \pm \chi^{(0)}_{2} \pm \chi^{(0)}_{3}  \right)^2 \ ,
\end{equation}
where the $\,\pm\,$ signs are not correlated. Note that the $(+,+,+)$ and $(-,-,-)$ eigenvalues in (\ref{analytic_m3/2_c=0_N8}) precisely reproduce the $\,\mathcal{N}=1\,$ gravitino mass (\ref{analytic_m3/2_c=0}) belonging to the $\,\mathbb{Z}_{2}^{3}$-invariant sector of the maximal supergravity by virtue of the constraint (\ref{chi1+chi_2+chi_3=0}). However, such an algebraic constraint does \textit{not} eliminate the dependence of the six remaining gravitino masses in (\ref{analytic_m3/2_c=0_N8}) on the axions $\,\chi^{(0)}_{1,2,3}\,$. And we have explicitly verified that the analytic flow in (\ref{analytic_sol_c=0}) with $\,\chi^{(0)}_{1,2,3} \neq 0\,$ is BPS only with respect to two gravitino masses (superpotentials), thus reducing the amount of supersymmetry of the solution by a factor of $1/4$.

\subsubsection{Semi-analytic flows at $\,c \neq 0\,$}
\label{sec:semi-analytic}

Let us now focus on the BPS equations (\ref{BPS_equations}) when the gauging in the maximal theory is of dyonic type, namely, $\,c \neq 0\,$. In this case there is no simple analytic solution of the BPS equations. However, we will be interested in perturbing the analytic solution (\ref{analytic_sol_c=0}) and solve the flow equations (\ref{BPS_equations}) order by order in powers of the deformation parameter $\,c\,$. We will refer to the resulting power series solution as the \textit{deformed} D3-brane solution.

At zeroth order the analytic solution in (\ref{analytic_sol_c=0}) is recovered, which depends on the arbitrary parameters $\,(\chi^{(0)}_{1,2,3} \,,\, \Phi_{0})\,$ subject to the constraint
\begin{equation}
\label{chi1+chi_2+chi_3=0_2}
\sum_{i=1}^{3} \chi^{(0)}_{1,2,3} = 0 \ .
\end{equation}
Following the discussion below (\ref{analytic_m3/2_c=0_N8}), we will set $\,\chi^{(0)}_{1,2,3} =0\,$ in the zeroth order solution so that the largest possible amount of supersymmetry is preserved at this order and the ten-dimensional $\,\textrm{AdS}_{5} \times \textrm{S}^{5}\,$ geometry of the D3-brane is globally recovered.

\subsubsection*{First order corrections and universality}

At first order in the deformation parameter $\,c\,$, an uneventful integration of the BPS equations in (\ref{BPS_equations}) yields
\begin{equation}
\label{analytic_sol_c}
\begin{array}{rll}
z_{1, 2, 3} &=&  c \,\, \chi^{(1)}_{1,2,3}(z)  \, + \, i \, \dfrac{(g \, z )^2}{8} \big[ \,  1 + c \,\, y^{(1)}_{1,2,3}(z) \, \big] \ ,  \\[4mm]
z_{4,5,6,7} &=&  c \,\, \chi^{(1)}_{4,5,6,7}(z)  \, + \,   i  \, e^{-\frac{1}{2} \Phi_{0}} \, \big[ \, 1 + c \,\, y^{(1)}_{4,5,6,7}(z) \, \big] \ ,  \\[4mm]
e^{A} &=& (g \, z )^3 \, \big[ \, 1 + c \,\, j(z) \, \big] \ ,
\end{array}
\end{equation}
in terms of a set of $z$-dependent functions
\begin{equation}
\label{functions_1st_order}
\begin{array}{rll}
\chi^{(1)}_{1,2,3}(z) &=&\frac{1}{3} \, \sinh\Phi_{0} - \rho_{1,2,3} - \dfrac{\lambda_{4}}{(g z)^{4}} \ ,   \\[4mm]
y^{(1)}_{1,2,3}(z) &=& \dfrac{\lambda_{1}}{g z} + \dfrac{\kappa_{1,2,3}}{(g z)^4} \ ,   \\[6mm]
\chi^{(1)}_{4,5,6,7}(z) &=&  e^{ - \frac{1}{2} \Phi_{0}}  \, \cosh\Phi_{0} \,\, \dfrac{4}{(g z)^2}  -  \dfrac{\rho_{4,5,6,7}}{(g z)^2}  -  \dfrac{\lambda_{6}}{(g z)^6} \ , \\[4mm]
y^{(1)}_{4,5,6,7}(z) &=& \lambda_{0} + \dfrac{\kappa_{4,5,6,7}}{(g z)^4} \ ,   \\[6mm]
j(z) &=& \tilde{\lambda}_{0} + \dfrac{3}{2}  \, \dfrac{\lambda_{1}}{g z} \ ,  
\end{array}
\end{equation}
which in turn depend on a set of integration constants 
\begin{equation}
\label{int_constant}
( \, \rho_{1,\dots,7}\,,\, \kappa_{1,\dots,7} \,)
\hspace{8mm} \textrm{ and } \hspace{8mm}
(\,   \tilde{\lambda}_{0} \,,\, \lambda_{0,1,4,6}   \,  ) \ .
\end{equation}
These are subject to the following constraints
\begin{equation}
\label{rho&kappa_constraints}
\sum_{i=1}^{3} \rho_{i} =  0 
\hspace{5mm} , \hspace{5mm} 
\sum_{i=4}^{7} \rho_{i} =  0 
\hspace{10mm} \textrm{ and } \hspace{10mm} 
\sum_{i=1}^{3} \kappa_{i} =  0
\hspace{5mm} , \hspace{5mm} 
\sum_{i=4}^{7} \kappa_{i} =  0 \ .
\end{equation}
It is worth noticing that the $c$-deformed solution in (\ref{analytic_sol_c})-(\ref{functions_1st_order}) does \textit{not} reduce to the purely electric ($c=0$) solution in (\ref{analytic_sol_c=0})-(\ref{chi1+chi_2+chi_3=0}) upon adjustment of the integration constants (\ref{int_constant}). Note the obstruction to have $\,\chi^{(1)}_{4,5,6,7}(z)=0\,$ and $\,\chi^{(1)}_{1,2,3}(z)=0\,$ (the latter whenever $\Phi_{0} \neq 0$) caused by the first two constraints in (\ref{rho&kappa_constraints}). This in turn implies a generic breaking of the $\,\textrm{SO}(6)\,$ symmetry of (\ref{analytic_sol_c=0}) when $\,\chi^{(0)}_{1,2,3}=0\,$ down to an $\,\textrm{SU}(3) \subset \textrm{SO}(6)\,$ subgroup. Moreover, while the parameters $\,(\tilde{\lambda}_{0} \, , \, \lambda_{0,1,4,6} )\,$ respect such an $\,\textrm{SU}(3)\,$ symmetry, the parameters $\,(\rho_{1,\dots,7}\,,\, \kappa_{1,\dots,7})\,$ do not, thus causing a further breaking of flavour symmetries in the dual field theory.

In the following we will analyse in more detail the case where all the integration constants are set to zero, both flavour breaking and SU(3)-preserving constants in (\ref{int_constant}). Then the solution (\ref{analytic_sol_c})-(\ref{functions_1st_order}) acquires a \textit{universal} form (to first order in the parameter $\,c\,$) given by
\begin{equation}
\label{analytic_sol_c_no_integration_constants}
\begin{array}{rll}
z_{1, 2, 3} &=&  \frac{1}{3} \, c \, \sinh\Phi_{0} \, + \, i \, \dfrac{(g \, z )^2}{8} \ ,  \\[4mm]
z_{4,5,6,7} &=&  4 \, e^{ - \frac{1}{2} \Phi_{0}}  \, \cosh\Phi_{0} \,\, \dfrac{c}{(g z)^2} \, + \,   i  \, e^{-\frac{1}{2} \Phi_{0}} \ ,  \\[4mm]
e^{A} &=& (g \, z )^3  \ ,
\end{array}
\end{equation} 
which necessarily induces a deviation from (\ref{analytic_sol_c=0}) that is linear in the parameter $\,c\,$ and sub-leading around $\,(gz) \rightarrow \infty\,$ (UV). However, (\ref{analytic_sol_c_no_integration_constants}) does not capture corrections in $\,\textrm{Im}z_{1,2,3}\,$, $\,\textrm{Im}z_{4,5,6,7}\,$ or the scale factor $\,e^{A}\,$. To get those one must go to higher-orders in $\,c\,$. Finally, it also follows from (\ref{analytic_sol_c_no_integration_constants}) that
\begin{equation}
\label{Re_chi1,2,3_c}
\sum_{i=1}^{3} \textrm{Re} z_{i} = c \, \sinh\Phi_{0}   \ ,
\end{equation}
in contrast to the relation (\ref{chi1+chi_2+chi_3=0}) obtained at $\,c=0\,$.

\subsubsection*{Higher-order universal corrections}

The power series procedure can be iterated to solve the BPS equations (\ref{BPS_equations}) to any desired order in the deformation parameter $\,c\,$. Setting all the integration constants that appear to zero, the general structure of the \textit{universal} $n$th-order solution is
\begin{equation}
\label{analytic_sol_cn_no_integration_constants}
\begin{array}{rll}
z_{1, 2, 3} &=&  \frac{1}{3} \, c \, \sinh\Phi_{0} \, \left(  1-  384 \, \cosh^2\Phi_{0}  \, \mu^2 \, \log(g z)  + \displaystyle\sum_{n=2}^{\infty}  \displaystyle\sum_{m=0}^{n-1}  \tilde{f}_{n,m}(\Phi_{0}) \, \, \mu^{2n}  \, \log^{m}(g z)\right)  \\[4mm]
& + &  i \, \dfrac{(g \, z )^2}{8} \left( 1 + 32 \, \cosh^2\Phi_{0}  \, \mu^2 + \displaystyle\sum_{n=2}^{\infty}  \displaystyle\sum_{m=0}^{n-1}  f_{n,m}(\Phi_{0}) \, \, \mu^{2n}  \, \log^{m}(g z)   \right) \ ,  \\[8mm]
z_{4,5,6,7} &=&  4 \, e^{ - \frac{1}{2} \Phi_{0}}  \, \cosh\Phi_{0} \,\, \mu \, \left(  1 + 64 \, \Big(1-3 \cosh(2\Phi_{0}) \Big)  \, \mu^2 \, \log(g z)  \right. \\
&& \left. \quad\quad\quad\quad\quad\quad\quad\quad\quad\quad \,\,\,\,\,\,\, + \displaystyle\sum_{n=2}^{\infty}  \displaystyle\sum_{m=0}^{n-1}  \tilde{g}_{n,m}(\Phi_{0}) \, \, \mu^{2n}  \, \log^{m}(g z) \right) \\[4mm] 
&+&    i  \, e^{-\frac{1}{2} \Phi_{0}}  \left( 1 -  8 \, \left(\cosh^2\Phi_{0} - 2 \sinh(2 \, \Phi_{0}) \, \right)  \, \mu^2
 + \displaystyle\sum_{n=2}^{\infty}  \displaystyle\sum_{m=0}^{n-1}  g_{n,m}(\Phi_{0}) \, \, \mu^{2n}  \, \log^{m}(g z) \right) \ ,  \\[8mm]
e^{A} &=& (g \, z )^3 \, \left( 1 + 16 \, \cosh^2\Phi_{0}  \, \mu^2  
+ \displaystyle\sum_{n=2}^{\infty}  \displaystyle\sum_{m=0}^{n-1}  j_{n,m}(\Phi_{0}) \, \, \mu^{2n}  \, \log^{m}(g z)  \right)  \ .
\end{array}
\end{equation} 
Note that (\ref{analytic_sol_cn_no_integration_constants}) actually becomes an expansion in powers of the quantity
\begin{equation}
\mu \equiv \frac{c}{(gz)^2} \ ,
\end{equation}
around $\,\mu=0\,$. Therefore, the original expansion in powers of $\,c\,$ around $\,c=0\,$ can be re-interpreted as an expansion in powers of $\,c/(gz)^2\,$ around the UV ($\,z \rightarrow \infty\,$).

Although we are not displaying the $\,\mu^{2n}\,$ higher-order corrections with $\,n \ge 2\,$ (it turns out that some of them vanish identically), we have explicitly computed the solution up to $\,n=6\,$. At quadratic order, two corrections of the form $\,\mu^2 \, \log{(gz)}\,$
involving non-vanishing functions of $\,\Phi_{0}\,$ appear in the axions $\,\textrm{Re}z_{1,2,3}\,$ and $\,\textrm{Re}z_{4,5,6,7}\,$. At quartic order, three corrections of the form $\,\mu^4 \, \log{(gz)}\,$ involving three different non-vanishing functions $\,f_{2,1}(\Phi_0)\,$, $\,g_{2,1}(\Phi_{0})\,$ and $\,j_{2,1}(\Phi_0) \,$ appear in $\,\textrm{Im}z_{1,2,3}\,$, $\,\textrm{Im}z_{4,5,6,7}\,$ and the scale factor $\,e^{A}\,$, respectively. These are the relevant orders at which the logarithms enter the universal solution to (\ref{BPS_equations}) when $\,c \neq 0\,$.

In what follows we will consider a truncation of the universal solution (\ref{analytic_sol_cn_no_integration_constants}) to cubic order in the deformation parameter $\,c\,$, namely,
\begin{equation}
\label{analytic_sol_c3_no_integration_constants}
\begin{array}{rll}
z_{1, 2, 3} &=&  \frac{1}{3} \, c \, \sinh\Phi_{0} \, \left(  1- 384 \, \cosh^2\Phi_{0}  \, \dfrac{c^2}{(g \, z)^4} \, \log(g z)\right)  \\[4mm]
& + &  i \, \dfrac{(g \, z )^2}{8} \left( 1 + 32 \, \cosh^2\Phi_{0}  \, \dfrac{c^2}{(g \, z)^4} \right) \ ,  \\[6mm]
z_{4,5,6,7} &=&  4 \, e^{ - \frac{1}{2} \Phi_{0}}  \, \cosh\Phi_{0} \,\, \dfrac{c}{(g z)^2} \, \left(  1 + 64 \, \Big(1-3 \cosh(2\Phi_{0}) \Big)  \, \dfrac{c^2}{(g \, z)^4} \, \log(g z) \right) \\[4mm] 
&+&    i  \, e^{-\frac{1}{2} \Phi_{0}}  \left( 1 -  8 \, \left(\cosh^2\Phi_{0} - 2 \sinh(2 \, \Phi_{0}) \, \right)  \, \dfrac{c^2}{(g \, z)^4} \right) \ ,  \\[6mm]
e^{A} &=& (g \, z )^3 \, \left( 1 + 16 \, \cosh^2\Phi_{0}  \, \dfrac{c^2}{(g \, z)^4} \right)  \ .
\end{array}
\end{equation} 
This order suffices to capture the first relevant terms in each of the scalar fields as well as in the scale factor for the deformed D3-brane solution. From (\ref{analytic_sol_c3_no_integration_constants}) one has that
\begin{equation}
\label{Re_chi1,2,3_c3}
\sum_{i=1}^{3} \textrm{Re} z_{i} = c \, \sinh\Phi_{0} \, \left(  1- 384 \, \cosh^2\Phi_{0}  \, \dfrac{c^2}{(g \, z)^4} \, \log(g z)\right)   \ ,
\end{equation}
which picks up a dependence on the coordinate $\,z\,$ in contrast to the relation (\ref{Re_chi1,2,3_c}) obtained at linear order in $\,c=0\,$.

\section{Holographic RG flows}
\label{sec:RG flows}

In this section we numerically construct BPS domain-wall solutions that interpolate between the supersymmetric AdS$_4$ vacua of Section~\ref{sec:AdS_vacua} in the IR ($\,z \rightarrow -\infty\,$) and the non-AdS$_4$ solution of Section~\ref{sec:Non-AdS_vacua} with $\,c \neq 0\,$ in the UV  ($\,z \rightarrow \infty\,$). We will also present an example of a domain-wall that interpolates between the AdS$_4$ vacuum with $\,\mathcal{N}=2 \, \& \, \textrm{SU}(2) \times \textrm{U}(1)\,$ symmetry in the IR and the AdS$_4$ vacuum with $\,\mathcal{N}=1 \, \& \, \textrm{SU}(3)\,$ symmetry in the UV. All these domain-walls in supergravity correspond to holographic RG flows in the field theory side.

\subsubsection*{Boundary conditions}

Let us start discussing the system of first-order and non-linear differential equations in (\ref{BPS_equations}). The set of equations for the scalars can be solved independently of the one for the scale factor, which can be readily integrated once the profiles for the scalars are known. This means that we must set one boundary condition per (real) scalar field. Moreover, it proves more efficient to numerically shoot from the IR and flow up to the UV.

Perturbing around an AdS$_4$ configuration dual to the J-fold CFT$_{3}$ in the deep IR with scalar VEVs $\,\Sigma^{(0)  I} \,$ translates into a choice of boundary conditions of the form\footnote{To be fully consistent with the forthcoming sections, we are denoting by $\,\Sigma^{I}\,$ either an axion $\,\chi_{i}\,$ or its corresponding dilaton $\,\log(y_{i})\,$ in the parameterisation (\ref{Phi_def_7}).}
\begin{equation}
\label{boundary_cond_IR}
\Sigma^I = \Sigma^{(0)  I} \, \left( \, 1 + \lambda^{I}{}_{J} \, e^{-\Delta_J \, \frac{z}{L}} \, \right) \ ,
\end{equation}
restricted to the set of modes with $\,\Delta_J <0\,$ in the AdS$_4$ spectrum, as demanded by regularity of the flow in the deep IR ($z\rightarrow -\infty$). After imposing \eqref{boundary_cond_IR}, the BPS equations will determine the set of permitted $\,\lambda^{I}{}_{J}\,$. However, generic values of the permitted $\,\lambda^{I}{}_{J}\,$ will end up in a singular flow for which some scalars diverge at a finite radial distance. We will thus have to perform a scanning of the $\,\lambda^{I}{}_{J}\,$ parameter space in order to determine the region yielding regular flows between the IR and the UV. In summary, a given choice of parameters $\,\lambda^{I}{}_{J}\,$ in the IR boundary conditions (\ref{boundary_cond_IR}) will translate into a specific choice of parameters $\,(\chi^{(0)}_{i}\,,\,\Phi_{0}\,,\,\rho_{1,2,3}\,,\,\lambda_{0}\,,\,\tilde{\lambda}_{0})\,$, as well as of subleading ones $\,(\kappa_{1,\ldots,7}\,,\,\rho_{4,\ldots,7}\,,\,\lambda_{1}\,,\,\lambda_{4}\,,\,\lambda_{6})\,$, in the first-order deformed D3-brane solution (\ref{analytic_sol_c})-(\ref{functions_1st_order}) around the deep UV.

\subsection{SYM$_{4}$ to CFT$_{3}$ with $\,\mathcal{N}=1 \, \& \, \textrm{SU}(3)\,$}
\label{sec:N=1_SYM}

We will solve the BPS equations (\ref{BPS_equations}) numerically by perturbing around the $\,\mathcal{N}=1 \, \& \, \textrm{SU}(3)\,$ AdS$_{4}$ vacuum (\ref{VEVs_z_N1}) in the IR ($z\rightarrow -\infty$). This will generate generic flows towards a non-conformal behaviour in the UV ($z\rightarrow \infty$). 

\subsubsection*{IR boundary conditions}

Around the $\,\mathcal{N}=1 \, \& \, \textrm{SU}(3)\,$ solution, there are two irrelevant modes in the spectrum (\ref{Deltas_SU3_scalars_chi=0})  
\begin{equation}
\Delta_{-} =  1 - \sqrt{6}  \,\, ( \times 2 ) \ ,
\end{equation}
that are compatible with regularity of the flows in the IR ($\,z\rightarrow -\infty\,$). The linearised BPS equations then allow for two arbitrary real parameters $\,(\Lambda \,,\,\lambda)\,$ specifying the IR boundary conditions  (\ref{boundary_cond_IR}), which read
\begin{equation}
\label{boundary_cond_N1}
\begin{array}{rlll}
\textrm{Im}z_{1,2,3} &=& c  \, \frac{\sqrt{5}}{3} \left( 1 - \Lambda \, e^{-(1-\sqrt{6}) \frac{z}{L}} \right) & ,  \\[3mm]
\textrm{Im}z_{4,5,6,7} &=& \sqrt{\frac{5}{6}} \left( 1 - \frac{1}{4} \big( 3 \, (2+\sqrt{6}) \, \lambda + ( \sqrt{6} -2 ) \, \Lambda\big)  e^{-(1-\sqrt{6}) \frac{z}{L}} \right) & , \\[3mm]
\textrm{Re}z_{1,2,3} &=& c \, \lambda \,e^{-(1-\sqrt{6}) \frac{z}{L}} & , \\[3mm]
\textrm{Re}z_{4,5,6,7} &=& \frac{1}{\sqrt{6}} - \frac{1}{12} \, \Big( 3 \left(\sqrt{6}+3\right) \lambda +5 \left(\sqrt{6}-3\right) \Lambda  \Big)   e^{-(1-\sqrt{6}) \frac{z}{L}} & .
\end{array}
\end{equation}
Note that, whenever non-vanishing, one of the parameters $\,\Lambda\,$ or $\,\lambda\,$ can be set at will by a shift on the coordinate $\,z\,$. We will set $\,\Lambda = -1\,$ without loss of generality\footnote{Setting $\,\Lambda = 0,1\,$ does not produce regular flows.}, which translates into a one-dimensional parameter space to be scanned.

\subsubsection*{Behaviour of the flows}

\begin{figure}[t]
\begin{center}
\mbox{\includegraphics[width=0.45\textwidth]{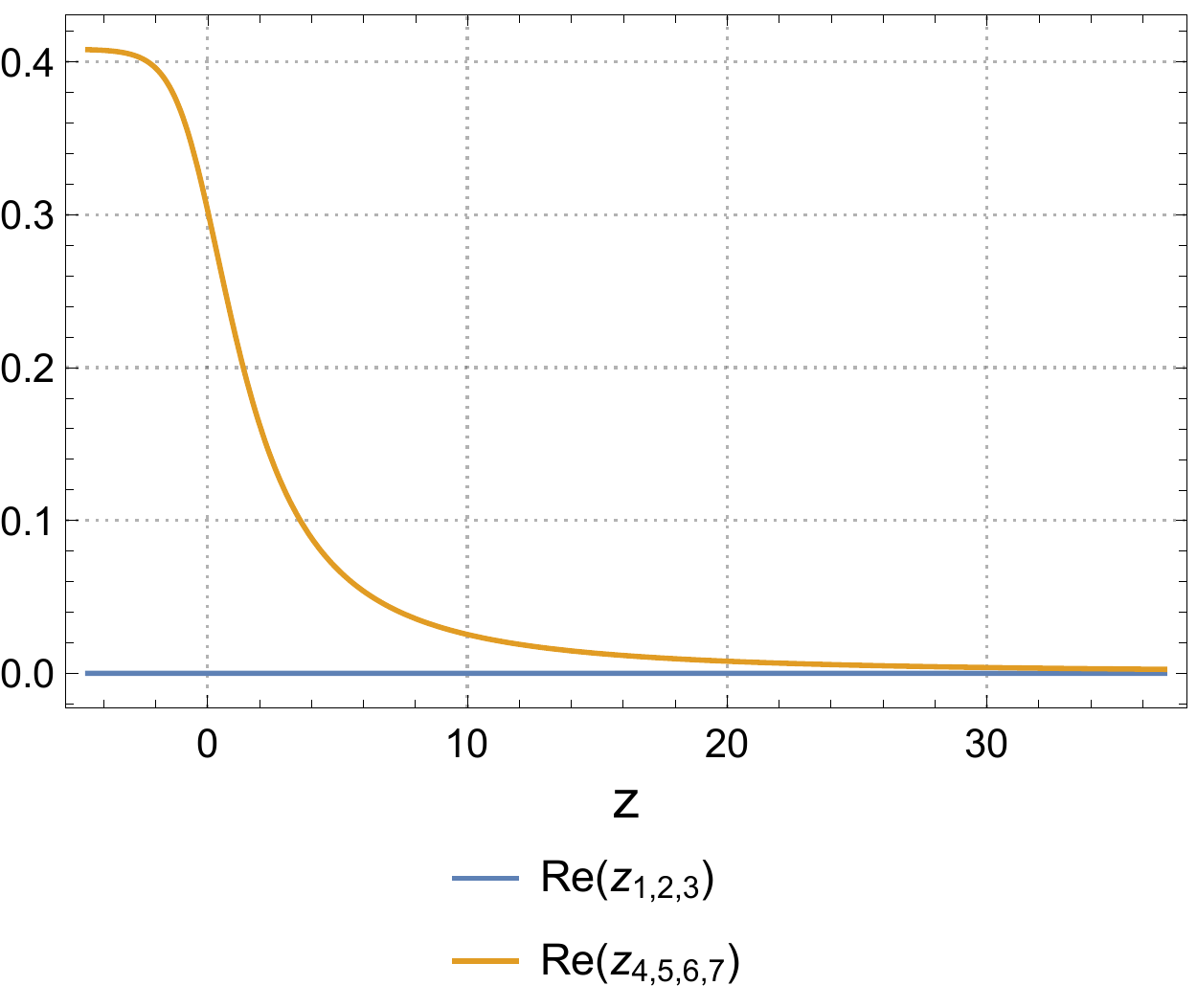}
\hspace{6mm}
\includegraphics[width=0.45\textwidth]{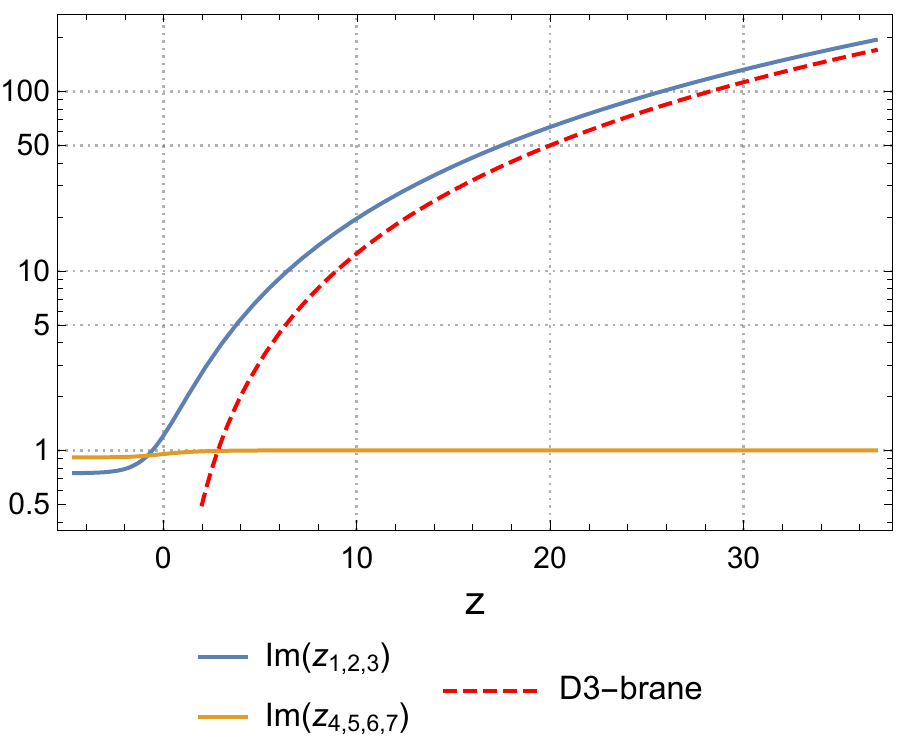}}
\caption{Holographic RG flow from $\,\mathcal{N}=4\,$ SYM$_{4}$ (UV, right) to $\,\mathcal{N}=1\, \& \, \textrm{SU}(3)\,$  J-fold CFT$_{3}$ (IR, left) with $\,\Lambda = -1\,$ and $\,\lambda = 0\,$.}
	\label{fig:Plot_N1_sym}
\end{center}
\end{figure}
\begin{figure}[t]
\begin{center}
\mbox{\includegraphics[width=0.45\textwidth]{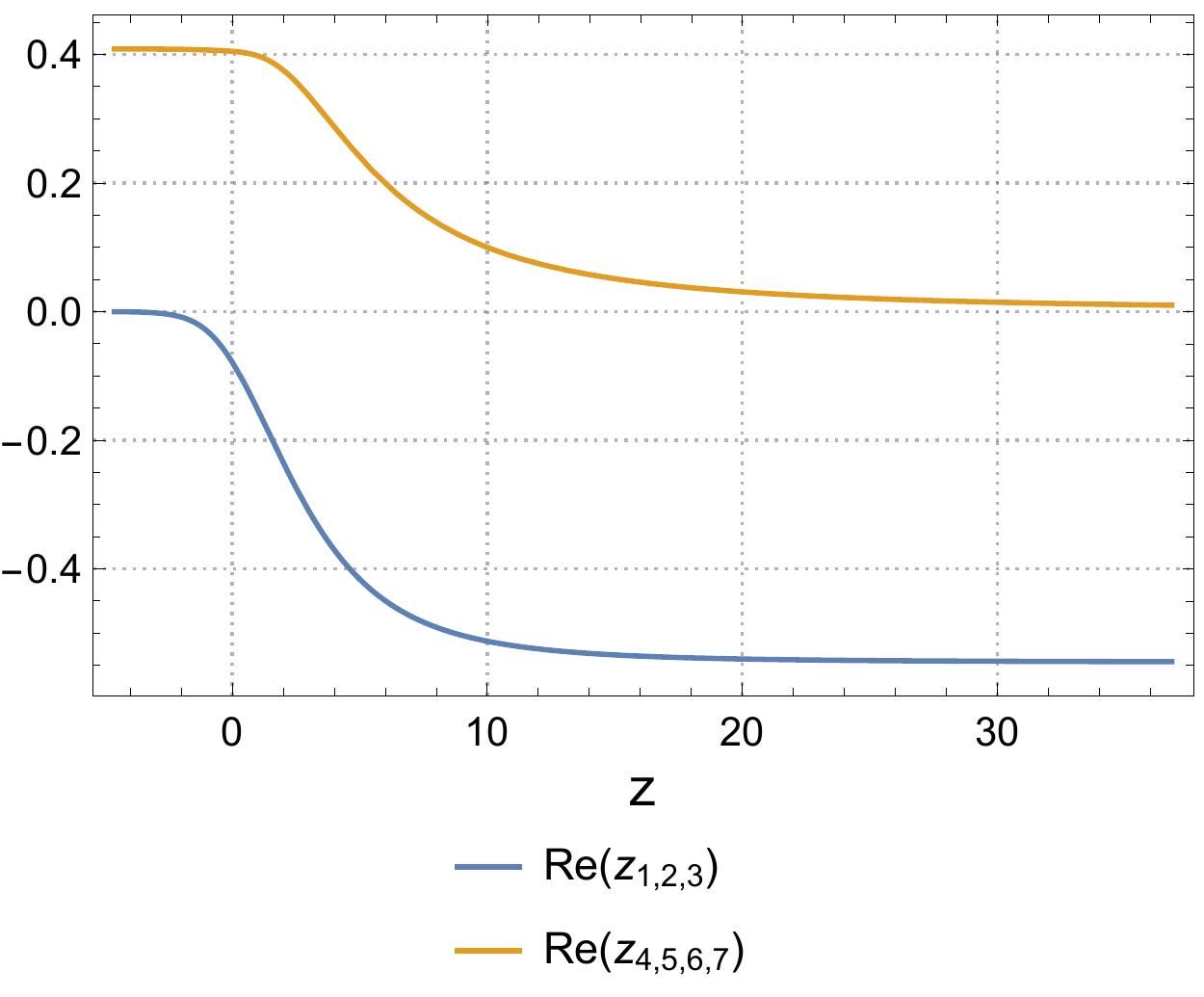}
\hspace{6mm}
\includegraphics[width=0.45\textwidth]{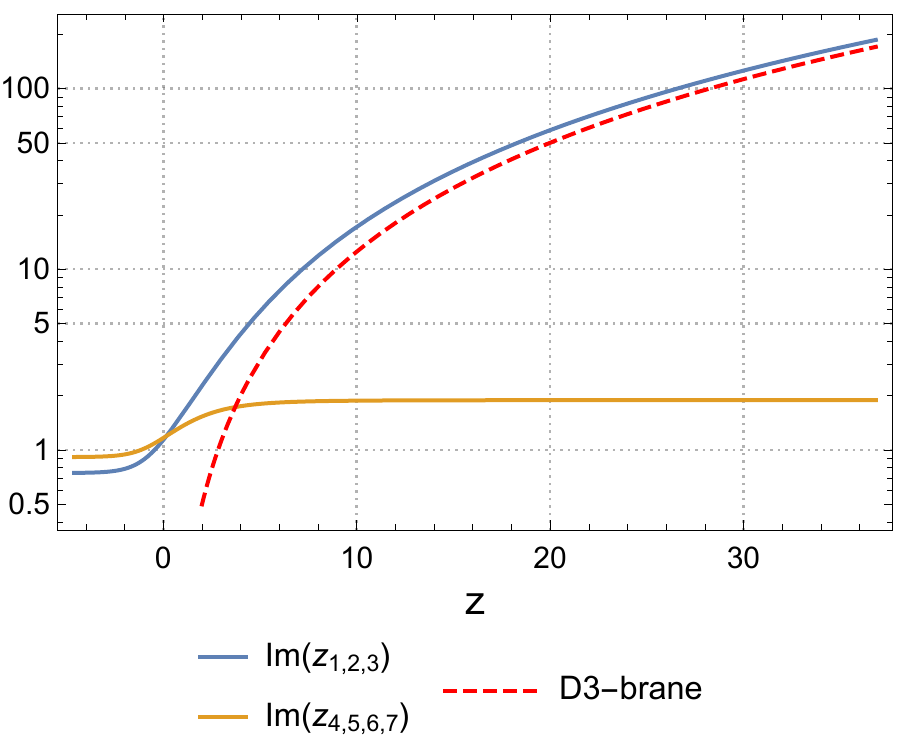}}
\caption{Holographic RG flow from $\,\mathcal{N}=4\,$ SYM$_{4}$ (UV, right) to $\,\mathcal{N}=1\, \& \, \textrm{SU}(3)\,$  J-fold CFT$_{3}$ (IR, left) with $\,\Lambda = -1\,$ and $\,\lambda = -0.16\,$.}	
\label{fig:Plot_N1_gen}
\end{center}
\end{figure}

Fixing $\,\lambda=0\,$ implies $\,{\textrm{Re}z_{1,2,3}}=0\,$ in the IR boundary conditions (\ref{boundary_cond_N1}). In this case we obtain the numerical flow\footnote{All the figures presented in this work are produced by setting the initial value of the radial coordinate to $\,z_{\textrm{ini}}=\log(10^{-2})\,$. Note that this value can be set at will by virtue of (\ref{boundary_cond_IR}).} depicted in Figure~\ref{fig:Plot_N1_sym} that approaches the deformed D3-brane solution in the UV ($z \rightarrow \infty$). As previously discussed in Section~\ref{sec:semi-analytic}, the UV behaviour of this flow is understood as a sub-leading correction in the electromagnetic deformation $\,c\,$ of the D3-brane solution in (\ref{analytic_sol_c=0}) with 
\begin{equation}
\textrm{Re}z_{1,2,3} = 0
\hspace{10mm} \textrm{ and } \hspace{10mm}
\Phi_{0} = 0 \ .
\end{equation}
Activating the parameter $\,\lambda\,$ makes the axions $\,\textrm{Re}z_{1,2,3}\,$ run along the flow. In this case, the UV region is reached with
\begin{equation}
\label{new_UV_N1}
\displaystyle\sum_{i=1}^{3} \textrm{Re}z_{i}  \approx   c \, \sinh \Phi_{0} 
\hspace{5mm} , \hspace{5mm}
\textrm{Im}z_{4,5,6,7}  \approx   e^{- \frac{1}{2} \Phi_{0}}
\hspace{5mm} \textrm{ and } \hspace{5mm}
\Phi_{0} \neq 0 \ .
\end{equation}
This agrees with (\ref{Re_chi1,2,3_c}) obtained at first-order in the deformation parameter $\,c\,$. One such generic flows is depicted in Figure~\ref{fig:Plot_N1_gen}.

\subsubsection*{Study of the parameter space}

We have performed a numerical scanning of values for the parameter $\,\lambda\,$, and found regular flows only within the interval
\begin{equation}
\label{lambda_range_N1}
\lambda \in \left[ \, -0.171 \,,\,   0.171  \, \right] \ .
\end{equation}
Outside this range, singular flows occur with some scalar fields diverging at a finite radial distance.

\subsubsection*{Gravitino masses and supersymmetry}

The $\,\mathcal{N}=1\,\&\,\textrm{SU}(3)\,$ AdS$_{4}$ vacuum in the IR realises an $\,\textrm{SU}(3)\,$ flavour symmetry group in the dual J-fold CFT$_{3}$. Under this $\,\textrm{SU}(3)\,$ symmetry the eight gravitini of the maximal theory decompose as in (\ref{Branching_SU3}), namely,
\begin{equation}
\label{Branching_4D_gravitini_SU3_flows}
\small{
\begin{array}{ccl}
\textrm{SU}(8)  &\supset & \textrm{SU}(3) \\[2mm]
\textbf{8} & \rightarrow &   \textbf{1}   \, + \, \textbf{3} \, + \, \bar{\textbf{3}}   \, + \, \textbf{1}
\end{array}
}
\end{equation}
For generic BPS flows with the parameter $\,\lambda\,$ within the range (\ref{lambda_range_N1}), an explicit evaluation of the eight gravitino masses, namely the eigenvalues of $\,A_{IJ} \, A^{JK}\,$, shows that only one of them is compatible with the IR value of the $\,\mathcal{N}=1\,$ gravitino mass (\ref{N1_gravitino_mass}) belonging to the $\,\mathbb{Z}_{2}^{3}$-invariant sector. More concretely, the eight gravitino masses of the maximal theory turn out to split as
\begin{equation}
\label{8_decomposition_N1}
\begin{array}{ccl}
\lambda \neq 0 & \hspace{5mm} : & \hspace{5mm}
8 \rightarrow 1 + 3 + 3 + \boxed{\color{blue}{1}} \ , \\[2mm]
\lambda = 0 & \hspace{5mm} : & \hspace{5mm}
8 \rightarrow 1 + 6 + \boxed{\color{blue}{1}} \ ,
\end{array}
\end{equation}
when evaluated along the numerical flows. In (\ref{8_decomposition_N1}) we have boxed the $\,\mathcal{N}=1\,$ supersymmetry realised at the AdS$_{4}$ vacuum in the deep IR and highlighted (in blue) those gravitino masses with respect to which the numerical flows are BPS. Note that the IR boundary conditions in (\ref{boundary_cond_N1}) are compatible with an $\,\textrm{SO}(6)\,$ symmetry when $\,\lambda=0\,$ (so that $\,\textrm{Re}z_{1,2,3}=0\,$), thus yielding the decomposition in (\ref{8_decomposition_N1}).

\subsection{SYM$_{4}$ to CFT$_{3}$ with $\,\mathcal{N}=2 \, \& \, \textrm{SU}(2)\,$}
\label{sec:N=2_SYM}

\begin{figure}[t]
\begin{center}
\mbox{\includegraphics[width=0.45\textwidth]{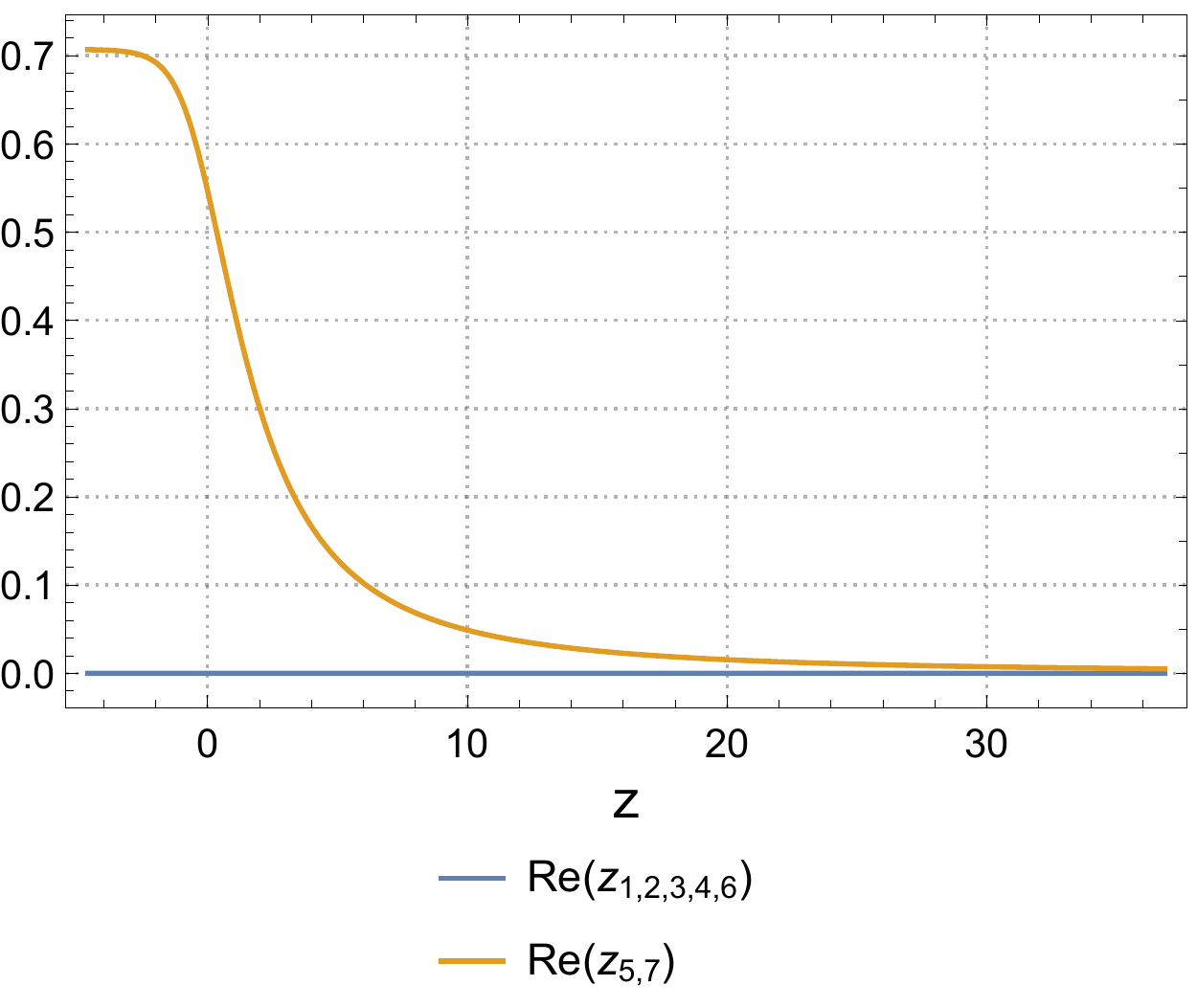}
\hspace{6mm}
\includegraphics[width=0.45\textwidth]{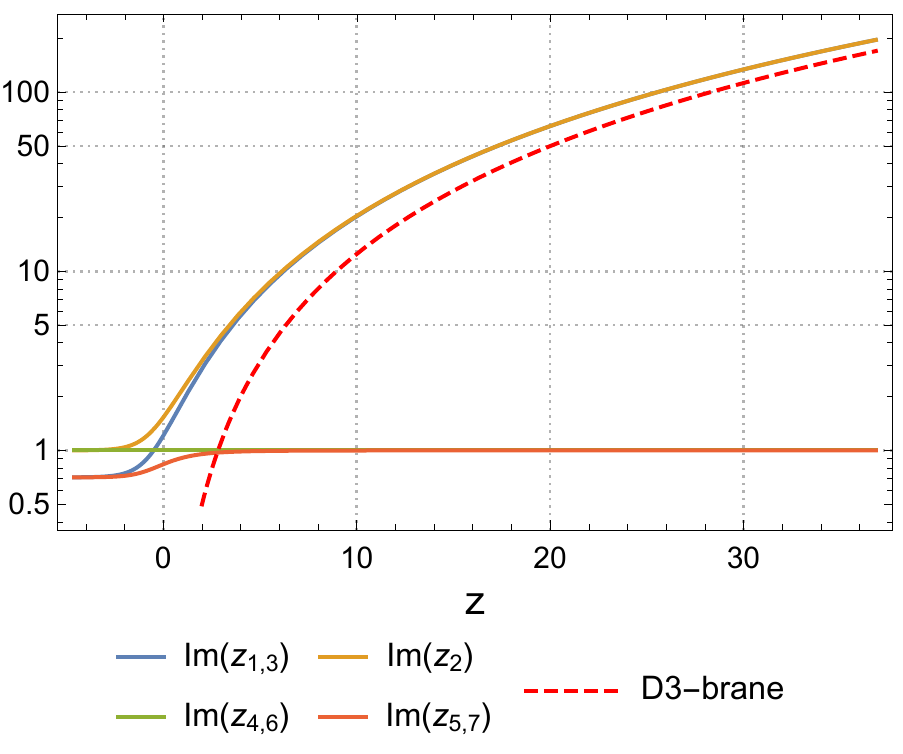}}	
	\caption{Holographic RG flow from $\,\mathcal{N}=4\,$ SYM$_{4}$ (UV) to $\,\mathcal{N}=2\, \&\,\textrm{SU}(2)\,$ J-fold CFT$_{3}$ (IR) with $\,\Lambda_1 = -1\,$ , $\,\Lambda_{2} = 0\,$ and $\,(\lambda_{1},\lambda_{2}) = (0,0)\,$.}
	\label{fig:Plot_N2_sym}
\end{center}
\end{figure}

Let us now solve the BPS equations (\ref{BPS_equations}) by perturbing around the $\,\mathcal{N}=2 \, \& \, \textrm{SU}(2) \times \textrm{U}(1)\,$ AdS$_{4}$ vacuum (\ref{VEVs_z_N2}) in the IR ($z\rightarrow -\infty$). This will cause again the appearance of generic flows towards a non-conformal behaviour in the UV ($z\rightarrow \infty$).

\subsubsection*{IR boundary conditions}

Around the $\,\mathcal{N}=2 \, \& \, \textrm{SU}(2) \times \textrm{U}(1)\,$ solution, there are four irrelevant modes in the spectrum (\ref{Deltas_SU2xU1_scalars_chi=0})
\begin{equation}
\Delta_{-} =  {\tfrac{1}{2} \, (1 - \sqrt{17}) } \,\, ( \times 2 )
\hspace{8mm} \textrm{ and } \hspace{8mm}
\Delta_{-} =  {\tfrac{1}{2} \, (3 - \sqrt{17}) } \,\, ( \times 2 ) \ ,
\end{equation}
that are compatible with regularity of the flows in the IR ($\,z\rightarrow -\infty\,$). The linearised BPS equations then allow for four parameters $\,(\Lambda_1,\,\Lambda_{2})\,$ and $\,(\lambda_{1},\,\lambda_{2})\,$  specifying the IR boundary conditions  (\ref{boundary_cond_IR}) which read
\begin{equation}
\label{boundary_cond_N2}
\begin{array}{rlll}
\textrm{Im}z_{1,3}    &=& c \, \frac{1}{\sqrt{2}} \left(1 - \frac{1+\sqrt{17}}{4}  \, \Lambda_1\, e^{-\frac{1}{2} (1-\sqrt{17}) \frac{z}{L}}\right) & , \\[3mm]
\textrm{Im}z_{2}    &=& c \left(1-\Lambda_1 \, e^{-\frac{1}{2} \left(1-\sqrt{17}\right) \frac{z}{L}} - \lambda_2\, e^{-\frac{1}{2}\left(3-\sqrt{17}\right) \frac{z}{L}}\right) & , \\[3mm]
\textrm{Im}z_{4,6}   &=& 1 + \frac{1+\sqrt{17}}{4} \, \Lambda_2 \,  e^{-\frac{1}{2}\left(1-\sqrt{17}\right)\frac{z}{L}} & , \\[2mm]
\textrm{Im}z_{5,7}   &=& \frac{1}{2\sqrt{2}}\left(2+(\Lambda_2-\Lambda_1 ) \, e^{-\frac{1}{2} \left(1-\sqrt{17}\right)\frac{z}{L}}+(\lambda_2-\lambda_1) \, e^{-\frac{1}{2}\left(3-\sqrt{17}\right) \frac{z}{L}}\right) & , \\[3mm]
\textrm{Re}z_{1,3} &=& - c \,\, \frac{1-\sqrt{17}}{4 \, \sqrt{2}} \,\, \lambda_1 \,\, e^{-\frac{1}{2} \left(3-\sqrt{17}\right)\frac{z}{L}} & , \\[3mm]
\textrm{Re}z_{2}  &=& - c \left(\Lambda_2\, e^{-\frac{1}{2} \left(1-\sqrt{17}\right) \frac{z}{L}}+\lambda _1\, e^{-\frac{1}{2} \left(3-\sqrt{17}\right)\frac{ z}{L}}\right) & , \\[3mm]
\textrm{Re}z_{4,6} &=& -\frac{1-\sqrt{17}}{4} \,\,  \lambda_2 \,\, e^{-\frac{1}{2} \left(3-\sqrt{17}\right)\frac{ z}{L}} & , \\[3mm]
\textrm{Re}z_{5,7} &=& \frac{1}{2\sqrt{2}} \left(2 + (\Lambda_1 + \Lambda_2) \, e^{-\frac{1}{2} \left(1-\sqrt{17}\right) \frac{z}{L}} - (\lambda_1 + \lambda_2) \,  e^{-\frac{1}{2} \left(3-\sqrt{17}\right)\frac{ z}{L}} \right) & .
\end{array}
\end{equation}
As before, and whenever non-vanishing, one of the parameters $\,\Lambda_{1,2}\,$ or $\,\lambda_{1,2}\,$ can be set at will by a shift on the coordinate $\,z\,$. We will set $\,\Lambda_{1} = -1\,$ without loss of generality\footnote{Setting $\,\Lambda_{1} = 0,1\,$ does not produce regular flows.}, which leaves us this time with a three-dimensional parameter space to be scanned.

\subsubsection*{Behaviour of the flows}

\begin{figure}[t]
\begin{center}
\includegraphics[width=0.45\textwidth]{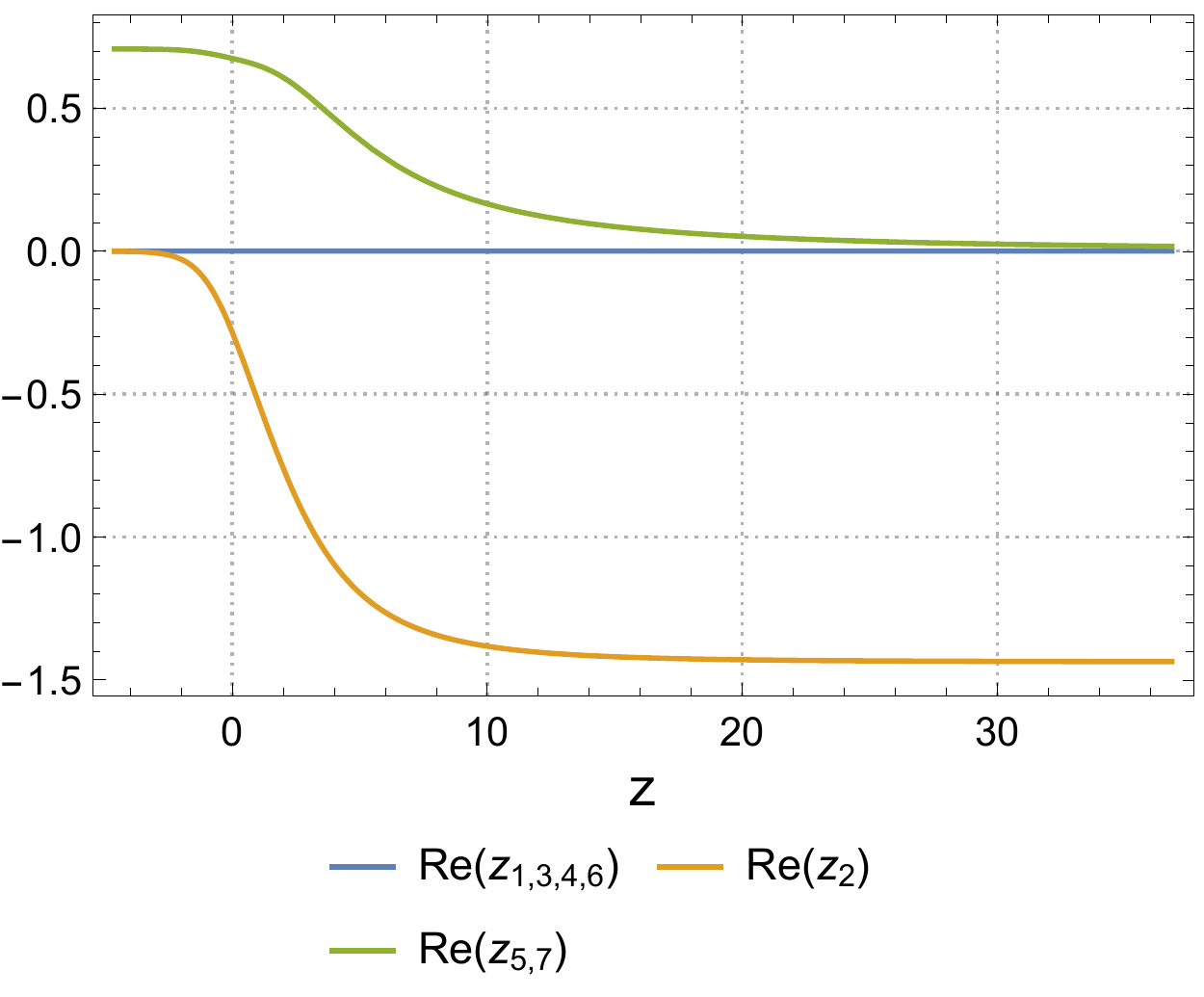}
\hspace{6mm}
\includegraphics[width=0.45\textwidth]{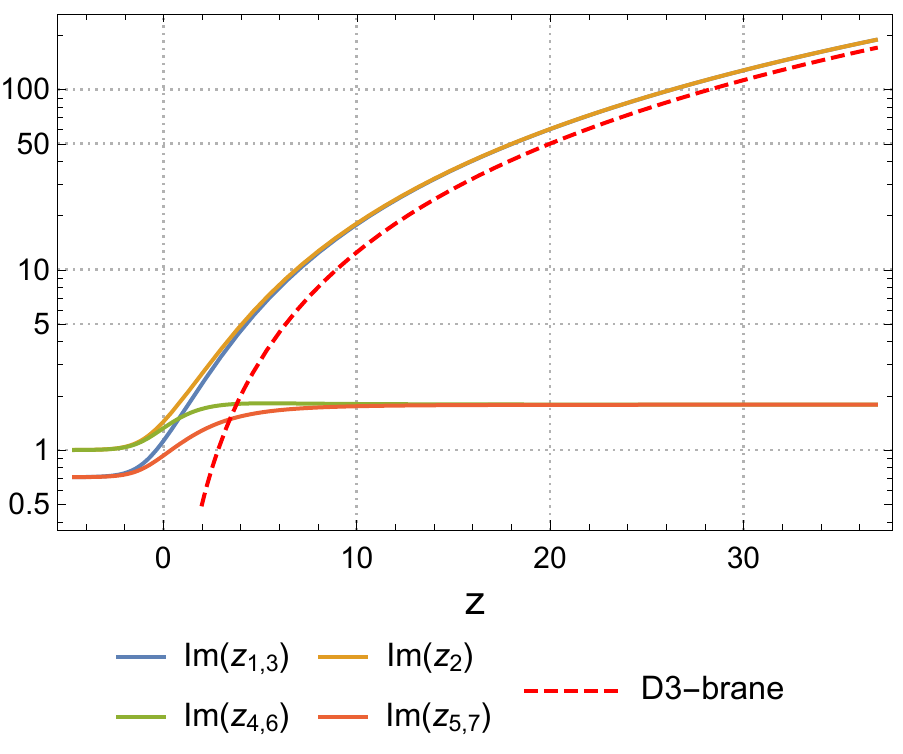}
\\[8mm]
\includegraphics[width=0.45\textwidth]{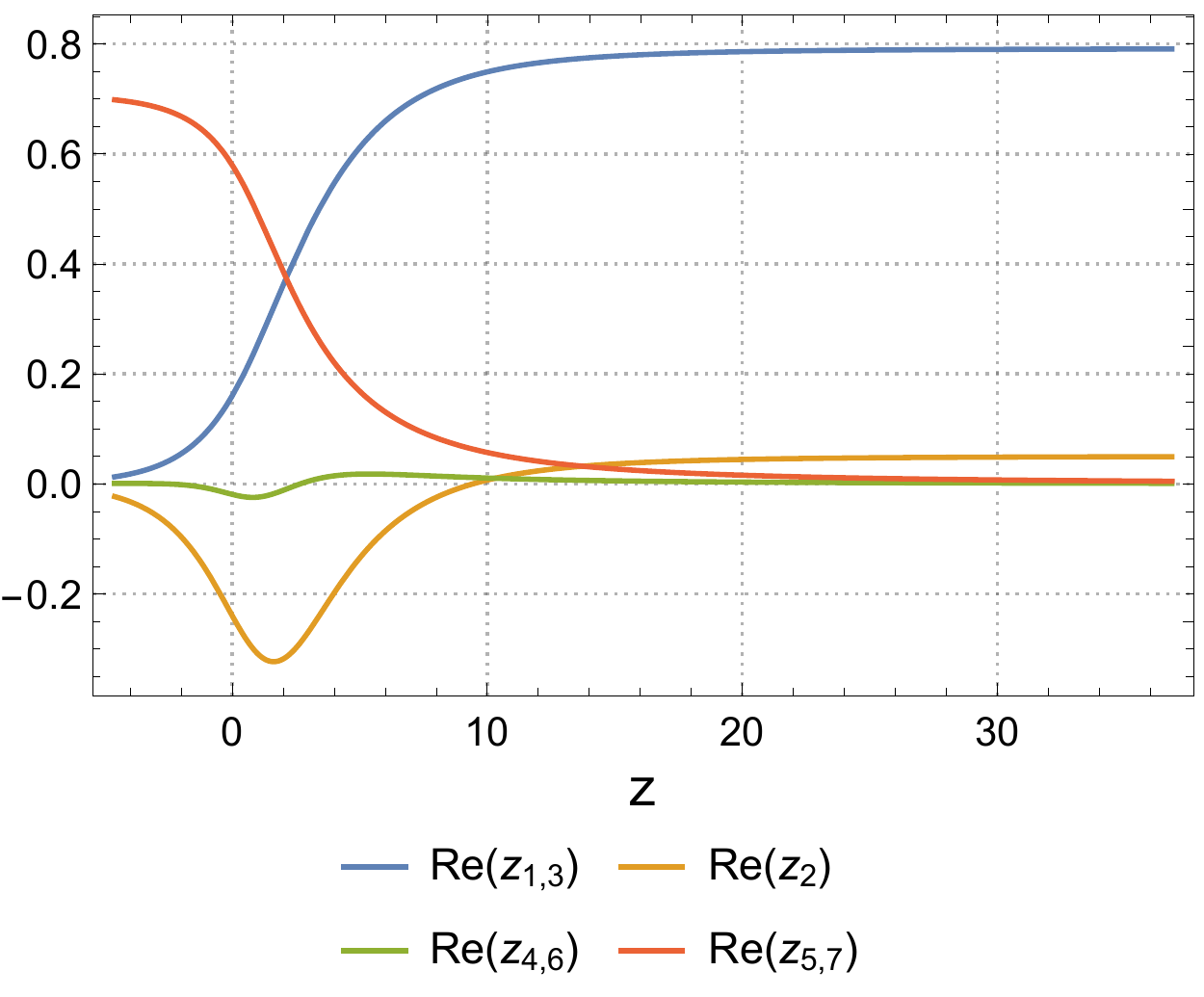}
\hspace{6mm}
\includegraphics[width=0.45\textwidth]{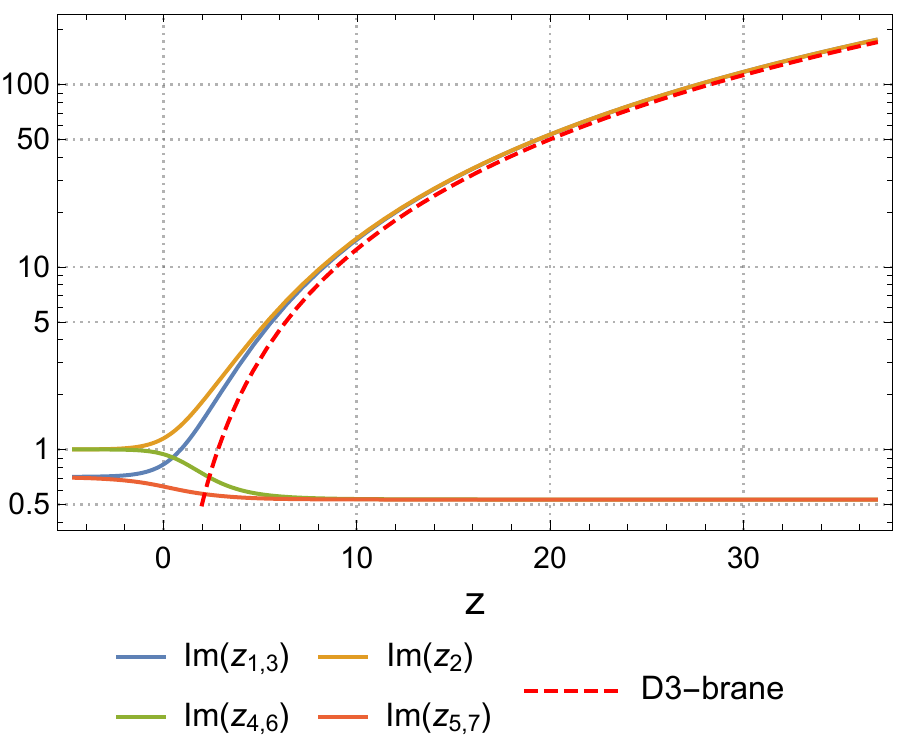}
\caption{Holographic RG flow from $\,\mathcal{N}=4\,$ SYM$_{4}$ (UV) to $\,\mathcal{N}=2\, \&\,\textrm{SU}(2)\,$ J-fold CFT$_{3}$ (IR). Top plots: $\,(\Lambda_1,\Lambda_{2}) = (-1 , 0.7)\,$ and $\,(\lambda_{1},\lambda_{2}) = (0,0)\,$. Bottom plots: $\,(\Lambda_1,\Lambda_2) = (-1,0)\,$ and $\,(\lambda_1,\lambda_{2})=(0.307, 0.011)\,$.}
\label{fig:Plot_N2_gen}
\end{center}
\end{figure}

Fixing $\,\Lambda_{2} = 0\,$ and $\,\lambda_{1,2} = 0\,$ implies $\,{\textrm{Re}z_{1,2,3}}=0\,$ and $\,{\textrm{Re}z_{4,6}}=0\,$ in the IR boundary conditions (\ref{boundary_cond_N2}). In this case we obtain the flow depicted in Figure~\ref{fig:Plot_N2_sym}. The UV ($\,z \rightarrow \infty\,$) behaviour of this flow is again understood as a sub-leading correction in the electromagnetic deformation $\,c\,$ of the D3-brane solution in (\ref{analytic_sol_c=0}) with
\begin{equation}
\textrm{Re}z_{1,2,3} = 0
\hspace{10mm} \textrm{ and } \hspace{10mm}
\Phi_{0} = 0 \ .
\end{equation}
Now we can explore the UV behaviour of the flows when activating the parameters $\,\Lambda_{2}\,$ and $\,\lambda_{1,2}\,$ in the boundary conditions (\ref{boundary_cond_N2}). As it can be directly seen from there, the parameter $\,\lambda_1\,$ controls the pair-wise identified perturbation of $\,\textrm{Re}z_1=\textrm{Re}z_3\,$ whereas $\,\Lambda_2\,$ subsequently controls the perturbation of $\,\textrm{Re}z_2\,$. Turning on just the parameter $\,\Lambda_2\,$ generates flows reaching the UV with 
\begin{equation}
\label{new_UV_N2}
\sum\limits_{i=1}^3 \textrm{Re}z_{i} \approx   c \, \sinh \Phi_{0}  
\hspace{5mm} , \hspace{5mm}
\textrm{Im}z_{4,5,6,7}  \approx   e^{- \frac{1}{2} \Phi_{0}} 
\hspace{5mm} \textrm{ and } \hspace{5mm}
\Phi_{0} \neq 0 \ ,
\end{equation}
in agreement with (\ref{Re_chi1,2,3_c}). Turning on the remaining parameters $\,\lambda_{1,2}\,$ makes more scalars run along the flow. Flows of these types are presented in Figure~\ref{fig:Plot_N2_gen}.

\subsubsection*{Study of the parameter space}

\begin{figure}[t!]
\begin{center}
\includegraphics[width=0.40\textwidth]{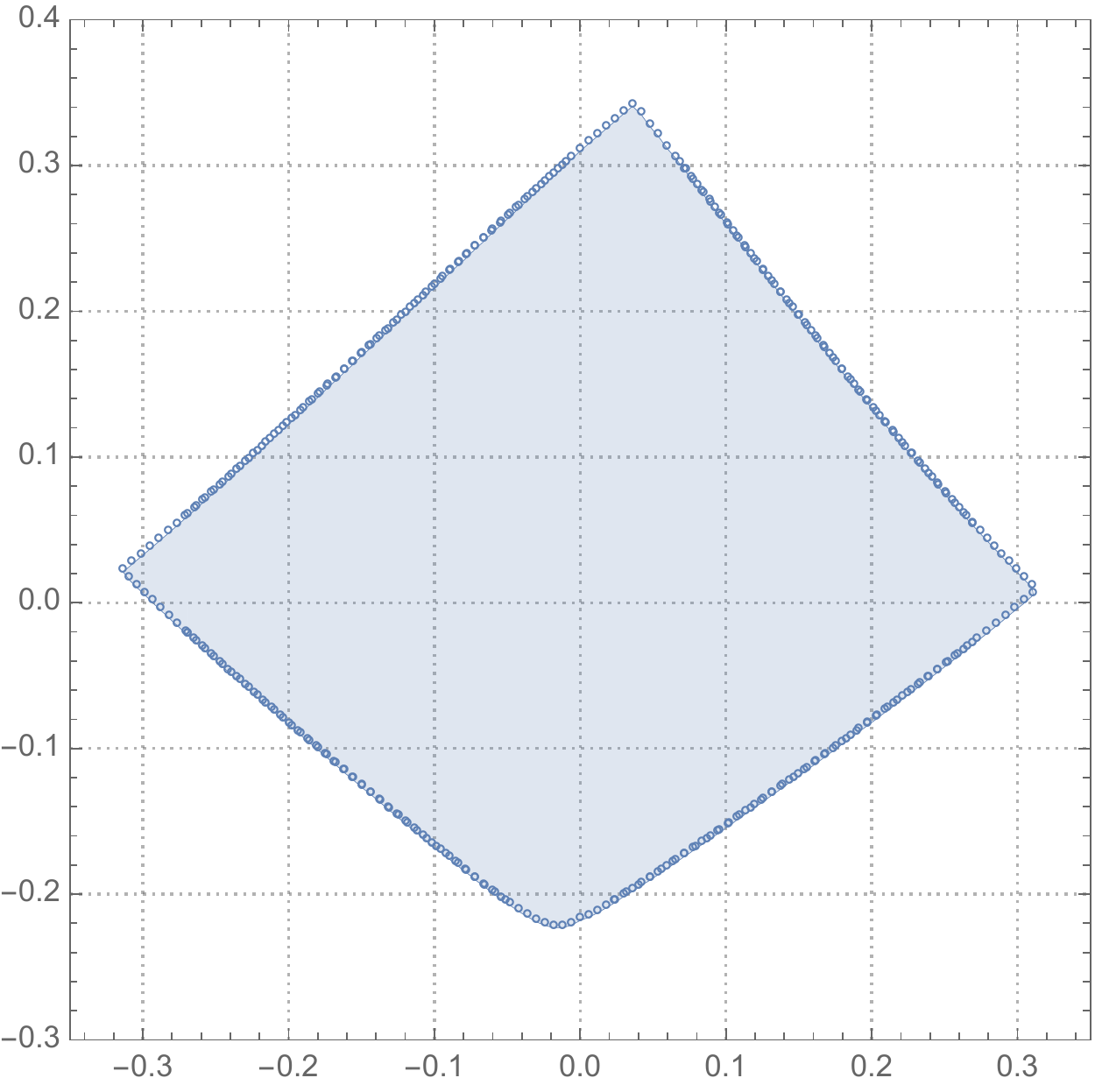}
\put(-88,-12){$\lambda_{1}$}
\put(-192,90){$\lambda_{2}$}
\\[2mm]
\includegraphics[width=0.40\textwidth]{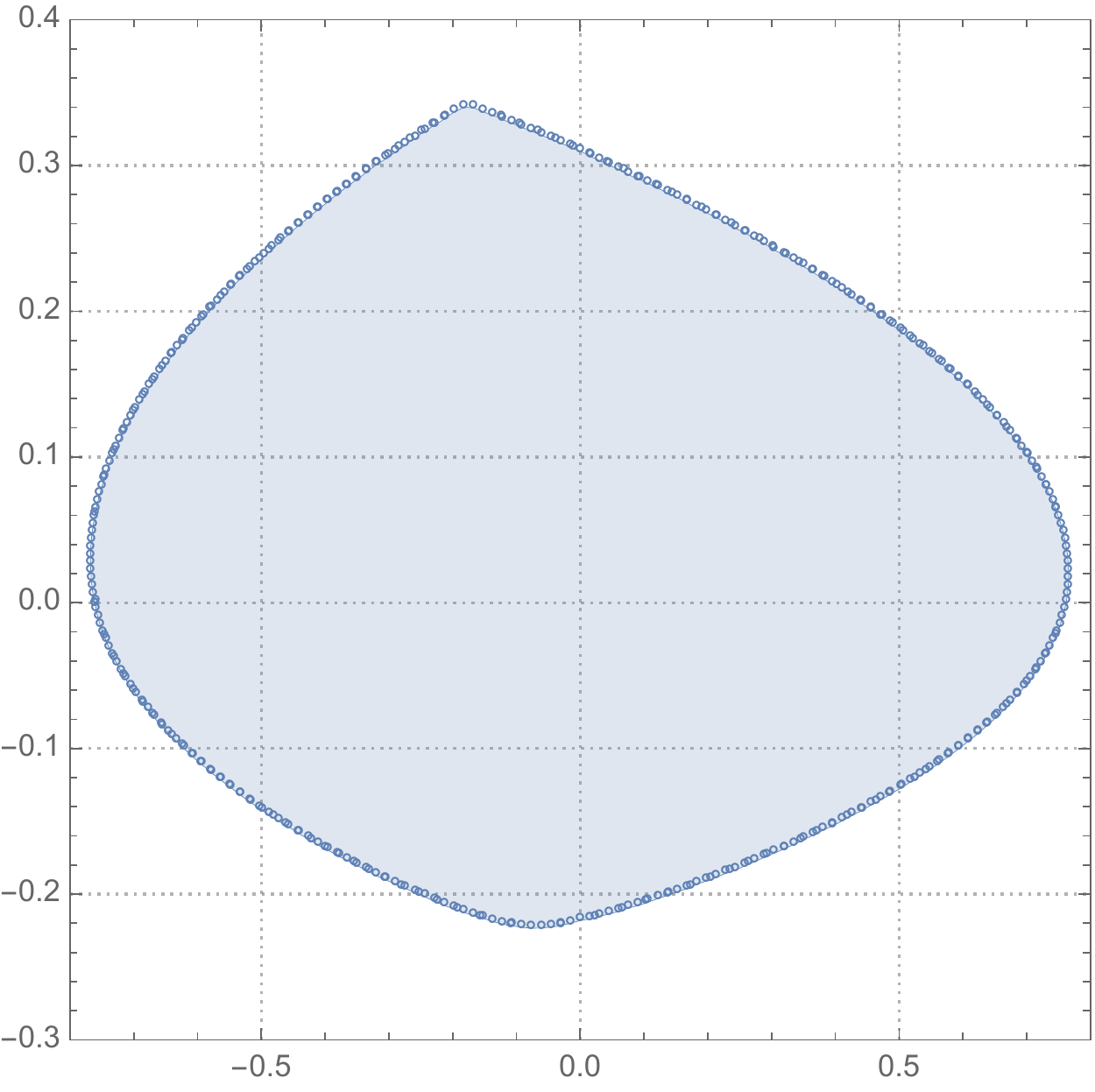}
\put(-88,-12){$\Lambda_{2}$}
\put(-192,90){$\lambda_{2}$}
\hspace{15mm}
\includegraphics[width=0.40\textwidth]{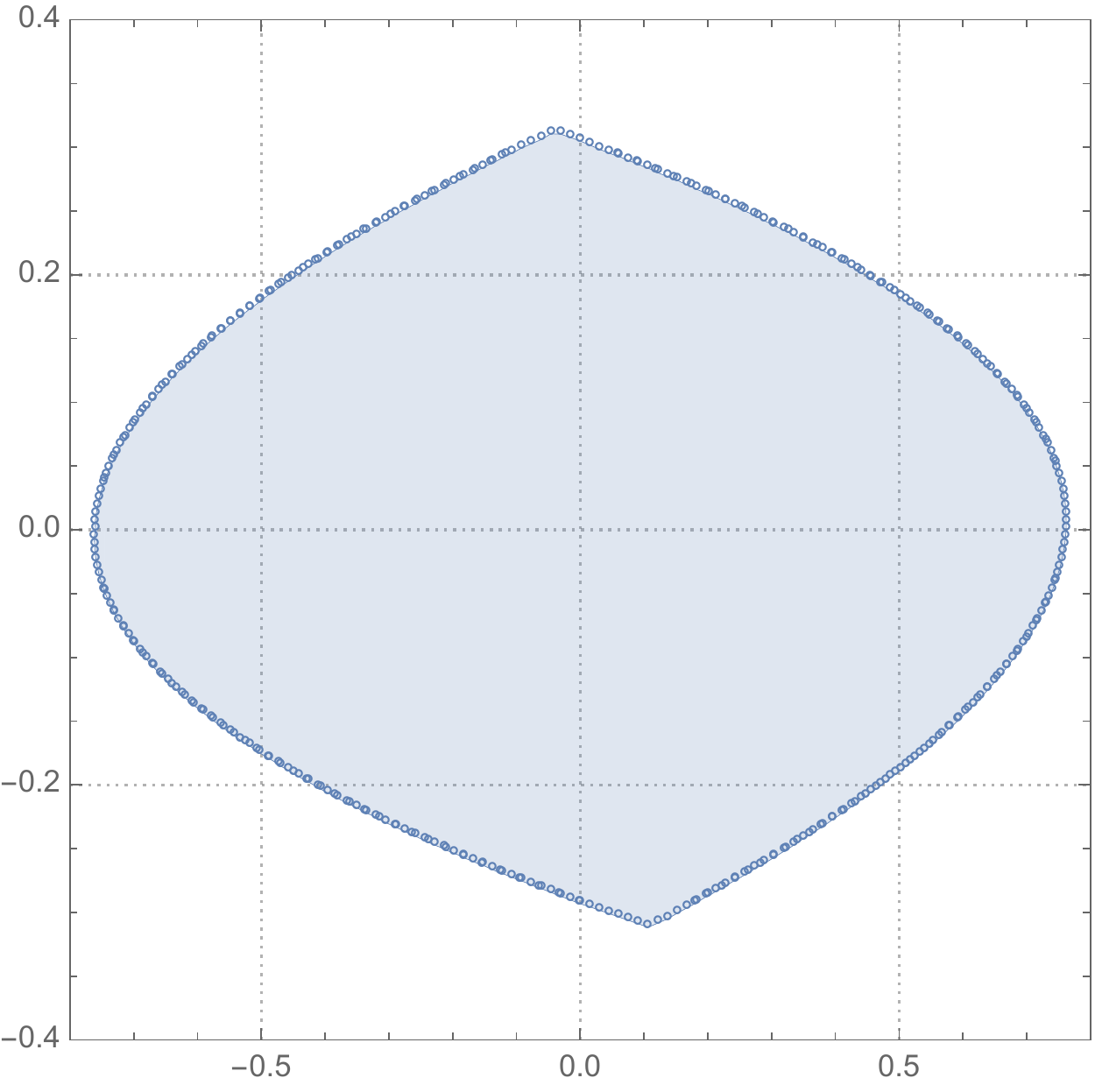}
\put(-88,-12){$\Lambda_{2}$}
\put(-192,90){$\lambda_{1}$}
\caption{Three sections of the parameter space showing the region accommodating regular holographic RG flows involving the $\,{\mathcal{N}=2\, \&\,\textrm{SU}(2)}\,$ J-fold CFT$_{3}$ in the IR: $\,\Lambda_2 =0\,$ (top), $\,\lambda_1=0\,$ (bottom-left) and $\,\lambda_2=0\,$ (bottom-right).}
\label{fig:Plot_parameter_space_N2}
\end{center}
\end{figure}

This time we must perform a numerical scan of flows in a three-dimensional parameter space $\,(\Lambda_{2} \, ; \,\lambda_{1} \,,\, \lambda_{2})\,$. Various sections of the parameter space can be taken which are depicted in Figure~\ref{fig:Plot_parameter_space_N2}. The three parameters $\,\Lambda_2\,$ and $\,\lambda_{1,2}\,$ control the values of the axions $\,\textrm{Re}z_{1,2,3}\,$ when generically approaching the D$3$-brane solution in the UV.

Let us discuss in more detail some features of the parameter space depicted in Figure~\ref{fig:Plot_parameter_space_N2}. The borders of the various sections delimit the region of the three-dimensional parameter space producing regular flows. Outside this region the flows have some scalar field diverging at a finite radial distance. Importantly, the region around the upper corner in the $\,(\lambda_{1},\lambda_{2})$-projection is very special as it produces flows passing arbitrarily close to the $\,\mathcal{N}=1\,$ J-fold CFT$_{3}$'s before continue flowing to $\,\textrm{SYM}_{4}\,$ in the deep UV. This limiting CFT$_{3}$ to CFT$_{3}$ holographic RG flows are presented separately in Section~\ref{sec:N=1_N=2}.

\subsubsection*{Gravitino masses and supersymmetry}

For the $\,\mathcal{N}=1\,\&\,\textrm{SU}(2)\times\textrm{U}(1)\,$ AdS$_{4}$ vacuum in the IR the flavour symmetry group realised in the dual J-fold CFT$_{3}$ is $\,\textrm{SU}(2)\,$. Under this $\,\textrm{SU}(2)\,$ symmetry the eight gravitini of the maximal theory decompose this time as in (\ref{Branching_4D_gravitini_SU2xU1}), namely,
\begin{equation}
\label{Branching_4D_gravitini_SU2_flows}
\small{
\begin{array}{ccl}
\textrm{SU}(8)  &\supset & \textrm{SU}(2) \\[2mm]
\textbf{8} & \rightarrow &   \textbf{2}  \, + \,   \textbf{2}    \, + \, \textbf{1}  \, + \, \textbf{1}  \, + \, \textbf{1}  \, + \, \textbf{1}
\end{array}
}
\end{equation}
For generic BPS flows with parameters $\,\Lambda_{2}\,$ and $\,(\lambda_{1} \,,\, \lambda_{2})\,$ in the regions shown in Figure~\ref{fig:Plot_parameter_space_N2}, the evaluation of the eight eigenvalues of $\,A_{IJ} \, A^{JK}\,$ shows that, as before, only one of them is generically compatible with the IR value of the $\,\mathcal{N}=1\,$ gravitino mass (\ref{N1_gravitino_mass}) belonging to the $\,\mathbb{Z}_{2}^{3}$-invariant sector. However, specific choices of the parameters $\,(\lambda_{1} \,,\, \lambda_{2})\,$ this time yield different splittings of the eight gravitino masses
\begin{equation}
\label{8_decomposition_N2}
\begin{array}{lll}
\lambda_{1}^2 + \lambda_{2}^2 \neq 0 & \hspace{5mm} : & \hspace{5mm} 8 \rightarrow 4 + 1 + 1 + \boxed{1 + \color{blue}{1}} \ , \\[2mm]
\lambda_{1}^2 + \lambda_{2}^2 = 0 & \hspace{5mm} : & \hspace{5mm} 8 \rightarrow 4 + 2 + \boxed{\color{blue}{2}}  \ ,
\end{array}
\end{equation}
when evaluated along the numerical flows. In (\ref{8_decomposition_N2}) we have boxed the $\,\mathcal{N}=2\,$ supersymmetry realised at the AdS$_{4}$ vacuum in the deep IR and highlighted (in blue) those gravitino masses with respect to which the numerical flows are BPS.

\subsection{SYM$_{4}$ to CFT$_{3}$ with $\,\mathcal{N}=4\,$}
\label{sec:N=4_SYM}

Lastly we will solve the BPS equations (\ref{BPS_equations}) by perturbing around the $\,\mathcal{N}=4 \, \& \, \textrm{SO}(4)\,$ AdS$_{4}$ vacuum (\ref{VEVs_z_N4}) in the IR ($z\rightarrow -\infty$). This will trigger again the appearance of generic flows towards a non-conformal behaviour in the UV ($z\rightarrow \infty$).

\subsubsection*{IR boundary conditions}

Around the $\,\mathcal{N}=4 \, \& \, \textrm{SO}(4)\,$ solution, there are again four irrelevant modes in the spectrum (\ref{Deltas_SO4_scalars})
\begin{equation}
\Delta_{-} = -2 
\hspace{8mm} \textrm{ and } \hspace{8mm}
\Delta_{-} = -1 \,\, ( \times 3 ) \ ,
\end{equation}
that are compatible with regularity of the flow in the IR ($\,z\rightarrow -\infty\,$). The linearised BPS equations then allow for four parameters $\,\Lambda\,$ and $\,\lambda_{i}\,$, with $\,i=1,2,3\,$, specifying the IR boundary conditions  (\ref{boundary_cond_IR}). These read
\begin{equation}
\label{IR_boundary_conds_N4}
\begin{array}{rlll}
\textrm{Im}z_{i} &=& c\left(1-2  \Lambda\,  e^{2\frac{z}{L}}\right) & ,  \\[2mm]
\textrm{Im}z_{3+i} &=& \frac{1}{\sqrt{2}} \left(1-\Lambda\,  e^{2\frac{z}{L}} - \frac{1}{4} \, (\lambda + \lambda_{i}) \, e^{\frac{z}{L}} \right) & ,  \\[2mm]
\textrm{Im}z_{7} &=& \frac{1}{\sqrt{2}} \left(1-\Lambda\,  e^{2\frac{z}{L}}- \frac{1}{2} \, \lambda \,  e^{\frac{z}{L}} \right) & ,  \\[2mm]
\textrm{Re}z_i &=& \frac{1}{2} \, c \,  \lambda_i  \, e^{\frac{z}{L}} & ,  \\[2mm]
\textrm{Re}z_{3+i} &=& \frac{1}{\sqrt{2}} \left(1+ \Lambda \,  e^{2\frac{z}{L}}- \frac{1}{4} \, (\lambda + \lambda_i) e^{\frac{z}{L}} \right) & ,  \\[2mm]
\textrm{Re}z_{7} &=& -\frac{1}{\sqrt{2}} \left(1 + \Lambda\,  e^{2\frac{z}{L}} - \frac{1}{2} \, \lambda \, e^{\frac{z}{L}} \right) & ,
\end{array}
\end{equation}
with $\,\lambda \equiv \lambda_{1} + \lambda_{2} + \lambda_{3}\,$. Note that the parameters $\,\lambda_{i}\,$ enter the IR boundary conditions (\ref{IR_boundary_conds_N4}) in a symmetric manner and fully specify $\,\textrm{Re}z_{i}\,$. As before, one of the parameters, either $\,\Lambda\,$ or $\,\lambda_{i}\,$, can be set at will by a shift on the coordinate $\,z\,$. We will set $\,\Lambda = -1\,$ without loss of generality\footnote{Setting $\,\Lambda = 0,1\,$ does not produce regular flows.}, which leaves us also this time with a three-dimensional parameter space to scan.

\subsubsection*{Behaviour of the flows}

\begin{figure}[t]
\begin{center}
\mbox{\includegraphics[width=0.45\textwidth]{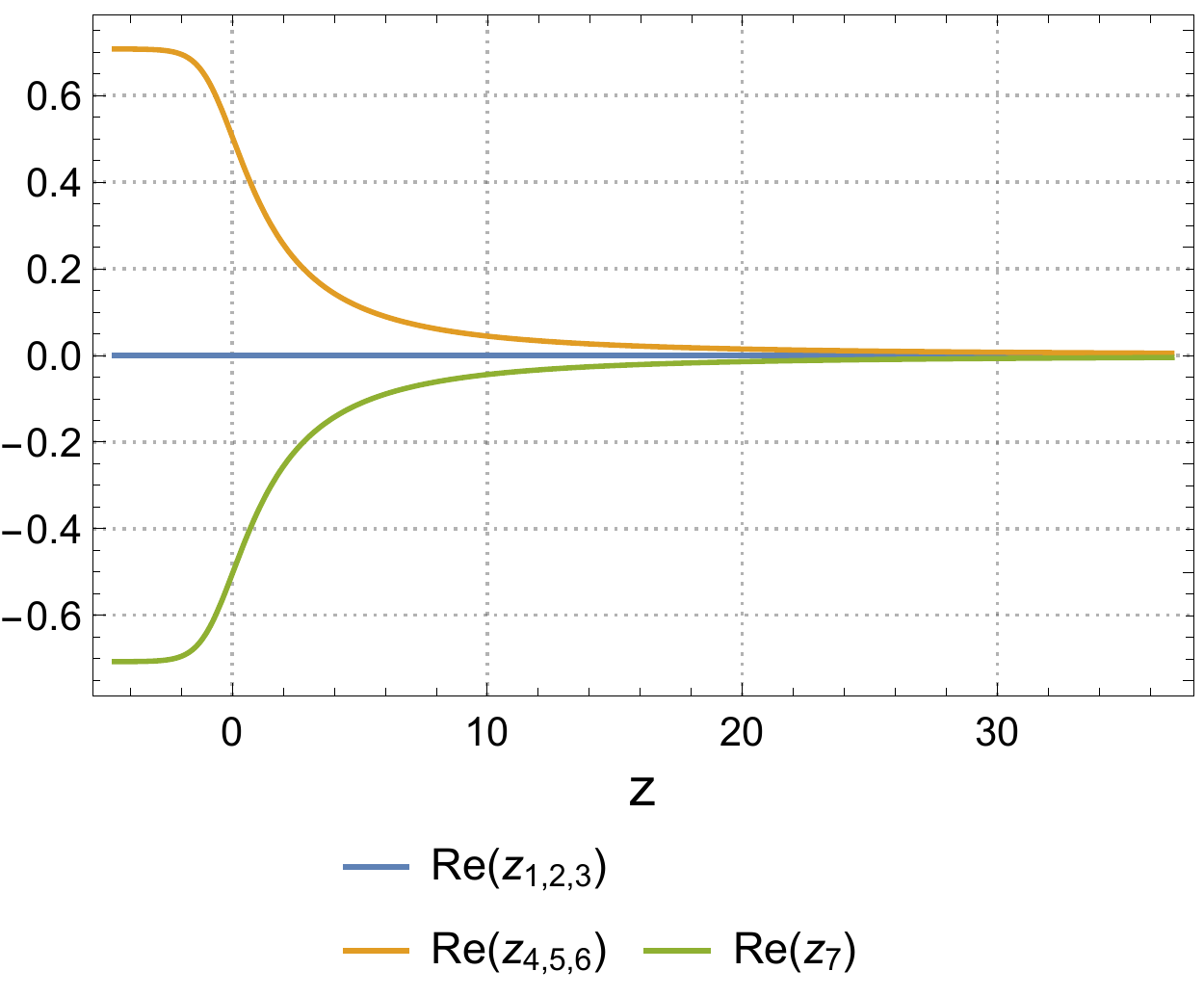}
\hspace{6mm}
\includegraphics[width=0.45\textwidth]{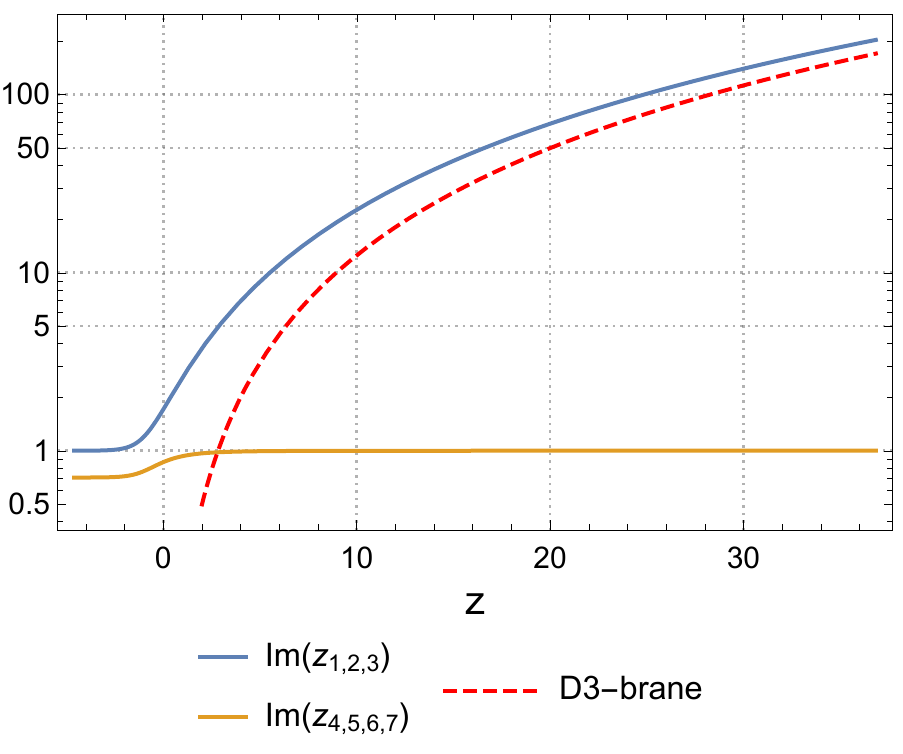}}
\caption{Holographic RG flow from $\,\mathcal{N}=4\,$ SYM$_{4}$ (UV) to $\,\mathcal{N}=4\,$ J-fold CFT$_{3}$ (IR) with $\,\Lambda = -1\,$ and $\,\lambda_{i} = 0\,$.}
\label{fig:Plot_N4_sym}
\end{center}
\end{figure}
\begin{figure}[t]
\begin{center}
\mbox{\includegraphics[width=0.45\textwidth]{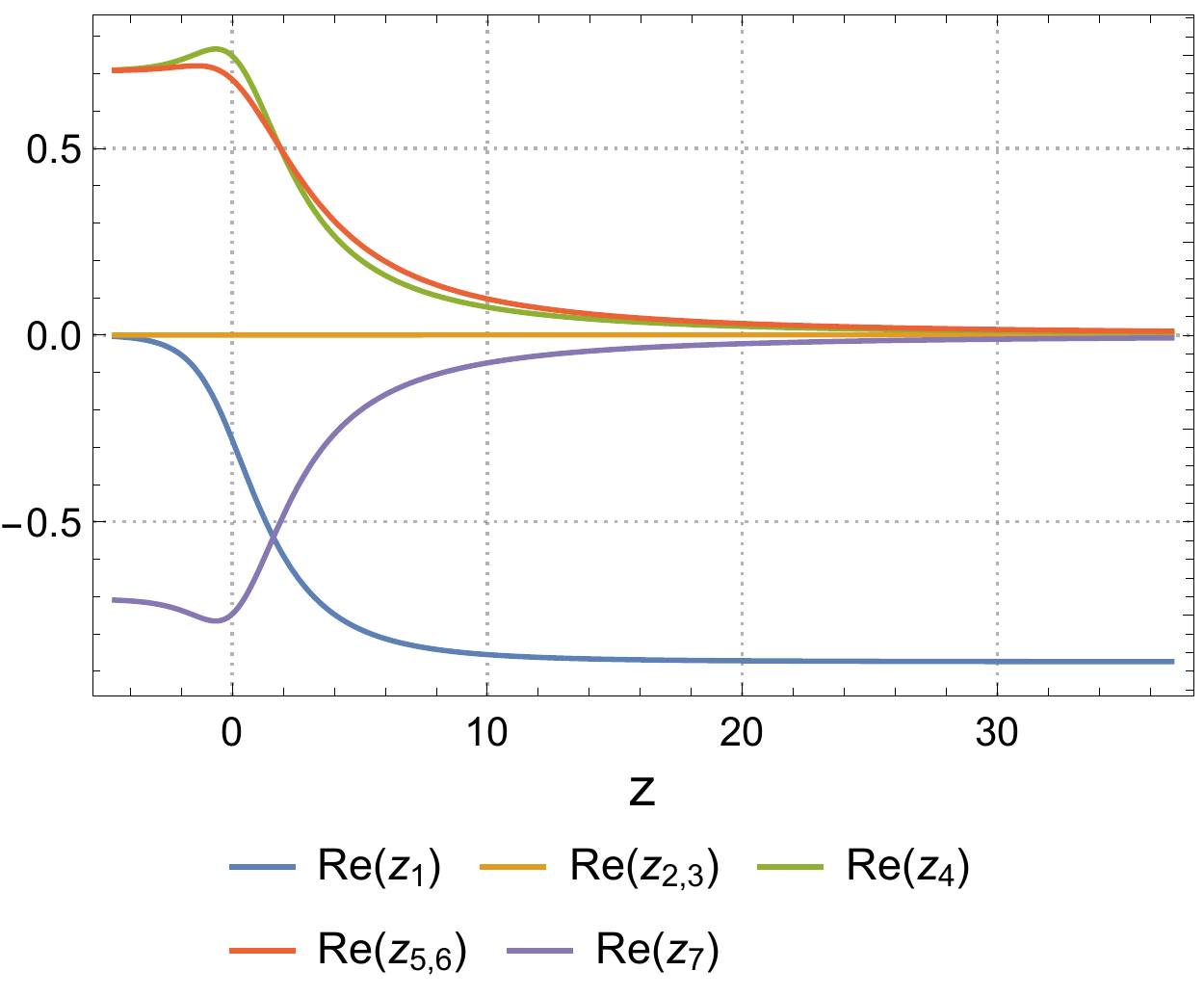}
\hspace{6mm}
\includegraphics[width=0.45\textwidth]{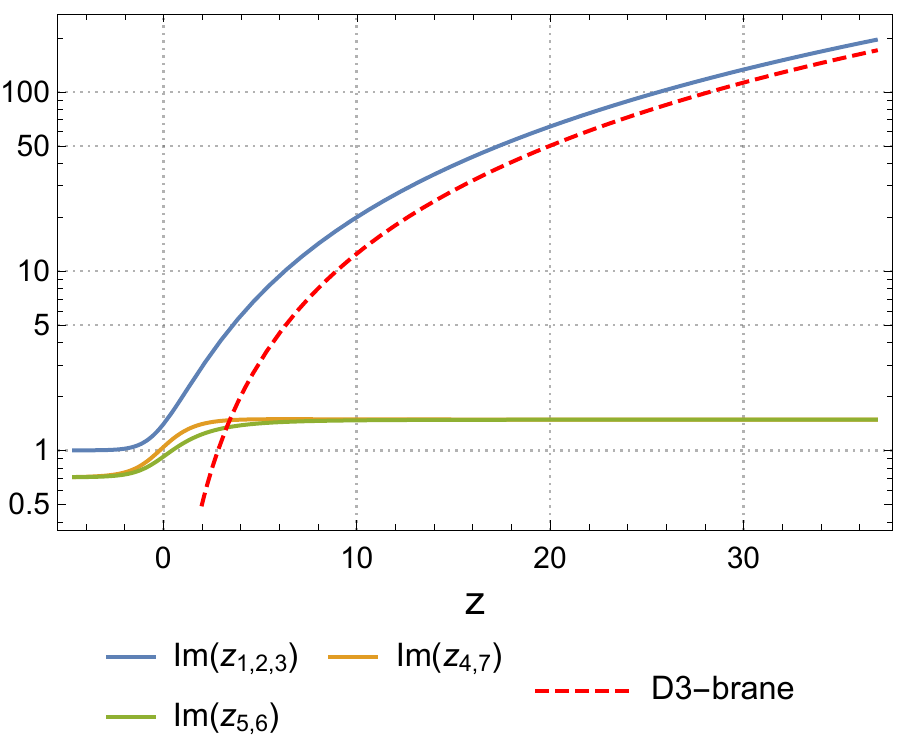}}
\caption{Holographic RG flow from $\,\mathcal{N}=4\,$ SYM$_{4}$ (UV) to $\,\mathcal{N}=4\,$ J-fold CFT$_{3}$ (IR) with $\,\Lambda = -1\,$ and $\,(\lambda_{1},\lambda_{2},\lambda_{3}) = (-0.8,0,0)\,$.}
	\label{fig:Plot_N4_gen}
\end{center}
\end{figure}

The IR boundary conditions (\ref{IR_boundary_conds_N4}) are highly symmetric. Fixing $\,\lambda_{i} = 0\,$ in (\ref{IR_boundary_conds_N4}) implies $\,{\textrm{Re}z_{1,2,3}}=0\,$. In this case we obtain the flow depicted in Figure~\ref{fig:Plot_N4_sym}. The UV ($\,z \rightarrow \infty\,$) behaviour of this flow is again understood as a sub-leading correction in the electromagnetic deformation $\,c\,$ of the D3-brane solution in (\ref{analytic_sol_c=0}) with
\begin{equation}
\textrm{Re}z_{1,2,3} = 0
\hspace{10mm} \textrm{ and } \hspace{10mm}
\Phi_{0} = 0 \ .
\end{equation}
Turning on the parameters $\,\lambda_{i}\,$ activates the asymptotic values of the axions $\,\textrm{Re}z_{i}\,$ in the UV. More concretely, the UV region is again reached as
\begin{equation}
\label{new_UV_N4}
\sum\limits_{i=1}^3\, \textrm{Re}z_{i}  \approx  c \, \sinh \Phi_{0} 
\hspace{5mm} , \hspace{5mm}
\textrm{Im}z_{4,5,6,7}  \approx  e^{- \frac{1}{2} \Phi_{0}}
\hspace{5mm} \textrm{ and } \hspace{5mm}
\Phi_{0} \neq 0 \ .
\end{equation}
This agrees again with (\ref{Re_chi1,2,3_c}) obtained at first-order in the deformation parameter $\,c\,$. A generic flow with $\,\lambda_{1} \neq 0\,$ and $\,\textrm{Re}z_{1} \neq 0\,$ in the UV is depicted in Figure~\ref{fig:Plot_N4_gen}.

\begin{figure}[t]
\begin{center}
\includegraphics[width=0.40\textwidth]{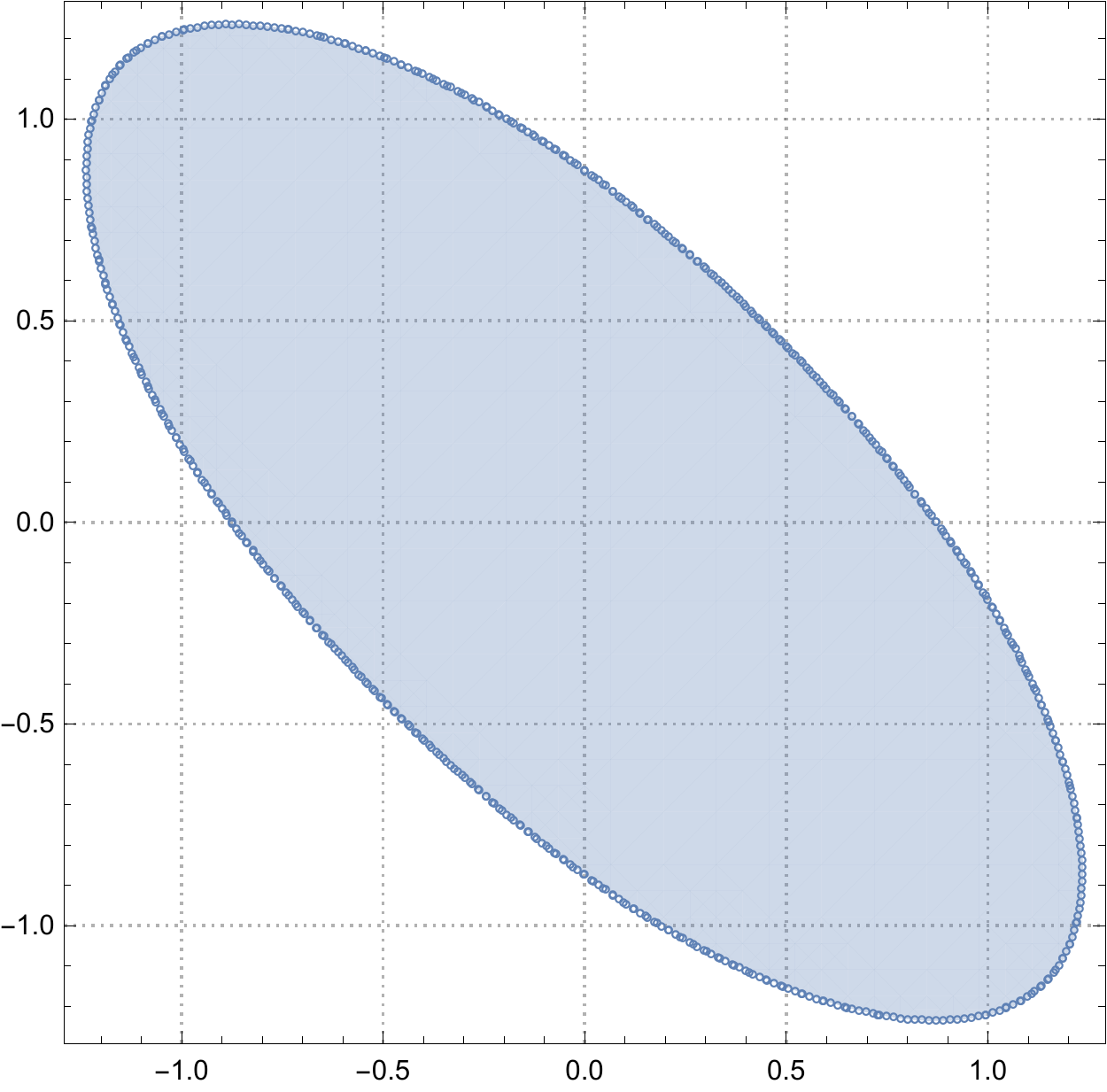}
\put(-88,-12){$\lambda_{2}$}
\put(-192,90){$\lambda_{3}$}
\caption{Section of the parameter space with $\,\lambda_{1}=0\,$ showing the region accommodating regular holographic RG flows involving the $\,\mathcal{N}=4\,$ J-fold CFT$_{3}$ in the IR. Sections with $\,\lambda_{2}=0\,$ or $\,\lambda_{3}=0\,$ are equivalent due to the exchangeability of the parameters $\,\lambda_{i}\,$.}
	\label{fig:Plot_Param_L1zero}
\end{center}
\end{figure}

\subsubsection*{Study of the parameter space}

The fact that $\,\lambda_{1,2,3}\,$ enter the IR boundary conditions (\ref{IR_boundary_conds_N4}) symmetrically renders the three parameters completely interchangeable as far as the induced flows are concerned. In Figure~\ref{fig:Plot_Param_L1zero} the section of the parameter space allowing for regular holographic RG flows with $\,\lambda_{1}=0\,$ is depicted. Similar figures are obtained upon setting $\,\lambda_{2}=0\,$ or $\,\lambda_3=0\,$. Finally, within our numerical precision, we do not observe flows reaching the $\,\mathcal{N}=1\,$ family of AdS$_{4}$ vacua (\ref{VEVs_z_N1}) in the UV. The three parameters $\,\lambda_{1,2,3}\,$ control the values of the axions $\,\textrm{Re}z_{1,2,3}\,$ when approaching the D$3$-brane solution in the UV.\\

\subsubsection*{Gravitino masses and supersymmetry}

The $\,\mathcal{N}=4\,\&\,\textrm{SO}(4)\,$ AdS$_{4}$ vacuum in the IR realises a trivial flavour symmetry group in the dual J-fold CFT$_{3}$. For generic BPS flows with parameters $\,\lambda_{i}\,$ in the parameter space of Figure~\ref{fig:Plot_Param_L1zero}, the evaluation of the eight eigenvalues of $\,A_{IJ} \, A^{JK}\,$ shows that, as in the previous cases, only one of them is generically compatible with the IR value of the $\,\mathcal{N}=1\,$ gravitino mass (\ref{N1_gravitino_mass}) belonging to the $\,\mathbb{Z}_{2}^{3}$-invariant sector. Specific choices of the parameters $\,\lambda_{i}\,$ yield again a different splitting of the eight gravitino masses
\begin{equation}
\label{8_decomposition_N4}
\begin{array}{lll}
\lambda \neq \lambda_i \,\,\,\,\,  \forall i & \hspace{5mm} : & \hspace{5mm} 8 \rightarrow 1 + 3 + \boxed{{\color{blue}{1}}+3} \ , \\[2mm]
\lambda = \lambda_i \,\,\,\, (\textrm{for one } \lambda_i)& \hspace{5mm} : & \hspace{5mm} 8 \rightarrow 4 +  \boxed{ 2 +\color{blue}{2}}  \ , \\[2mm]
\lambda = \lambda_i \,\,\,\, (\textrm{for two } \lambda_i \textrm{ and }  \lambda_j  )& \hspace{5mm} : & \hspace{5mm} 8 \rightarrow 1 + 3 + \boxed{1 + {\color{blue}{3}}} \ , \\[2mm]
\lambda = \lambda_i \,\,\,\,  \forall i  \,\,\, \Leftrightarrow \,\,\, \lambda_{i} = 0 \,\,\,\,\,  \forall i \,\,  & \hspace{5mm} : & \hspace{5mm} 8 \rightarrow 4 + \boxed{\color{blue}{4}}  \ ,
\end{array}
\end{equation}
with $\,\lambda \equiv \lambda_{1} + \lambda_{2} + \lambda_{3}\,$ when evaluated along the numerical flows. In (\ref{8_decomposition_N4}) we have boxed the $\,\mathcal{N}=4\,$ supersymmetry realised at the AdS$_{4}$ vacuum in the deep IR and highlighted (in blue) those gravitino masses with respect to which the numerical flows are BPS. Note that the IR boundary conditions in (\ref{IR_boundary_conds_N4}) are compatible with a symmetry that ranges from $\,\textrm{SO(3)}\,$ ($\,\lambda \neq \lambda_i\,$) to $\,\textrm{SO(4)}\,$ ($\,\lambda_{i} = 0\,$).

\subsection{CFT$_{3}$ to CFT$_{3}$}
\label{sec:N=1_N=2}

\begin{figure}[t!]
\begin{center}
\mbox{\includegraphics[width=0.45\textwidth]{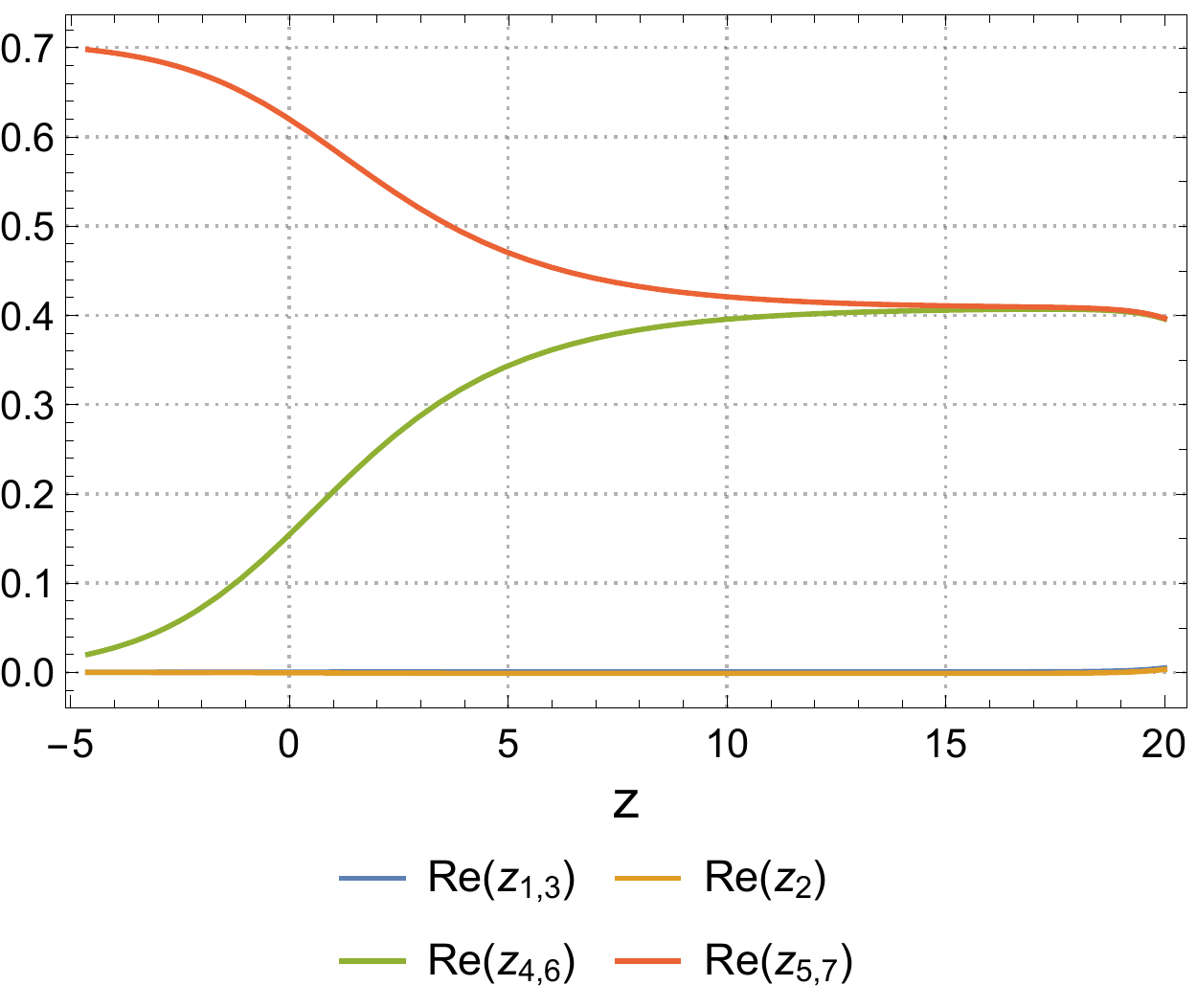}
\hspace{6mm}
\includegraphics[width=0.45\textwidth]{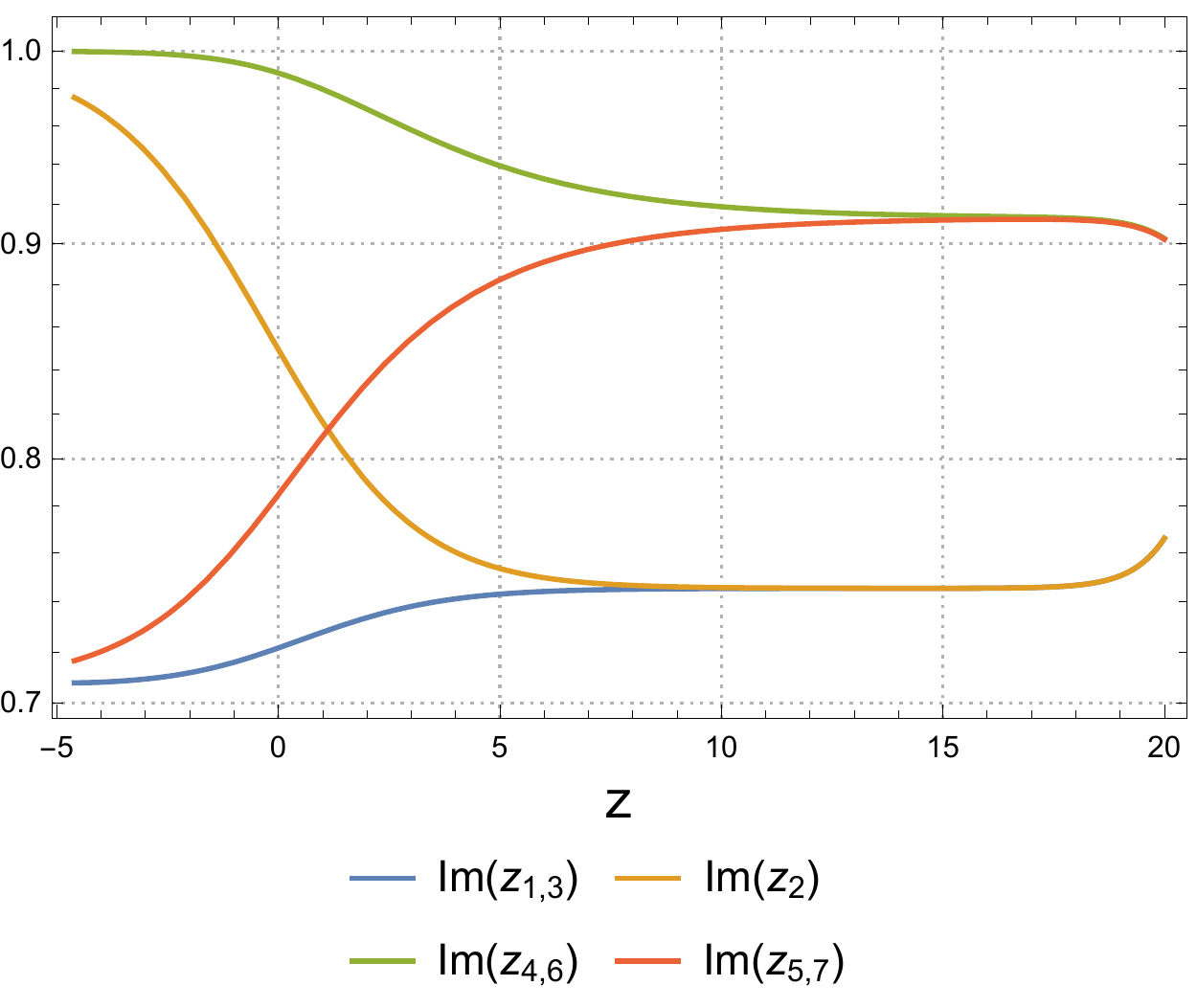}}
\caption{Holographic RG flow from $\,\mathcal{N}=1\, \& \, \textrm{SU}(3)\,$  J-fold CFT$_{3}$ (UV, right) to $\,{\mathcal{N}=2\, \& \, \textrm{SU}(2)}\,$ J-fold CFT$_{3}$ (IR, left) with $\,(\Lambda_1,\Lambda_{2}) = (-1 , -0.1566939789)\,$ and $\,(\lambda_{1},\lambda_{2}) = (0.0042958950\, , \,  0.3421361222)\,$.}
	\label{fig:Plot_CFTtoCFT_su3}
\end{center}
\end{figure}

We now present an example of CFT$_{3}$ to CFT$_{3}$ holographic RG flow that connects the J-fold CFT$_{3}$ with $\,\mathcal{N}=1 \, \& \, \textrm{SU}(3)\,$ symmetry in the UV to the J-fold CFT$_{3}$ with $\,\mathcal{N}=2 \, \& \, \textrm{SU}(2)\,$ symmetry in the IR (see Figure~\ref{fig:net_DW}). This flow requires an extreme fine tuning of the IR boundary conditions in (\ref{boundary_cond_N2}) and is depicted in Figure~\ref{fig:Plot_CFTtoCFT_su3}.

\begin{figure}[t!]
\begin{center}
\mbox{\includegraphics[width=0.45\textwidth]{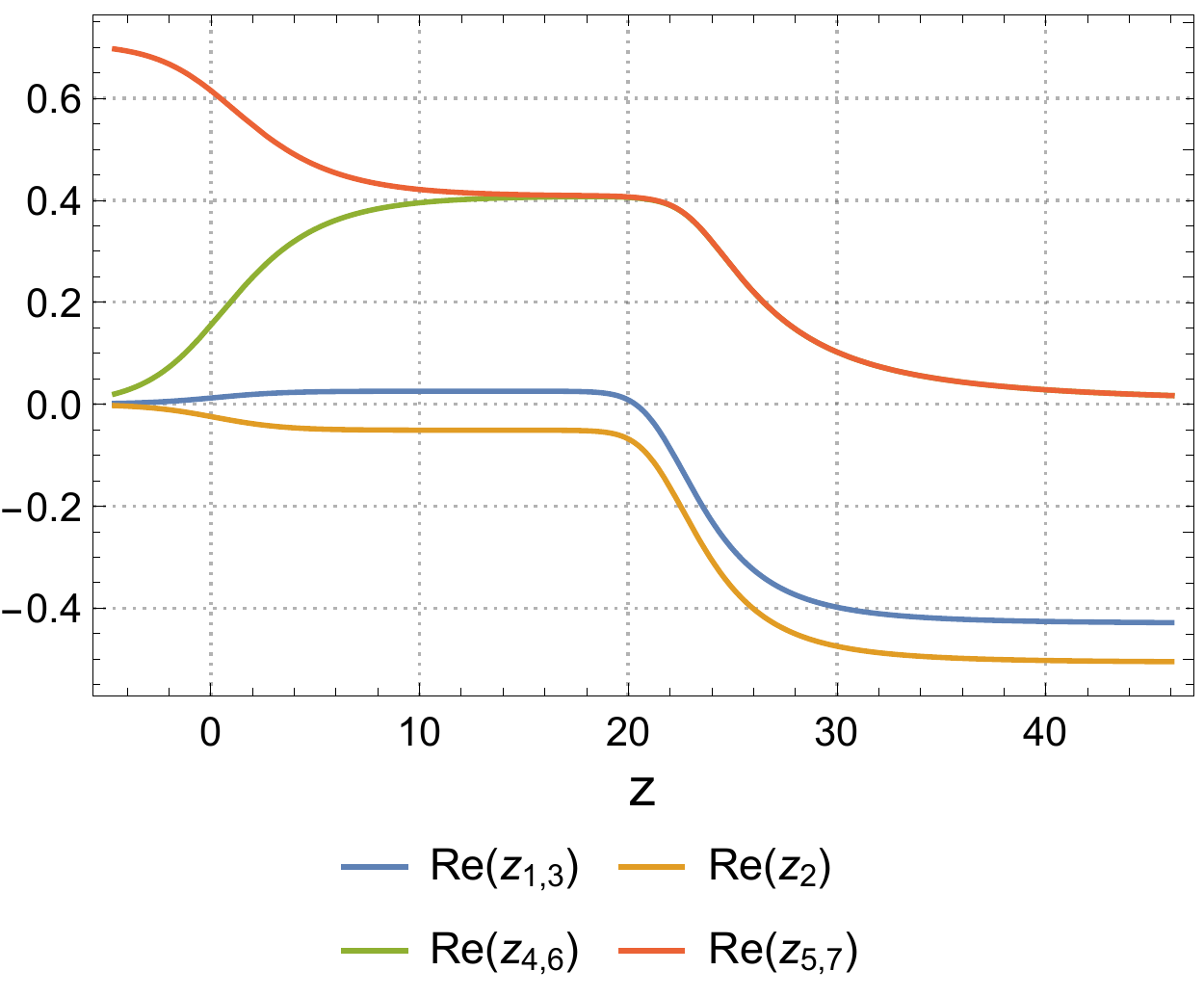}
\hspace{6mm}
\includegraphics[width=0.45\textwidth]{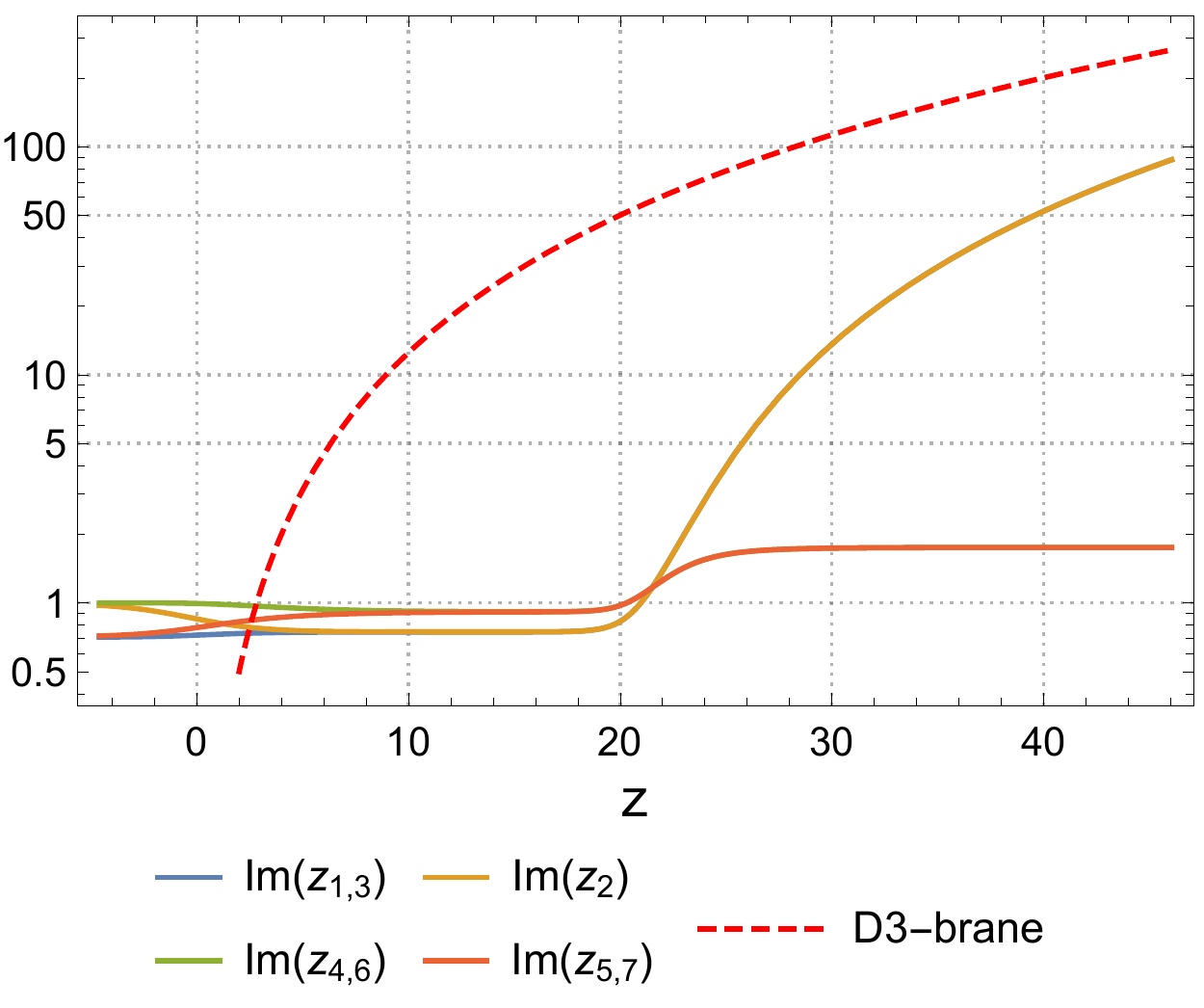}}
\caption{Holographic RG flow from $\,\mathcal{N}=1\, \& \, \textrm{SU}(2) \times \textrm{U}(1)\,$  J-fold CFT$_{3}$ (UV, right) to $\,{\mathcal{N}=2\, \& \, \textrm{SU}(2)}\,$ J-fold CFT$_{3}$ (IR, left) with $\,(\Lambda_1,\Lambda_{2}) = (-1 , 0)\,$ and $\,(\lambda_{1},\lambda_{2}) = ( 0.0364510166 \, , \, 0.3417265522)\,$.}
	\label{fig:Plot_CFTtoCFT_su2xu1}
\end{center}
\end{figure}

This extremely fine-tuned flow is actually a limiting case within a more general class of flows connecting the $\,\mathcal{N}=1 \, \& \, \textrm{SU}(2)\times\textrm{U}(1)\,$ CFT$_3$'s in the UV to the $\,\mathcal{N}=2 \, \& \, \textrm{SU}(2)\,$ CFT$_3$ in the IR. Such flows are described by domain-walls connecting the associated AdS$_4$ vacua in (\ref{VEVs_z_N1}) and (\ref{VEVs_z_N2_2}), and approach the UV with non-vanishing axions satisfying (\ref{chi_SU3_enhancement}) with 
\begin{equation}
\label{pair-wise_identification}
\textrm{Re}z_{1}=\textrm{Re}z_{3}=-\tfrac{1}{2}\textrm{Re}z_2 \ ,
\end{equation}
as shown in Figure~\ref{fig:Plot_CFTtoCFT_su2xu1}. Therefore, the UV symmetry enhancement to $\,\textrm{SU}(3)\,$ does not take place.


From a ten-dimensional perspective, having the pair-wise identification (\ref{pair-wise_identification}) between axions translates into reaching the $\,\mathcal{N}=1\,$ type IIB \mbox{S-folds} with $\,\textrm{SU}(2) \times \textrm{U}(1) \subset \textrm{SU}(3)\,$ symmetry \cite{Guarino:2020gfe} in the UV. The precise manner in which the axions $\, \textrm{Re}z_{1,2,3} \,$ trigger such a symmetry breaking geometrically is investigated in Section~\ref{sec:axions_SU3} by looking at the local coordinate redefinitions they induce on the internal geometry.

Finally, a more detailed study of CFT$_{3}$ to CFT$_{3}$ holographic RG flows including the axions $\, \textrm{Re}z_{1,2,3} \,$ dual to exactly marginal deformations in the corresponding J-fold CFT$_{3}$'s goes beyond the scope of this work, and will be presented elsewhere together with their type IIB embedding.

\section{Ten dimensions}
\label{sec:10D}

In Sections~\ref{sec:N=1_SYM}, \ref{sec:N=2_SYM}  and \ref{sec:N=4_SYM} we numerically constructed holographic RG flows across dimensions connecting ($c$-dependent deformations of) $\,\mathcal{N}=4\,$ SYM$_{4}$ in the UV to different J-fold CFT$_{3}$'s, the latter with various amounts of supersymmetry and the largest possible flavour symmetry: $\,{\mathcal{N}=1 \, \& \, \textrm{SU}(3)}\,$, $\,{\mathcal{N}=2 \, \& \, \textrm{SU}(2) }\,$ and $\,{\mathcal{N}=4}\,$. Featuring the largest possible flavour symmetry in the field theory side translates into the requirement that 
\begin{equation}
\textrm{Re}z_{1,2,3} = 0  \ ,
\end{equation}
in the IR. As shown in \cite{Inverso:2016eet,Guarino:2019oct,Guarino:2020gfe}, this amounts to having a factorised internal geometry of the form $\,\textrm{M}_{6}=\textrm{S}^{1} \times \textrm{S}^{5}\,$. However, the axions $\,\textrm{Re}z_{1,2,3}\,$ were shown in the previous section to generically run over the RG flow and to reach the deep UV with a non-zero value.

\begin{figure}
\centering
$\begin{array}{ccc}
\boxed{\textrm{IR}} & & \boxed{\textrm{UV}} \\[3mm]
\textrm{AdS}_{4} \times \textrm{S}^1 \times \textrm{S}^5   
&  \Longleftrightarrow  &
\textrm{AdS}_{5} \times  \textrm{S}^5 \,\,\,\, (c\textrm{-deformed})    \\[3mm]
ds_{10}^2 = \frac{1}{2}  \, \Delta_{\textrm{IR}}^{-1} \, \left( \, ds_{\textrm{AdS}_{4}}^2 + 2 \, f(z_{i}) \, d\eta^2 \,  \right)  & & ds_{10}^2 = \tfrac{1}{2} \,  \Delta_{\textrm{UV}}^{-1} \, \left( ds^2_{\textrm{DW}_{4}}  +  2 \, \Delta_{\textrm{UV}}\,H(z) \,  d\eta^{2} \, \right)  \\[3mm]
\hspace{-22mm} + \,  g^{-2} \, ds_{\textrm{S}^5}^2 &  & \hspace{-22mm} + \,  g^{-2} \,  d{s}_{\textrm{S}^5}^2 
\end{array}$
\caption{Type IIB geometry describing an holographic RG flows across dimensions from ($c$-dependent deformations of) $\,\mathcal{N}=4\,$ SYM$_{4}$ in the UV to different J-fold CFT$_{3}$'s in the IR.}
\label{fig:diagram_flows_10D}
\end{figure}

On the gravity side, the above RG flows are described by ten-dimensional type IIB \mbox{backgrounds} with an interpolating geometry of the form sketched in Figure~\ref{fig:diagram_flows_10D}. In the UV, the solution asymptotes to a $c$-dependent subleading correction of the $\,\textrm{AdS}_{5} \times  \textrm{S}^5\,$ geometry (only locally if the axions $\,\textrm{Re}z_{1,2,3} \neq 0\,$) whereas, in the IR, an S-fold geometry emerges \cite{Inverso:2016eet,Guarino:2019oct,Guarino:2020gfe}. The function $\,\Delta_{\textrm{IR}}\,$ is the warping function at the S-fold solution in the deep IR, possibly depending on the $\,\textrm{S}^5\,$ coordinates. This is to be distinguished from the $z$-dependent scaling behaviour of $\,\Delta_{\textrm{UV}}\,$ in the deep UV which is generically governed, as we saw in Section~\ref{sec:RG flows}, by the \textit{deformed} D3-brane solution in Section~\ref{sec:semi-analytic}.

\subsection{Uplifting the IR: S-folds}

In the deep IR, the RG flows reach the type IIB S-fold solutions originally presented in \cite{Guarino:2019oct} ($\mathcal{N}=1$), \cite{Guarino:2020gfe} ($\mathcal{N}=2$) and \cite{Inverso:2016eet} ($\mathcal{N}=4$). We concentrate on the S-fold solutions with $\,{\textrm{Re}z_{1,2,3}=0}\,$ allowing for the largest possible flavour symmetries (no exactly marginal deformations) in the dual J-fold CFT$_{3}$'s. In this case, the function $\,f(z_{i})\,$ in Figure~\ref{fig:diagram_flows_10D} turns out to take the simple form
\begin{equation}
f(z_{i}) = \prod_{i=1}^{3} \textrm{Im}z_{i}  \ , 
\end{equation}
and depends only on the scalars $\,\textrm{Im}z_{1,2,3}\,$. In the deep IR these scalars must be evaluated at their VEVs for the three S-folds under consideration (see Table~\ref{vacua_summary}), so that 
\begin{equation}
f(z_{i}) = \textrm{cst} \propto c^3 \ .
\end{equation}
Therefore, the electromagnetic deformation $\,c\,$ is essential for the existence of the J-fold CFT$_{3}$'s serving as IR fixed points in the RG flows from the deformed $\,\mathcal{N}=4\,$ SYM$_{4}$.

\subsection{Uplifting the UV: D3-brane at $\,c=0\,$}
\label{sec:D3-brane_c=0}

In this section we uplift the four-dimensional solution (\ref{analytic_sol_c=0}) obtained at $\,c=0\,$ to ten-dimensional type IIB supergravity and connect it (locally) to the D3-brane solution.

\subsubsection*{Vanishing axions}

Let us first set the three axions $\,\chi^{(0)}_{1,2,3}=0\,$ so that the largest amount of supersymmetry is preserved (see discussion below (\ref{analytic_m3/2_c=0_N8})). Then the four-dimensional solution contains two arbitrary parameters $\,(g,\Phi_{0})\,$ and uplifts to a ten-dimensional type IIB background with a factorised internal geometry of the form $\,\textrm{S}^1 \times \textrm{S}^5\,$ in the limit of an infinite radius for $\,\textrm{S}^1\,$ so that $\,\textrm{S}^1 \rightarrow \mathbb{R}\,$. The five-sphere is \textit{round} and displays its largest possible $\,\textrm{SO}(6)\,$ symmetry. The various ten-dimensional fields are given by
\begin{equation}
\label{10D_solution_c=0}
\begin{array}{rll}
ds_{10}^2 &=& \tfrac{1}{2} \,  \Delta^{-1} \, ds^2_{\textrm{DW}_{4}}  +  \Delta^{2} \,  d\eta^{2} +  g^{-2} \, d\mathring{s}_{\textrm{S}^5}^2   \ , \\[6mm]
\widetilde{F}_{5}  &=&  4 \,g \, (1 + \star) \, \textrm{vol}_{5} \ , \\[4mm]
m_{\alpha \beta} &=& 
\left(  
\begin{array}{cc}
e^{- \Phi_{0}} & 0 \\[1mm]
0 & e^{\Phi_{0}}
\end{array} 
\right) 
\hspace{10mm} \textrm{ with } \hspace{10mm}  \Phi_{0} = cst \ , \\[6mm] 
\mathbb{H}^{\alpha} &=& 0 \ ,
\end{array}
\end{equation}
where $\,ds^2_{\textrm{DW}_{4}}\,$ is the domain-wall metric displayed in (\ref{DW4_metric}) and $\, \textrm{vol}_{5}= g^{-5} \, \mathring{\textrm{vol}}_{5}\,$. The metric on the round $\,\textrm{S}^5\,$ of unit radius is given by
\begin{equation}
\label{metric_round_S5}
d\mathring{s}_{\textrm{S}^5}^2 = G_{ij} \,  dy^{i} \, dy^{j}
\hspace{8mm} \textrm{ with } \hspace{8mm}
G_{ij} = \hat{G}_{ij} = \delta_{ij} \, + \, \delta_{ik} \, \delta_{jl} \,  \frac{y^{k} \, y^{l}}{\big(1- |\vec{y}\,|^2 \big)} \ ,
\end{equation}
and is normalised as $\,R_{ij} = 4 \, G_{ij}\,$. Lastly, the warping function $\,\Delta(z)\,$ takes the simple form
\begin{equation}
\label{Delta_10D}
\Delta = \frac{(g \, z)^2}{8}  = \textrm{Im}z_{1,2,3} \ .
\end{equation}
At first sight the DW$_{4}$ metric in (\ref{10D_solution_c=0}) seems to break conformal invariance. However, the five-dimensional piece of the metric (\ref{10D_solution_c=0}) spanned by the domain-wall and the coordinate $\,\eta\,$ can be recast as an AdS$_{5}$ metric
\begin{equation}
\label{AdS5_metric}
\begin{array}{llll}
 \tfrac{1}{2}  \, \Delta^{-1} \,\,  ds^2_{\textrm{DW}_{4}}  + \Delta^{2} \, d\eta^{2} & = &  \, (g z)^{4} \, \big( 4 \, \eta_{\alpha\beta} \, dx^{\alpha} \,  dx^{\beta} +   2^{-6} \, d\eta^{2}  \big) + \dfrac{4}{(g z)^2}  \, dz^2 \\[2mm]
& = &  \dfrac{g^{-2}}{r^2} \, \left( -dt^2 + dx^2 + dy^2 +  dw^{2} + dr^2  \right)\\[4mm]
&  = & ds_{\textrm{AdS}_{5}}^2 \ ,
\end{array}
\end{equation}
upon a change of coordinates 
\begin{equation}
t= 2  \, x^0
\hspace{2mm} , \hspace{2mm} 
x= 2 \, x^1
\hspace{2mm} , \hspace{2mm} 
y= 2 \, x^2
\hspace{2mm} , \hspace{2mm} 
w = 2^{-3} \,  \eta
\hspace{2mm} , \hspace{2mm} 
r =  \frac{g^{-1}}{(gz)^{2}}  \ ,
\end{equation}
thus recovering the maximally supersymmetric $\,\textrm{AdS}_{5} \times \textrm{S}^{5}\,$ near-horizon geometry of the D3-brane with $\,L_{\textrm{AdS}_5}=g^{-1}\,$. This is nothing but the holographic dual of $\,\mathcal{N}=4\,$ SYM$_{4}$.

\subsubsection*{Non-vanishing axions $=$ local coordinate redefinitions}

Turning on a constant value for the axions $\,\chi^{(0)}_{1,2,3} \neq 0\,$ in (\ref{analytic_sol_c=0}) does not affect the metric on the round $\,\textrm{S}^5\,$ so it is still given by $\,\hat{G}_{ij}\,$ in (\ref{metric_round_S5}). However, the $\,G_{\eta \eta}\,$ component of the metric reads
\begin{equation}
\label{G_etaeta_mod}
G_{\eta \eta} =  \Delta^2  \, + \,  \left(\chi^{(0)}_{1}\right)^2 \,\, (\mathcal{Y}_{\underline{2}}{}^2 + \mathcal{Y}_{\underline{3}}{}^2) \, + \,  \left(\chi^{(0)}_{2}\right)^2 \,\, (\mathcal{Y}_{\underline{4}}{}^2 + \mathcal{Y}_{\underline{5}}{}^2) \, + \,\,  \left(\chi^{(0)}_{3}\right)^2 \,\, (\mathcal{Y}_{\underline{6}}{}^2 + \mathcal{Y}_{\underline{7}}{}^2) \ ,
\end{equation}
and the $\,G_{i\eta}$ components of the metric take the form
\begin{equation}
G_{i\eta} = \big(1- |\vec{y}\,|^2 \big)^{-\frac{1}{2}}   \left( 
\begin{array}{c}
\chi^{(0)}_3 \, \mathcal{Y}_{\underline{2}} \, \mathcal{Y}_{\underline{6}}  +  \chi^{(0)}_1 \, \mathcal{Y}_{\underline{3}} \, \mathcal{Y}_{\underline{7}} \\[2mm]
\chi^{(0)}_3 \, \mathcal{Y}_{\underline{3}} \,  \mathcal{Y}_{\underline{6}} -  \chi^{(0)}_1 \, \mathcal{Y}_{\underline{2}} \, \mathcal{Y}_{\underline{7}} \\[3mm]
\chi^{(0)}_3 \, \mathcal{Y}_{\underline{4}} \, \mathcal{Y}_{\underline{6}} + \chi^{(0)}_2 \, \mathcal{Y}_{\underline{5}} \, \mathcal{Y}_{\underline{7}} \\[2mm]
\chi^{(0)}_3 \, \mathcal{Y}_{\underline{5}} \, \mathcal{Y}_{\underline{6}} - \chi^{(0)}_2 \, \mathcal{Y}_{\underline{4}} \, \mathcal{Y}_{\underline{7}} \\[3mm]
\chi^{(0)}_3 \, (\mathcal{Y}_{\underline{6}}{}^2 + \mathcal{Y}_{\underline{7}}{}^2) 
\end{array}
\right) \ ,
\end{equation}
in terms of the embedding coordinates $\,\mathcal{Y}_{\underline{m}} \,$ on $\,\mathbb{R}^6\,$ ($\,\underline{m}=\underline{2},\ldots,\underline{7}\,$) for the five-sphere of unit radius
\begin{equation}
\mathcal{Y}_{\underline{m}}=\left\lbrace  y^{i} \, , \,  \mathcal{Y}_{7} \equiv \big(1- |\vec{y}\,|^2 \big)^{\frac{1}{2}} \right\rbrace 
\hspace{8mm} \textrm{ so that } \hspace{8mm}
\delta^{\underline{mn}} \, \mathcal{Y}_{\underline{m}} \, \mathcal{Y}_{\underline{n}} = 1 \ .
\end{equation}
Therefore, the reconstruction of a direct product metric $\,\textrm{AdS}_{5} \times \textrm{S}^{5}\,$, as performed when $\,\chi^{(0)}_{1,2,3}=0\,$, is no longer obvious due to the axion-induced terms in (\ref{G_etaeta_mod}). Nonetheless, for arbitrary (constant) values of $\,\chi^{(0)}_{1,2,3}\,$, the ten-dimensional metric
\begin{equation}
\begin{array}{lll}
ds_{10}^2 &=& \tfrac{1}{2}  \, \Delta^{-1} \, ds^2_{\textrm{DW}_{4}}  +  G_{\eta\eta} \,  d\eta^{2} + 2 \, g^{-1} \, G_{i\eta} \,\,  dy^{i} \, d\eta + g^{-2} \, G_{ij} \,\,  dy^{i} \, dy^{j}   \ , 
\end{array}
\end{equation}
with $\,\Delta\,$ given in (\ref{Delta_10D}), can be related to the one with $\,\chi^{(0)}_{1,2,3}=0\,$ by a local change of coordinates as we show now.

Firstly, it is possible to get a better understanding of the internal geometry by moving to an angular parameterisation exploiting the $\,\textrm{SO}(2)^3\,$ symmetry that independently rotates the planes $\,(\mathcal{Y}_2,\mathcal{Y}_3)\,$, $\,(\mathcal{Y}_4,\mathcal{Y}_5)\,$ and $\,(\mathcal{Y}_6,\mathcal{Y}_7)\,$. This parameterisation is given by
\begin{equation}
\label{Y_coords_SO(2)^3}
\begin{array}{lllllllllll}
\mathcal{Y}_2 &=& \cos\alpha \, \sin\theta_1 & , & 
\mathcal{Y}_4 &=& \sin\alpha \, \cos\beta \, \sin\theta_2 & , & \mathcal{Y}_6 &=& \sin\alpha \, \sin\beta \, \cos\theta_3 \ , \\[2mm]
\mathcal{Y}_3 &=& \cos\alpha \, \cos\theta_1 & , &
\mathcal{Y}_5 &=& \sin\alpha \, \cos\beta \, \cos\theta_2 & .
\end{array}
\end{equation}
Using the angular variables $\,(\alpha \,,\, \beta \,,\, \theta_{1} \,,\, \theta_{2} \,,\, \theta_{3})\,$ the metric on the round $\,\textrm{S}^{5}\,$ of unit radius 	displays its three commuting translational isometries (shifts along $\,\theta_{1,2,3}\,$) and takes the form
\begin{equation}
\label{metric_S5_angles}
d\mathring{s}_{\textrm{S}^5}^2 = d\alpha^2  +  \cos^{2} \alpha \,  d\theta_{1}^2 + \sin^2 \alpha \,  \left( \,  d\beta^2 + \cos^2\beta  \, d\theta_{2}^2 + \sin^2 \beta \, d\theta_{3}^2 \, \right) \ .
\end{equation}
The mixed terms $\,G_{i\eta} \,\,  dy^{i} \, d\eta\,$ in the internal geometry are expressed in terms of an axion-dependent one-form
\begin{equation}
\begin{array}{lll}
\Xi_{(1)}  \equiv G_{i\eta} \,\,  dy^{i} =   \chi^{(0)}_1 \, \cos^2\alpha \, d\theta_1  +  \sin^2\alpha \left(  \chi^{(0)}_2 \, \cos^2\beta \, d\theta_{2}  -  \chi^{(0)}_3  \, \sin^2\beta \, d\theta_3 \right) \ .
\end{array}
\end{equation}
Finally, when expressed in terms of the angular variables, the $\,G_{\eta\eta}\,$ component of the internal metric in (\ref{G_etaeta_mod}) reduces to
\begin{equation}
G_{\eta\eta} = \Delta^2  \, + \,  \left(\chi^{(0)}_{1}\right)^{2} \, \cos^{2}\alpha \, + \,  \sin^{2}\alpha \left(     \left(\chi^{(0)}_{2}\right)^{2} \, \cos^{2}\beta  +   \left(\chi^{(0)}_{3}\right)^{2} \, \sin^{2}\beta \right) \ . 
\end{equation}
Bringing the various pieces of the external and internal geometry together, and using the angular variables to describe the latter, one finds a ten-dimensional metric of the form
\begin{equation}
\label{10D_metric_c=0_chi_123}
\begin{array}{lll}
ds_{10}^2 &=& \tfrac{1}{2} \, \Delta^{-1} \, ds^2_{\textrm{DW}_{4}}  + \Delta^2 \,  d\eta^2  +  g^{-2} \left[  d\alpha^2 +  \cos^{2}\alpha \, \left( d \theta_{1} + g \, \chi^{(0)}_{1} \, d\eta \right)^2 \right.
\\[2mm]
&& \left. \quad \quad + \, \sin^{2}\alpha \,  \Big(  d\beta^2 +  \cos^{2}\beta \, \left( d \theta_{2} + g \, \chi^{(0)}_{2} \, d\eta \right)^2 + \sin^{2}\beta \, \left( d \theta_{3} - g \, \chi^{(0)}_{3} \, d\eta \right)^2  \Big)  \right] \ ,
\end{array}
\end{equation}
with $\,\Delta\,$ given in (\ref{Delta_10D}). It now becomes obvious that redefining the angular variables as
\begin{equation}
\label{cob_angles}
\theta_{1}' = \theta_{1} + g \, \chi^{(0)}_{1} \, \eta
\hspace{8mm} , \hspace{8mm}
\theta_{2}' = \theta_{2} + g \, \chi^{(0)}_{2} \, \eta
\hspace{8mm} , \hspace{8mm}
\theta'_{3} = \theta_{3} - g \, \chi^{(0)}_{3} \, \eta  \ ,
\end{equation}
with arbitrary (constant) axions $\,\chi^{(0)}_{i}\,$ makes the $\,\textrm{S}^5\,$ geometry in (\ref{10D_metric_c=0_chi_123}) go back to its form in (\ref{metric_S5_angles}), namely,
\begin{equation}
\label{metric_S5_angles_prime}
d{s}_{\textrm{S}^5}'^2 = d\alpha^2  +  \cos^{2} \alpha \,  d\theta_{1}'^2 + \sin^2 \alpha \,  \left( \,  d\beta^2 + \cos^2\beta  \, d\theta_{2}'^2 + \sin^2 \beta \, d\theta_{3}'^2 \, \right)  \ .
\end{equation}
As a result, the ten-dimensional geometry reduces \textit{locally} to $\,\textrm{AdS}_{5} \times \textrm{S}^{5}\,$ and the type IIB backgrounds are still given by (\ref{10D_solution_c=0}) in terms of the $\, \textrm{S}^{5}\,$ redefined volume $\,\textrm{vol}'_{5}\,$. Consequently, the resulting type IIB backgrounds with $\,\chi^{(0)}_{1,2,3} \neq 0\,$ are \textit{locally} equivalent to the one in (\ref{10D_solution_c=0}) with $\,{\chi^{(0)}_{1,2,3} = 0}\,$ upon the change of coordinates in (\ref{cob_angles}). However, a global obstruction to this equivalence arises as a consequence of the coordinate redefinitions in (\ref{cob_angles}) as we will see in detail in Section~\ref{sec:axions_SU3}. This gets reflected in the amount of supersymmetry preserved by the background which becomes maximal if $\,\chi^{(0)}_{1,2,3} =0\,$ (see analysis below (\ref{analytic_m3/2_c=0_N8})).

\subsection{Uplifting the UV: deformed D3-brane at $\,c \neq 0\,$}

In this section we investigate various aspects of the $\,\mathbb{Z}_{2} \times \textrm{SU}(3)\,$ invariant sector of the ${[\textrm{SO}(1,1) \times \textrm{SO}(6)] \ltimes \mathbb{R}^{12}} \,$ gauged supergravity which is obtained upon identifying the scalar fields in the $\,\mathbb{Z}_{2}^3\,$ invariant sector as
\begin{equation}
\label{Z2xSU(3)_truncation}
z_{1}=z_{2}=z_{3} \equiv z_{1,2,3} 
\hspace{10mm} \textrm{and} \hspace{10mm}
z_{4}=z_{5}=z_{6}=z_{7}\equiv z_{4,5,6,7}  \ .
\end{equation}
This sector of the maximal theory is of special interest (see Appendix~\ref{app:5Dvs4D} for details on its group theoretical embedding). Perturbing the four-dimensional incarnation of the D3-brane solution in (\ref{analytic_sol_c=0}), and solving the BPS equations perturbatively in the parameter $\,c\,$, we found the universal solution in (\ref{analytic_sol_cn_no_integration_constants}) which, as already emphasised, is compatible with (\ref{Z2xSU(3)_truncation}). It will also provide us with a starting point to discuss axion deformations of $\,\mathcal{N}=1\,$ S-folds later on in Section~\ref{sec:axions_SU3}.

\subsubsection{Type IIB uplift of the $\,\mathbb{Z}_{2} \times \textrm{SU}(3)$-invariant sector}
\label{sec:Uplift_IIB}

Fetching techniques from $\,\textrm{E}_{7(7)}\,$ Exceptional Field Theory ($\textrm{E}_{7(7)}$-ExFT) \cite{Hohm:2013uia,Hohm:2014qga}, the type IIB uplift of the $\,\mathbb{Z}_{2} \times \textrm{SU}(3)\,$ invariant sector of the $\,{[\textrm{SO}(1,1) \times \textrm{SO}(6)] \ltimes \mathbb{R}^{12}}\,$ maximal supergravity can be straightforwardly obtained. This sector is compatible with the scalar identifications in (\ref{Z2xSU(3)_truncation}). 

\subsubsection*{Ten-dimensional metric}

The ten-dimensional metric takes the form
\begin{equation}
\label{10D_metric_UV}
ds_{10}^2 =  \tfrac{1}{2} \,  \Delta^{-1} \, \left( ds^2_{\textrm{DW}_{4}}  + 2 \, (g c)^{-2} \, \Delta \, H(z_i) \,  d\eta^{2} \, \right) +  g^{-2} \, F(z_i) \, \left[ \, ds_{\mathbb{CP}_2}^2 + F(z_i)^{-2} \,  \boldsymbol{\eta}^2 \right]\ ,
\end{equation}
in terms of a four-dimensional space-time given by $\,ds^2_{\textrm{DW}_{4}}\,$ in (\ref{DW4_metric}) and an internal space $\,\textrm{M}_{6}=\textrm{S}_{\eta}^{1} \rtimes \textrm{S}^{5}\,$ with $\,\textrm{S}^{5} = \mathbb{CP}_{2} \rtimes \textrm{S}^{1}\,$. We refer the reader to Appendix A of \cite{Guarino:2019oct} for a detailed description of the SU(2)-structure on the five-sphere $\,\textrm{S}^{5} = \mathbb{CP}_{2} \rtimes \textrm{S}^{1}\,$ when viewed as a Sasaki-Einstein manifold. As we will discuss in detail in Section~\ref{sec:axions_SU3}, a non-trivial monodromy on $\,\textrm{M}_{6}\,$ is induced by the set of non-zero (constant) axions $\,\textrm{Re}z_{1,2,3}\,$ in the type IIB backgrounds so that the real one-form $\,\boldsymbol{\eta}\,$ in (\ref{10D_metric_UV}) is given by
\begin{equation}
\label{U(1)_fibration}
\boldsymbol{\eta} = d\beta + \textrm{Re}z_{1,2,3} \,\, d\eta + \boldsymbol{A}_{1}  \ .
\end{equation}
In (\ref{U(1)_fibration}), $\,\boldsymbol{A}_{1}=\frac{1}{2} \sin^2\alpha \, \sigma_{3}\,$ is the one-form potential on $\,\mathbb{CP}_2\,$ such that $\,d\boldsymbol{A}_{1} = 2\boldsymbol{J}\,$ with $\,\boldsymbol{J}\,$ being the real two-form specifying the SU(2)-structure of the five-sphere. In addition, the torsion conditions for the SU(2)-structure of $\,\textrm{S}^5\,$ further involve a complex two-form $\,\boldsymbol{\Omega}\,$ (together with $\,\boldsymbol{\eta}\,$ and $\,\boldsymbol{J}\,$) and read
\begin{equation}
\label{CP2_diff_geometry}
d  (\boldsymbol{\eta}-\textrm{Re}z_{1,2,3} \, d\eta) = \boldsymbol{J}
\hspace{5mm} , \hspace{5mm}
d\boldsymbol{J}=0
\hspace{5mm} , \hspace{5mm}
d\boldsymbol{\Omega}= 3 \,i \,  (\boldsymbol{\eta}-\textrm{Re}z_{1,2,3} \, d\eta) \,  \wedge \, \boldsymbol\Omega \ .
\end{equation}
The standard Fubini-Study metric on $\,\mathbb{CP}_2\,$ appears in (\ref{10D_metric_UV}) which takes the form
\begin{equation}
ds_{\mathbb{CP}_2}^2 = d\alpha^2 + \left( \frac{\sin\alpha }{2}\right)^2 \left( \sigma_{1}^2 + \sigma_{2}^2 + \cos^2\alpha \, \sigma_{3}^2 \right) \ ,
\end{equation}
in terms of a set of SU(2)-invariant one-forms
\begin{equation}
\label{sigma_CP2}
\begin{array}{rll}
\sigma_{1} & = & -\sin\psi \, d\theta + \cos\psi \, \sin\theta \, d\phi \ , \\[2mm]
\sigma_{2} & = &  \cos\psi \, d\theta + \sin\psi \, \sin\theta \, d\phi  \ , \\[2mm]
\sigma_{3} & = & d\psi + \cos\theta \, d\phi \ .
\end{array}
\end{equation}
It will also be useful to introduce a set of frame fields on $\,\mathbb{CP}_2\,$ such that $\,ds_{\mathbb{CP}_2}^2 =  \delta_{ab} \, e^{a} \, e^{b} \,$ with
\begin{equation}
\label{frames_CP2}
e^{0} = d\alpha 
\hspace{4mm} ,  \hspace{4mm}
e^{1} = \tfrac{1}{2}\sin\alpha\, \sigma_1
\hspace{4mm} ,  \hspace{4mm}
e^{2} = \tfrac{1}{2}\sin\alpha\, \sigma_2
\hspace{4mm} ,  \hspace{4mm}
e^{3} = \tfrac{1}{2}\sin\alpha\,\cos\alpha\, \sigma_3 \ .
\end{equation}
This assignment of frames on $\,\mathbb{CP}_2\,$ is compatible with a choice of embedding coordinates $\,\mathcal{Y}_{\underline{m}}\,$ on the five-sphere of the form
\begin{equation}
\label{Y_coords_SU(3)}
\begin{array}{lllllllllll}
\mathcal{Y}_2 &=& \sin\alpha \, \cos\frac{\theta}{2} \, \cos\left(  \frac{\psi+\phi}{2}+\beta\right) & , & 
\mathcal{Y}_4 &=& \sin\alpha \, \sin\frac{\theta}{2} \, \cos\left(  \frac{\psi-\phi}{2}+\beta\right) & , &  \\[2mm]
\mathcal{Y}_3 &=& \sin\alpha \, \cos\frac{\theta}{2} \, \sin\left(  \frac{\psi+\phi}{2}+\beta\right) & , &
\mathcal{Y}_5 &=& \sin\alpha \, \sin\frac{\theta}{2} \, \sin\left(  \frac{\psi-\phi}{2}+\beta\right) & ,\\[2mm]
\mathcal{Y}_6 &=& - \cos\alpha \, \sin\beta \ .
\end{array}
\end{equation}
As a result of the type IIB uplift, the metric in (\ref{10D_metric_UV}) depends on two scalar-dependent functions
\begin{equation}
\label{H_F_functions}
F(z_i) 
= |z_{4,5,6,7}|^{-1} \,\, \textrm{Im}z_{4,5,6,7}
\hspace{10mm} \textrm{and} \hspace{10mm}
H(z_i)  = F(z_i)^{-1}  \,\,  (\textrm{Im}z_{1,2,3})^2   \ ,
\end{equation}
and the warping factor
\begin{equation}
\label{Delta_UV}
\Delta =  F(z_i) \,\, \textrm{Im}z_{1,2,3}  \ .
\end{equation}
Note that the axions $\,\textrm{Re}z_{1,2,3}\,$ enter the ten-dimensional geometry (\ref{10D_metric_UV}) exclusively through the one-form $\,\boldsymbol{\eta}\,$ in (\ref{U(1)_fibration}).

\subsubsection*{Axion-dilaton and background fluxes}

The rest of the type IIB fields can systematically be obtained. The two-form potentials $\,\mathbb{B}^{\alpha} = (B_{2},C_{2})\,$ are given by
\begin{equation}
\label{twist_B}
\mathbb{B}^{\alpha} = A^{\alpha}{}_{\beta} \, \mathfrak{b}^{\beta} \ , 
\end{equation}
in terms of the $\eta$-dependent $\,\textrm{SL}(2)_{\textrm{IIB}}\,$ hyperbolic twist matrix
\begin{equation}
\label{A-twist}
A^{\alpha}{}_{\beta} = \left(
\begin{array}{rr}
\cosh\eta & \sinh\eta \\
\sinh\eta  &  \cosh\eta
\end{array}
\right) 
\hspace{5mm} , \hspace{5mm}
(A^{-1})^{\alpha}{}_{\beta} = \left(
\begin{array}{rr}
\cosh\eta & -\sinh\eta \\
-\sinh\eta  &  \cosh\eta
\end{array}
\right) 
\ ,
\end{equation}
and the complex combination of potentials
\begin{equation}
\label{mathfrakb}
\begin{array}{lll}
\mathfrak{b}^2 + i \, |z_{4,5,6,7}|^2\, \mathfrak{b}^1 &=& i \, g^{-2} \,  e^{3 i \beta} \,\, \text{Re}z_{4,5,6,7} \,\,  \left( \mathrm{d}\alpha + i\,\tfrac{1}{4} \sin(2\alpha)\,\sigma_3 \right)  \wedge  \left(  \tfrac{1}{2} \sin(\alpha) (\sigma_1 - i \, \sigma_2)  \right) \\[2mm]
&=&  -  i \, g^{-2} \,\text{Re}z_{4,5,6,7} \,\,  \boldsymbol{\Omega} \ .
\end{array}
\end{equation}
Equivalently,
\begin{equation}
\label{mathfrakb_components}
g^{2} \, \mathfrak{b}^{1} = -  |z_{4,5,6,7}|^{-2}\,\, \textrm{Re}z_{4,5,6,7} \,\, \textrm{Re}\boldsymbol{\Omega}
\hspace{8mm} ,  \hspace{8mm}
g^{2} \, \mathfrak{b}^{2} = \textrm{Re}z_{4,5,6,7}  \,\, \textrm{Im}\boldsymbol{\Omega} \ .
\end{equation}
The associated three-form field strengths $\,\mathbb{H}^{\alpha}=d\mathbb{B}^{\alpha}= (H_{3},F_{3})\,$ are directly computed from (\ref{twist_B}), (\ref{A-twist}) and (\ref{mathfrakb}). They take the form\footnote{We have used the relation $\,\partial_{\eta}{A}=A \, \theta^{t}=A \, \theta\,$.}
\begin{equation}
\label{H_SU3}
\mathbb{H}^{\alpha} 
=
A^{\alpha}{}_{\beta} \, \left( \,  d\eta \wedge
\mathfrak{b}^{\gamma} \, \theta_{\gamma}{}^{\beta}   +  d\mathfrak{b}^{\beta}  \, \right) \ ,
%
\end{equation}
with $\,\mathfrak{b}^{\alpha}\,$ given in (\ref{mathfrakb_components}) and
\begin{equation}
\label{dfrakb}
\begin{array}{lll}
g^{2} \, d\mathfrak{b}^{1} &=& -  d\left(\dfrac{\textrm{Re}z_{4,5,6,7}}{|z_{4,5,6,7}|^2}\right) \wedge  \textrm{Re}\boldsymbol{\Omega}  + 3 \, \dfrac{\textrm{Re}z_{4,5,6,7}}{|z_{4,5,6,7}|^2} \,\, (\boldsymbol{\eta}-\textrm{Re}z_{1,2,3} \, d\eta)  \wedge \textrm{Im}\boldsymbol\Omega \ , \\[6mm]
g^{2} \, d\mathfrak{b}^{2}  &=&  d\textrm{Re}z_{4,5,6,7}  \wedge \textrm{Im}\boldsymbol{\Omega}  + 3 \, \textrm{Re}z_{4,5,6,7}  \,\,  (\boldsymbol{\eta}-\textrm{Re}z_{1,2,3} \, d\eta)  \wedge \textrm{Re}\boldsymbol\Omega \ .
\end{array}
\end{equation}
In (\ref{H_SU3}) we have introduced the constant matrix 
\begin{equation}
\theta_{\gamma}{}^{\beta}  =\begin{pmatrix}
0 & 1 \\
1 & 0 
\end{pmatrix} \ ,
\end{equation}
and used the third of the SU(2)-structure relations in (\ref{CP2_diff_geometry}). Note that neither the gauge potentials in (\ref{mathfrakb_components}) nor the field strengths in (\ref{H_SU3}) depend explicitly on the axions $\,\textrm{Re}z_{1,2,3}\,$.

The axion-dilaton matrix is given by
\begin{equation}
\label{m_matrix_SU3_main}
m_{\alpha \beta} = (A^{-t})_{\alpha}{}^{\gamma}  \, \mathfrak{m}_{\gamma \delta}  \,  (A^{-1})^{\delta}{}_{\beta}       
\hspace{8mm} \textrm{ with } \hspace{8mm} 
\mathfrak{m}_{\gamma \delta} = \begin{pmatrix}|z_{4,5,6,7}|^2& 0\\0&|z_{4,5,6,7}|^{-2}\end{pmatrix} \ ,
\end{equation}
thus also being independent of the axions $\,\textrm{Re}z_{1,2,3}\,$. Finally the self-dual five-form field strength depends on the axions $\,\textrm{Re}z_{1,2,3}\,$ and is given by
\begin{equation}
\label{Ftilde5_SU3}
\begin{array}{lll}
\widetilde{F}_5 &=& g \,  (1+ \star)  \, \Big[ \, \Big(4 - 6 \left(1 - F(z_{i})^2 \right) \Big) \,  \textrm{vol}_{\mathbb{CP}_2}   \wedge (\boldsymbol{\eta}-\textrm{Re}z_{1,2,3} \, d\eta)  \\[4mm]
&+& \Big ( 4 \, \textrm{Re}z_{1,2,3}  + (\textrm{Re}z_{4,5,6,7})^2  \,  \left(1-|z_{4,5,6,7}|^{-4} \right)  \Big) \, \textrm{vol}_{\mathbb{CP}_2} \wedge  d\eta \\[4mm]
&-& d \textrm{Re}z_{1,2,3} \,  \wedge d\eta \wedge \boldsymbol{J} \wedge  (\boldsymbol{\eta}-\textrm{Re}z_{1,2,3} \, d\eta) \,  \Big] \ .
\end{array}
\end{equation}
This concludes the type IIB uplift of the $\,\mathbb{Z}_{2} \times \textrm{SU}(3)\,$ invariant sector of the $\,[\textrm{SO}(1,1) \times \textrm{SO}(6)] \ltimes \mathbb{R}^{12}\,$ maximal supergravity.

\subsubsection*{Recovering IR and UV geometries}

In order to recover the ten-dimensional geometries for the S-folds and the D3-brane, we will express the metric (\ref{10D_metric_UV}) as
\begin{equation}
\label{10D_metric_UV_2}
\begin{array}{lll}
ds_{10}^2 
&=& \tfrac{1}{2} \,  F(z_i)^{-1}  \, \left[  \, e^{B(\rho)} \, \eta_{\alpha \beta} \, dx^{\alpha} dx^{\beta}  + (g c)^{-2} \, e^{C(\rho)} \,  dw^{2} + d\rho^2\, \right] \\[4mm]
&+&  g^{-2} \,  \left[ \, F(z_i) \, ds_{\mathbb{CP}_2}^2 + F(z_i)^{-1} \,  \boldsymbol{\eta}^2 \right] \ ,
\end{array}
\end{equation}
with $\,\alpha=0,1,2\,$, in terms of a rescaled coordinate $\,w=2^{-3} \, \eta\,$, a new radial coordinate $\,\rho\,$ defined as
\begin{equation}
d\rho = (\textrm{Im}z_{1,2,3})^{-\frac{1}{2}} \, dz  \,\,\,\,\, \Rightarrow \,\,\,\,\, \rho = \int (\textrm{Im}z_{1,2,3})^{-\frac{1}{2}} \, dz \ ,
\end{equation}
and the functions
\begin{equation}
e^{B} = (\textrm{Im}z_{1,2,3})^{-1} \, e^{2 A}
\hspace{10mm} \textrm{ and } \hspace{10mm}
e^{C} =    2^7 \,  (\textrm{Im}z_{1,2,3})^2 \ .
\end{equation}

Let us first consider the deep IR region of the BPS domain-walls constructed in Section~\ref{sec:RG flows}. When approaching this region the  scalars get a constant value. Then $\,\rho = (\textrm{Im}z_{1,2,3})^{-\frac{1}{2}} \,  z\,$ and $\,A=L^{-1}\, z\,$ so that  
\begin{equation}
e^{B} \sim e^{2 \, (\textrm{Im}z_{1,2,3})^{\frac{1}{2}} \,  L^{-1} \rho}
\hspace{8mm} , \hspace{8mm}
e^{C} \sim \textrm{cst} \ ,
\end{equation}
and the metric (\ref{10D_metric_UV_2}) boils down to the $\,\textrm{AdS}_{4} \times \textrm{S}^{1} \times \textrm{S}^5\,$ metric for the $\,\mathcal{N}=1\,$ S-fold in \cite{Guarino:2019oct}. On the other hand, substituting the analytic D$3$-brane solution (\ref{analytic_sol_c=0}) obtained at $\,c=0\,$\footnote{Note that the factor $\,(gc)^{-2}\,$ in the metric (\ref{10D_metric_UV_2}) becomes pathological in this limit and must be just removed. Also, the $A$-twist matrix in (\ref{A-twist}) must be taken to be the identity when $\,c=0\,$.} yields $\,F(z_{i})=1\,$ together with
\begin{equation}
e^{B}= 8 \, e^{\sqrt{2} \, (g \rho)}
\hspace{5mm} , \hspace{5mm}
e^{C}= 2 \, e^{\sqrt{2} \, (g \rho)}
\hspace{8mm} \textrm{ with } \hspace{8mm}
e^{g \rho}=(g z)^{2 \sqrt{2}} \ .
\end{equation}
Then the expected $\,\textrm{AdS}_{5} \times \textrm{S}^{5}\,$ local geometry with a round $\,\textrm{S}^{5}\,$ metric $\,d\mathring{s}_{\textrm{S}^5}^2 =ds_{\mathbb{CP}_2}^2 + \boldsymbol{\eta}^2\,$ is recovered. Note also that, since $\,\textrm{Re}z_{1,2,3}\,$ take constant values in the D3-brane solution (\ref{analytic_sol_c=0}), one has that (\ref{U(1)_fibration}) can be re-expressed as $\,\boldsymbol{\eta} = d (\beta + \textrm{Re}z_{1,2,3} \,\, \eta)  + \boldsymbol{A}_{1} \,$, thus redefining the coordinate $\,\beta\,$ along the $\,\textrm{U}(1)\,$ fiber in $\,\textrm{S}^{5} = \mathbb{CP}_{2} \rtimes \textrm{S}^{1}\,$ as 
\begin{equation}
\label{betatobeta'}
\beta \rightarrow \beta + \textrm{Re}z_{1,2,3} \,\, \eta \ .
\end{equation}
This coordinate redefinition is the analogue of (\ref{cob_angles}) and, therefore, the same global \textit{vs} local issues regarding the periodicity of the new angular variable apply here. This will be discussed in more detail in Section~\ref{sec:axions_SU3}.

\subsubsection{Deformed D3-brane and anisotropic $\,\textrm{SYM}_{4}\,$}

Let us investigate the ten-dimensional type IIB uplift of the domain-walls constructed in Sections~\ref{sec:N=1_SYM}, \ref{sec:N=2_SYM} and \ref{sec:N=4_SYM} when approaching the UV ($\,z \rightarrow \infty\,$). We will holographically relate such a UV behaviour to having an anisotropic deformation of SYM$_4$.

\subsubsection*{Five-dimensional picture}

Setting $\,c\neq 0\,$ modifies the AdS$_{5}$ metric in (\ref{AdS5_metric}) so that it acquires a dependence on the scalar fields
\begin{equation}
\label{AdS5_metric_2}
ds_{5}^2 = \tfrac{1}{2} \,  \Delta_{\textrm{UV}}^{-1} \, \left( ds^2_{\textrm{DW}_{4}}  + 2 \, (gc)^{-2} \,  \Delta_{\textrm{UV}} \, H(z_i) \,  d\eta^{2} \, \right) \ ,
\end{equation}
in terms of the functions $\,F(z_i)\,$ and $\,H(z_i)\,$ in (\ref{H_F_functions}) and the warping factor $\,\Delta\,$ in (\ref{Delta_UV}).
Consistently, $\,F(z_i)=1\,$ and $\,H(z_i)=\Delta^2\,$ when $\,c=0\,$ so that the undeformed AdS$_{5}$ metric in (\ref{AdS5_metric}) is recovered. Note that the deformed $\,\textrm{AdS}_{5}\,$ metric in (\ref{AdS5_metric_2}) singles out the coordinate $\,\eta\,$ which, in the deep IR, will span the  $\,\textrm{S}^1$ factor of the $\,\textrm{AdS}_{4} \times \textrm{S}^{1} \times \textrm{S}^5\,$ S-fold geometry \cite{Inverso:2016eet,Guarino:2019oct,Guarino:2020gfe}. This is the same coordinate along which the type IIB fields acquire a dependence on as a consequence of the $\,\textrm{SL}(2)_{\textrm{IIB}}\,$ monodromy in (\ref{A-twist}). However, the RG flows constructed in Section~\ref{sec:RG flows} occur along the radial direction $\,gz\,$. Therefore, the dependence of the type IIB fields $\,m_{\alpha\beta}\,$ and $\,\mathbb{B}^{\alpha}\,$ on the $\,\eta\,$ coordinate induced by the S-folding (equivalently by the parameter $\,c\,$) is ultimately connected to having an anisotropic deformation of $\,\mathcal{N}=4\,$ SYM$_{4}$ (see, \textit{e.g.} \cite{Mateos:2011ix,Jain:2014vka}). Let us look at this issue in more detail from a ten-dimensional perspective.

\subsubsection*{Ten-dimensional picture}

In order to present the ten-dimensional metric (\ref{10D_metric_UV}) in a suitable form to be compared with previous results in the literature regarding holographic anisotropy in SYM$_{4}$ (see \cite{Conde:2016hbg,Hoyos:2020zeg}), it is convenient to first perform a redefinition of the radial coordinate as
\begin{equation}
g \, \zeta = (g z)^2 \ , 
\end{equation}
and then introduce a set of metric functions 
\begin{equation}
\label{Anisotropy_functions}
h = 4 \,  \Delta_{\textrm{UV}}^{2} \,  e^{-4 A}
\hspace{5mm} , \hspace{5mm}
e^{2m}  =  2 \, e^{-2 A} \, \Delta_{\textrm{UV}} \, H  
\hspace{5mm} , \hspace{5mm}
e^{2f} =    2^4 \,  \Delta_{\textrm{UV}}^{2} \,  e^{-2 A}  \, (g \zeta)^3   \ ,
\end{equation}
so that (\ref{10D_metric_UV}) conforms with
\begin{equation}
\label{10D_metric_anisotropy}
\begin{array}{lll}
ds_{10}^2 &=& h^{-\frac{1}{2}}\,  \left( \eta_{\alpha \beta} \, dx^{\alpha} dx^{\beta} 
 + (gc)^{-2} \,  e^{2m}  \, d\eta^{2}  \right) \\[4mm]
 & + & h^{\frac{1}{2}}  \left[ \, (g \zeta)^2 \,e^{-2f}  \, d\zeta^2 +   g^{-2} \, h^{-\frac{1}{2}} \, F \,\, ds_{\mathbb{CP}_2}^2 +  g^{-2} \,  h^{-\frac{1}{2}} \, F^{-1} \,\,  \boldsymbol{\eta}^2 \right] \ .
\end{array}
\end{equation}
It then becomes transparent that the function $\,e^{2m}\,$ in (\ref{10D_metric_anisotropy}) will be responsible for anisotropy in the dual 4D field theory. Using (\ref{H_F_functions}) and the four-dimensional universal solution in (\ref{analytic_sol_c3_no_integration_constants}) the three metric functions in (\ref{Anisotropy_functions}) read
\begin{equation}
\label{Anisotropy_functions_case_1}
\begin{array}{rll}
h &=& \dfrac{1}{16 \, (g \zeta)^4} \left( 1- 16 \cosh^2\Phi_0 \, \dfrac{c^2}{(g \zeta)^2} + \ldots \right) \ ,  \\[5mm]
e^{2m}  & = &   2 \, e^{-2 A} \, (\textrm{Im}z_{1,2,3})^3 \, = \,   2^{-8} \,  \left( 1+ 64 \cosh^2\Phi_0 \, \dfrac{c^2}{(g \zeta)^2} + \ldots \right)     \ , \\[5mm]
e^{2f} & = & \dfrac{(g \zeta)^2}{4} \,  \left( 1+ 16 \, \cosh^2\Phi_0 \, \dfrac{c^2}{(g \zeta)^2} + \ldots \right)      \ .
\end{array}
\end{equation}
As a result the anisotropy function $\,e^{2m}\,$ acquires a non-trivial $\,c^2/(g \zeta)^2\,$ correction in the UV. Note that this correction does \textit{not} occur at linear order in $\,c\,$, as can be checked by direct substitution of the four-dimensional solution in (\ref{analytic_sol_c_no_integration_constants}) instead of the one in (\ref{analytic_sol_c3_no_integration_constants})\footnote{This is also the case for the non-universal solution in (\ref{analytic_sol_c})-(\ref{functions_1st_order}) computed at linear order in $\,c\,$ provided the identifications in (\ref{Z2xSU(3)_truncation}) hold.}.\footnote{Similarly, the functions occurring in front of the internal $\,\textrm{S}^5\,$ metric in (\ref{10D_metric_anisotropy}) are given by
\begin{equation}
\label{Anisotropy_functions_case_2}
\begin{array}{rll}
h^{-\frac{1}{2}} \, F & = & 4 \, (g \, \zeta)^2 \,  \left( 1- \dfrac{384}{49}  \, \cosh^2\Phi_0 \,  \Big( \, f_{1}(\Phi_0) + f_{2}(\Phi_0) \, \log(g \zeta) \, \Big) \dfrac{c^4}{(g \zeta)^4} + \ldots \right)   \ , \\[4mm]
h^{-\frac{1}{2}} \, F^{-1} & = & 4 \, (g \, \zeta)^2 \,  \left( 1+ 16 \, \cosh^2\Phi_0 \, \dfrac{c^2}{(g \zeta)^2} + \ldots \right)   \ ,
\end{array}
\end{equation}
with
\begin{equation}
f_{1}(\Phi_{0}) = 9 + \cosh(2 \Phi_0) -28 \, \sinh(2 \Phi_0)
\hspace{8mm} \textrm{ and } \hspace{8mm}
f_{2}(\Phi_{0}) = 56 \, \big(1 -3 \, \cosh(2 \Phi_0) \big)  .
\end{equation}
}

Note that, upon dimensional reduction on the five-sphere $\,\textrm{S}^5\,$, the three-form field strengths $\,\mathbb{H}^{\alpha}\,$ in (\ref{H_SU3}) give rise to an $\,\textrm{SL}(2)_{\textrm{IIB}}\,$ doublet of one-form field strengths
\begin{equation}
\mathcal{F}_{(1)}{}^{\alpha} = 
A^{\alpha}{}_{\beta}   \,  d\eta \wedge
\underbrace{\mathfrak{b}^{\gamma} \,  \theta_{\gamma}{}^{\beta}}_{\textrm{purely along $\,\textrm{S}^5\,$}}  \ ,
\end{equation}
along the $\,d\eta\,$ direction in the (external) geometry (\ref{AdS5_metric_2}). From the expressions of $\,\mathfrak{b}^{\alpha}\,$ in (\ref{mathfrakb_components}), and using the four-dimensional solution in (\ref{analytic_sol_c3_no_integration_constants}), one finds
\begin{equation}
\label{mathfrakb_components_UV}
\begin{array}{llr}
g^{2} \,  \mathfrak{b}^{1} &=& - \left( 4 \, e^{ \frac{1}{2} \Phi_{0}}  \, \cosh\Phi_{0} \,\, \dfrac{c}{g \, \zeta} + \ldots   \right) \, \textrm{Re}\boldsymbol{\Omega} \ ,\\[4mm]
g^{2} \,  \mathfrak{b}^{2} &=& \Big( 4 \, e^{ - \frac{1}{2} \Phi_{0}}  \, \cosh\Phi_{0} \,\, \dfrac{c}{g \, \zeta}  + \ldots \Big)\,\, \textrm{Im}\boldsymbol{\Omega} \ .
\end{array}
\end{equation}
Importantly, the dependence on the coordinate $\,\eta\,$ through the twist matrix $\,A(\eta)\,$ will factorise out due to the contraction with the axion-dilaton matrix $\,m_{\alpha\beta}\,$ in the ten-dimensional kinetic term $\,m_{\alpha\beta} \, \mathbb{H}^{\alpha} \wedge \star \mathbb{H}^{\beta}\,$.

Similar one-form terms along the $\,d\eta\,$ direction in the (external) geometry (\ref{AdS5_metric_2}) appear from the axion-dilaton matrix in (\ref{m_matrix_SU3_main}) with
\begin{equation}
\label{frakm_matrix_SU3}
\mathfrak{m}_{\gamma \delta} = \begin{pmatrix}
\mathfrak{m}_{+} & 0 \\ 0 & \mathfrak{m}_{-} 
\end{pmatrix} 
\hspace{8mm} \textrm{ and } \hspace{8mm}
\mathfrak{m}_{\pm}=e^{\mp\Phi_0}\left(1 \pm  32 \, \sinh(2\Phi_{0}) \, \dfrac{c^2}{(g \, \zeta)^2}  + \ldots  \right) 
\ ,
\end{equation}
and also from $\,\widetilde{F}_{5}\,$ in (\ref{Ftilde5_SU3}). From the axion-dilaton one finds
\begin{equation}
\begin{array}{lll}
\mathcal{F}_{(1) \alpha \beta} &=& -  (|z_{4,5,6,7}|^2+|z_{4,5,6,7}|^{-2}) \, (A^{-t}  \, \theta \,  A^{-1})_{\alpha \beta}  \,\, d\eta \\[2mm]
&=& - 2 \, \cosh\Phi_{0}  \left(1 -  64 \, \sinh^2\Phi_{0} \, \dfrac{c^2}{(g \, \zeta)^2}  + \ldots  \right) \, (A^{-t}  \, \theta \,  A^{-1})_{\alpha \beta}  \,\, d\eta  \ ,
\end{array}
\end{equation}
whereas from the five-form field strength one gets
\begin{equation}
\begin{array}{lll}
\widetilde{\mathcal{F}}_{(1)} &=&  \Big[ \, \frac{4}{3} \, c \, \sinh\Phi_{0} \, \left(  1- 192 \, \cosh^2\Phi_{0}  \, \dfrac{c^2}{(g \, \zeta)^2} \, \log(g \, \zeta) + \ldots \right) \\[4mm]  
&-&  \left( 16 \, \sinh(2\Phi_{0}) \, \cosh\Phi_{0}  \, \dfrac{c^2}{(g \, \zeta)^2} + \ldots \right)  \, \Big] \, d\eta  \wedge \textrm{vol}_{\mathbb{CP}_2} \ .
\end{array}
\end{equation}
We have verified the stability of the above results using the four-dimensional solution computed up to twelfth order in the parameter $\,c\,$. 

Other type IIB constructions have been put forward to generate holographic anisotropy in $\,\textrm{SYM}_{4}\,$. For instance, \cite{Conde:2016hbg,Hoyos:2020zeg} employ backreacted geometries involving D3- and (smeared) D5-branes along $\,2+1\,$ dimensions. Our setup involves only closed strings (without source terms) and differs from the one in \cite{Conde:2016hbg,Hoyos:2020zeg} in that it generates anisotropy in a purely geometric manner by implementing the locally geometric $\,\textrm{SL}(2)_{\textrm{IIB}}\,$ twist $\,A(\eta)\,$ in (\ref{A-twist}).  As a consequence of the mechanism here, the various one-form sources of anisotropy, namely, $\,(\mathcal{F}_{(1)}{}^{\alpha}\,,\,\mathcal{F}_{(1)\alpha\beta}\,,\,\widetilde{\mathcal{F}}_{(1)})\,$, organise themselves into $\,\textrm{SL}(2)_{\textrm{IIB}}\,$ multiplets. The $\,\textrm{SL}(2)_{\textrm{IIB}}\,$ covariance here could explain the difference between the $\zeta$-powers appearing in (\ref{Anisotropy_functions_case_1})-(\ref{Anisotropy_functions_case_2}) and in \cite{Hoyos:2020zeg}. Also along these lines, it would be interesting to characterise the operators triggering $\,\textrm{SYM}_{4}\,$ anisotropy in an $\,\textrm{SL}(2)_{\textrm{IIB}}\,$ covariant setup of the type investigated here. To this end, it would be helpful to have the oxidation of the RG flows presented in this work to five dimensions. Some group theoretical considerations on the $\,\textrm{4D} \leftrightarrow \textrm{5D}\,$ dictionary are presented in Appendix~\ref{app:5Dvs4D}.

\subsection{Axions and symmetry breaking}
\label{sec:axions_SU3}

The UV behaviour of the RG flows presented in Section~\ref{sec:RG flows} was shown to be controlled by the asymptotic values of the axions $\,\textrm{Re}z_{1,2,3}\,$. This was so for both $\,\textrm{SYM}_{4}\,$ and $\,\mathcal{N}=1\,$ \mbox{J-fold} CFT$_{3}$ cases. In order to have a ten-dimensional geometric description of such asymptotics, we must uplift four-dimensional solutions with arbitrary (constant) axions $\,\chi_{1,2,3}\,$. To this end we will follow the geometric construction in \cite{1852633}, and provide the uplift of a generic $\,\mathcal{N}=1\,$ S-fold with $\,\textrm{U}(1)^2\,$ symmetry. This will lead us to propose a generalisation of the mechanism in \cite{1852633} which allows us to understand the holographic breaking of a symmetry group $\,\textrm{G}\,$: $\,\textrm{G}=\textrm{SU}(3)\,$ for the $\,\mathcal{N}=1\,$ S-folds, $\,\textrm{G}=\textrm{SO}(6)\,$ for the D3-brane (and also for the non-supersymmetric S-folds in \cite{Guarino:2020gfe}), and $\,\textrm{G}=\textrm{SU}(2)\,$ for the $\,\mathcal{N} = 2\,$ S-folds. This last case was the one studied in detail in \cite{1852633}, where the implications for the structure of the Kaluza-Klein spectrum were also analysed. It would then be interesting to perform a similar analysis of the KK spectrum for the $\,\mathcal{N}=1\,$ and non-supersymmetric S-folds.

\subsubsection{$\mathcal{N}=1$ S-folds and SU(3) symmetry}
\label{sec:axions_su(3)_example}

The mechanism builds upon the morphism $\,\textrm{S}^5 \cong \textrm{SU}(3)/\textrm{SU}(2)\,$ and the largest  possible flavour symmetry group $\,\textrm{G}=\textrm{SU}(3)\,$ of the corresponding $\,\mathcal{N}=1\,$ J-fold CFT$_{3}$'s. Choosing a point $\,p_0 \in \textrm{S}^5 \subset \mathbb{C}^3\,$ we can map $\,\textrm{SU}(3) \rightarrow \textrm{S}^5 : g \rightarrow g(p_0)\,$ with isotropy group $\,\textrm{SU}(2)\,$. This map allows us to construct a right-action of $\,\textrm{SU}(3)\,$ over $\,\textrm{S}^5$ and to relate a coordinate system on $\,\textrm{SU}(3)\,$ to a coordinate system on $\textrm{S}^5\,$. There is also, a priori, a left-action of $\,\textrm{U}(1)\,$ -- the commutant of $\,\textrm{SU}(2)\,$ in $\,\textrm{SU}(3)\,$ -- on $\,\textrm{S}^5$.

When the axions are set to zero, $\,\chi_{1,2,3} = 0\,$, the internal space is the direct product $\,\textrm{M}_6 = \textrm{S}^5 \times \textrm{S}^1\,$. The coordinate $\,\eta\,$ along $\,\textrm{S}^1$ will then be periodic with $\,\eta \in [0,\,T)\,$. On the contrary, when $\,\chi_{i} \neq 0\,$, the axions induce a fibration of $\,\textrm{S}^5\,$ over $\,\textrm{S}^1\,$ characterised by a non-trivial monodromy. In order to identify such a monodromy, let us first introduce a twist element
\begin{equation}
\label{h_matrix_su3}
h(\eta) = \begin{pmatrix}e^{i\chi_1 \eta} & 0 & 0\\ 0& e^{i \chi_2 \eta}& 0 \\ 0 & 0 & e^{i\chi_3 \eta}\end{pmatrix} \in \textrm{SU}(3) \ .
\end{equation}
For this matrix to be an element of $\,\textrm{SU}(3)\,$ $\, \forall \eta \in [0,\,T)\,$ the algebraic constraint
\begin{equation}
\label{chi_SU3_monodromy}
\sum_{i=1}^{3} \chi_{i} = 0 \ ,
\end{equation}
must be imposed. Then the monodromy $\,h(T)\,$ induced by the twist element (\ref{h_matrix_su3}) is simple given by
\begin{equation}
\label{h(T)_monodromy}
h(T) = \begin{pmatrix}e^{i\chi_1 T} & 0 & 0\\ 0& e^{i \chi_2 T}& 0 \\ 0 & 0 & e^{i\chi_3 T}\end{pmatrix} \in  \textrm{U}(1)^2 \subset \textrm{SU}(3) \ ,
\end{equation}
provided (\ref{chi_SU3_monodromy}) holds. Note that (\ref{chi_SU3_monodromy}) precisely matches the constraint (\ref{chi_SU3_enhancement}) found for the axions in the $\,\mathcal{N}=1\,$ family of S-folds.

To each point $\,\eta \in \textrm{S}^1\,$ we will now associate an $\,\textrm{S}^5(\eta)\,$ parameterised by an element $\,\hat{g}  \in  \textrm{SU}(3)/\textrm{SU}(2)\,$ of the form
\begin{equation}
\hat{g} = h(\eta) \cdot g \ .
\end{equation}
This allows us to define an application
\begin{equation}
\label{phi_COB}
\phi_h : \,\,\,\, \textrm{S}^5\times \textrm{S}^1 \,\, \rightarrow \,\, \textrm{S}^5\times \textrm{S}^1/\sim \,\, : \,\,  (g \,,\, \eta) \, \rightarrow \, (h(\eta) \cdot  g \,,\, \eta) \ ,
\end{equation}
that is globally well-defined if we have the equivalence relation $\,\sim\,$ on $\,\textrm{S}^5\times \textrm{S}^1$ defined as
\begin{equation}
\left( g,\,\eta = 0\right) \sim \left(h(T)\cdot g,\,\eta = T \right) \ .
\end{equation}
Since $\,h(T) \in \textrm{G}\,$, we can use this application to map a solution with $\,\chi_{i}=0\,$ to a new solution with generic (constant) axions $\,\chi_{i}\,$. Even if we can reabsorb the axions locally, there is an obstruction to do it globally. Having a non-trivial monodromy $\,h(T)\,$ in (\ref{h(T)_monodromy}) reduces the action of $\,\textrm{G}\,$ to $\,\textrm{G}_{0} \subset \textrm{G}\,$ on $\,\textrm{S}^5\times \textrm{S}^1/\sim\,$ provided  $\,\textrm{G}_{0}\,$ commutes with $\,h(T)\,$. For the various type IIB backgrounds appearing in this work, and also for the non-supersymmetric S-folds in \cite{Guarino:2020gfe}\footnote{In passing we correct \textbf{[v1]} of \cite{Guarino:2020gfe}. In Section~$2.3.1$ there, it is said that a generic solution belonging to the non-supersymmetric family of S-folds will preserve an $\,\textrm{SU}(2)\,$ symmetry instead of a $\,\textrm{U}(1)^3\,$ symmetry.}, we have explicitly verified that the coordinate redefinitions induced by $\,\phi_h\,$ in (\ref{phi_COB}) precisely recovers the dependence of the ten-dimensional metric on the axions entering $\,h(\eta)\,$ in (\ref{h_matrix_su3}). The latter can be readily obtained by employing $\,\textrm{E}_{7(7)}$-ExFT techniques, as it was done in Section~\ref{sec:Uplift_IIB}.

As an example, let us consider the $\,\mathcal{N}=1\,$ S-fold with the largest possible symmetry $\,\textrm{G}=\textrm{SU(3)}\,$ and $\,\chi_{i}=0\,$. Turning on two axions $\,\chi_1 = -\chi_3 = n\frac{2\pi}{T}\,$ with $\,n\in \mathbb{N}\,$ yields $\,h(T) = \mathbb{I}\,$ so that $\,\textrm{G}_{0}=\textrm{SU}(3)\,$. However, turning on three axions $\,\chi_1 = \chi_2 = -\tfrac{1}{2}\chi_3\,$ as in (\ref{pair-wise_identification}) yields $\,\textrm{G}_{0}=\textrm{SU}(2) \times \textrm{U}(1)\,$. It then becomes clear that a general axion configuration will yield $\,\textrm{G}_{0}=\textrm{U}(1)^2\,$ so that $\,\textrm{G}=\textrm{SU}(3)\,$ gets broken to its Cartan subgroup. Note also that specific choices of $\,\chi_i\,$ might be of special interest, like for example fractional multiples $\,\frac{2\pi}{k T}\,$, as these generate $\,\mathbb{Z}_k\,$ monodromies.

Using the embedding coordinates $\,\mathcal{Y}_{\underline{m}} \,$ in (\ref{Y_coords_SU(3)}) the action of $\,\phi_{h}\,$ on the angular variables entering (\ref{U(1)_fibration})-(\ref{sigma_CP2}) reads
\begin{equation}
\label{cob_SU3}
\phi \rightarrow \phi \, - \,  (\chi_1 - \chi_2) \, \eta 
\hspace{3mm} , \hspace{3mm}
\psi \rightarrow \psi \,-\, (\chi_1+ \chi_2 - 2 \, \chi_3) \, \eta
\hspace{3mm} , \hspace{3mm}
\beta \rightarrow \beta - \chi_3 \, \eta \ .
\end{equation}
Imposing the axions identification $\,\chi_{1}=\chi_{2}=\chi_{3}\,$ as required by the $\,\mathbb{Z}_{2} \times \textrm{SU}(3)\,$ invariant sector in (\ref{Z2xSU(3)_truncation}) reduces (\ref{cob_SU3}) to
\begin{equation}
\beta \rightarrow \beta - \chi_3 \, \eta \ ,
\end{equation}
in perfect agreement with the ten-dimensional result obtained in (\ref{betatobeta'}). It now becomes clear that $\,\textrm{SU}(3)\,$ invariance restricts the coordinate redefinitions to those inducing a monodromy only on the U(1) fiber in $\,\textrm{S}^{5} = \mathbb{CP}_{2} \rtimes \textrm{S}^{1}\,$. Finally, note also that the change of coordinates in (\ref{cob_SU3}) is globally well-defined if $\,\chi_i = n_i \frac{2\pi}{T}\,$ with $\,\sum_i n_i = 0\,$. This will be shown in a moment to be equivalent to the case $\,\chi_{i}=0\,$.

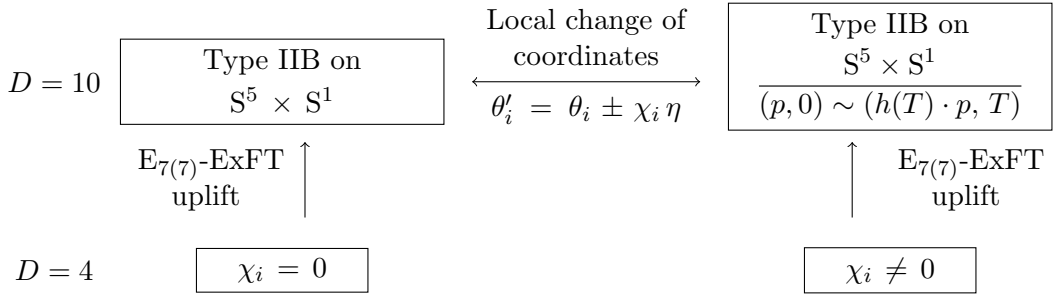
\begin{figure}[t]
\centering
\begin{tikzpicture}
	\node at(-1,4){$D=10$};
	\node at(-1,1.5){$D=4$};
	\node[draw,text width=4cm,text centered] at(2,4){Type IIB on\\[1mm ]$\textrm{S}^5 \times \textrm{S}^1$};
	\node[draw,text width=4cm,text centered] at(10,4.2){Type IIB on\\[1mm]	$\dfrac{\textrm{S}^5 \times \textrm{S}^1}{(p,0)\sim(h(T)\cdot p,\,T)}$};
	\node[draw,text width=2cm,text centered] at(10,1.5){$\chi_i \neq 0$};
	\node[draw,text width=2cm,text centered] at(2,1.5){$\chi_i = 0$};
\draw[<->](4.5,4)--(7.5,4);
\node[text width=4cm,text centered] at(6,4.2){Local change of\\ coordinates\\[2mm] $\theta_i' = \theta_i \pm \chi_i\, \eta$};
\draw[->](9.5,2.2)--(9.5,3.2);
\node[text width=1.8cm,text centered] at(11,2.7){\mbox{$\textrm{E}_{7(7)}$-ExFT}\\ uplift};
  \draw[->](2.3,2.2)--(2.3,3.2);
	\node[text width=1.8cm,text centered] at(1,2.7){\mbox{$\textrm{E}_{7(7)}$-ExFT}\\ uplift};
\end{tikzpicture}
\caption{Illustration of the symmetry breaking induced by the monodromy and connection with non-vanishing axions in four dimensions.}
\label{fig:monodromy_diagram}
\end{figure}

\subsubsection{Monodromies, mapping torus and symmetry breaking}
\label{sec:monodromies}

Let us start with a background involving an internal manifold of the form $\,\textrm{M}_n \times \textrm{S}^1\,$ where $\,\textrm{M}_n\,$ is an homogeneous space for a compact group $\,\textrm{G}\,$ and $\,\textrm{S}^1\,$ is a $T$-periodic circle parameterised by a coordinate $\,\eta\,$. We will show how to modify this factorised geometry by introducing a fibration of $\,\textrm{M}_n\,$ over $\,\textrm{S}^1\,$ in such a way that the original background is mapped to a new one. This new background is parameterised by an element $\,h \in \textrm{G}\,$ which allows us to define the \textit{mapping torus} $\,T(\textrm{M})_h\,$ as the quotient
\begin{equation}
\label{mapping_torus}
T(\textrm{M})_h = \frac{\textrm{M}_n \times \textrm{S}^1}{(p,0)\sim(h\cdot p,\,T) }
\hspace{6mm} \textrm{ with } \hspace{6mm}
h =\text{exp}\left(i \, T \, \sum\limits_{i=0}^r \chi_i \, \mathfrak{h}_i \right) \in \textrm{U}(1)^{r} \subset \textrm{G} \ ,
\end{equation}
for some $\,\chi_i \, ,\, T \in\mathbb{R}\,$ where $\,\left\lbrace\mathfrak{h}_i\right\rbrace\,$ denotes a normalised basis of the $r$-dimensional Cartan subalgebra $\,\mathfrak{h}\,$ of $\,\mathfrak{g}\,$. We must then find a map from the original background with $\,\textrm{M}_{n} \times \textrm{S}^1\,$ to a new one corresponding to a different element in $\,T(\textrm{M})_h\,$ in a consistent manner. In other words, as we go around $\,\textrm{S}^1\,$, we want to smoothly transport points on $\,\textrm{M}_n\,$ by elements of the isometry group, $\,h(\eta)\,$, such that $\,h(T)\, p = h \, p\,$ for $\,p \in \textrm{M}_n\,$. A way to do this is by choosing
\begin{equation}
h(\eta)=	\prod\limits_{i=1}^r h_i(\eta) = \text{exp}\left(i \sum\limits_{i=0}^r \chi_i \, \mathfrak{h}_i \, \eta\right) \in \textrm{U}(1)^{r} \subset \textrm{G}  \ ,
\end{equation}
so that the application
\begin{equation}
\phi_{h} : M\times S^1 \rightarrow T(\textrm{M})_h: (p,\,\eta) \rightarrow \left(h(\eta)\cdot p,\,\eta\right) \ ,
\end{equation}
consistently maps \textit{fields} on the original manifold to \textit{fields} on $\,T(\textrm{M})_h\,$. Some comments are now in order:
\begin{itemize}
\item The symmetry $\,\textrm{G}\,$ of the original background has been reduced to its subgroup $\,\textrm{G}_{0}\,$ that commutes with the element $\,h(T)\,$ for its action to be globally well-defined. This can break the symmetry $\,\textrm{G}\,$ down to a direct product of rank($\textrm{G}$) abelian $\,\textrm{U}(1)\,$ factors.

\item Locally, every operation can be reversed by a local change of coordinates so that the new background is also a solution of the (local) equations of motion.

\item The relevant information is encoded in the choice of $\,h(T)\,$. In particular, this means that the axions  $\,\chi_i\,$ have periodicity $\,\frac{2\pi}{T}\,$. Moreover, the mapping torus could be trivial in the sense that $\,T(M)_h \cong \textrm{M}_n \times \textrm{S}^1\,$ as for example when $\,h(T) = \mathbb{I}\,$.
\end{itemize}

Let us show how this construction works in simpler terms. We start by quotienting $\,\textrm{M}_n\,$ by the $\,\textrm{U}(1)^r \subset \textrm{G}\,$ generated by $\,\left\lbrace\mathfrak{h}_i \right\rbrace\,$. In this way we obtain $\,\textrm{M}_{n-r} = \textrm{M}_n/\textrm{U}(1)^r\,$ so that $\,\textrm{M}_n\,$ can be recast as $\,\mathbb{T}^r\ltimes \textrm{M}_{n-r}\,$. By choosing an appropriately normalised basis for the Cartan subalgebra, it is always possible to assign coordinates $\,\theta_i\,$ ($\,i=1,\ldots,r\,$) on $\,\mathbb{T}^{r}\,$ with periodicities of $\,2\pi\,$ such that the generators $\,h_i\,$ act on them as translations (or shifts). Then, the map
\begin{equation}
\label{phi_theta}
\phi_{h}(\theta_i) = \theta_i +\chi_i \, \eta \ , 
\end{equation}
can locally be seen as change of variables on $\,\textrm{M}_{n} \times \textrm{S}^1\,$. Note that (\ref{phi_theta}) is precisely the local coordinate transformations that appeared in (\ref{cob_angles}) and (\ref{cob_SU3}). Focussing now on the bigger torus $\,\mathbb{T}^r\ltimes \textrm{S}^1\,$, we can identify it as $\,\mathbb{R}^{r+1}/\Lambda\,$ where $\,\Lambda\,$ is the lattice
\begin{equation}
\Lambda = \left\{\sum\limits_{i=1}^r n_i\, 2\pi\, e_i + T \, n_0 \left(\sum\limits_{i=1}^r \chi_i\, e_i + e_\eta\right)\,\,\, \text{ with } \,\,\, n_i \in \mathbb{N}\right\} \ .
\end{equation}
Here $\,e_i\,$ are vectors with a non-zero entry only in the $\,i^{th}\,$ coordinate and $\,e_\eta\,$ can be thought of as $\,e_{r+1}\,$. This illustrates how $n$-tuples of $\,\chi_i\,$ producing the same lattice will give rise to the same monodromy and therefore equivalent fibrations and geometries. Examples are given by those changes of basis parameterised by elements of $\,\textrm{SL}(r+1,\,\mathbb{Z})\,$. Since we want to keep the general form of the basis in order to be able to identify the axions, we must restrict to the subgroup $\,\mathbb{Z}^r \in \textrm{SL}(r+1,\,\mathbb{Z})\,$ containing elements of the form 
\begin{equation}
\label{axions_coc}
\begin{pmatrix}
\mathbb{I}_{r\times r}&n_i\\0& 1
\end{pmatrix} 
\hspace{5mm} \text{  with  } \hspace{5mm}
n_i \in \mathbb{Z} \ .
\end{equation}
These elements act on the axions $\,\chi_i\,$ as $\,\chi_i \rightarrow \chi_i + n_i \, \frac{2 \pi}{T}\,$ which is in line with the periodic nature of the axions when viewed as ingredients of the internal geometry. Note that the change of coordinates in (\ref{axions_coc}) is always well-defined. Then, the fibration allows to replace $\,d\theta_i \rightarrow d\theta_i + \chi_i \,d\eta\,$ in the original background with vanishing axions and vice-versa.

\subsubsection{D3-brane and SO(6) symmetry}
\label{sec:axions_so(6)_example}

As we showed in Section~\ref{sec:D3-brane_c=0}, the D3-brane at $\,c=0\,$ features an $\,\textrm{SO}(6)\,$ symmetry that suggests this time to think of $\,\textrm{S}^5\,$ as the quotient $\,\textrm{SO}(6)/\textrm{SO}(5)\,$. The relevant symmetry group is now $\,\textrm{G}=\textrm{SO}(6)\,$. We will investigate the action of $\,\textrm{SO}(6)$ on the geometry and the relation with turning on the axions $\,\chi^{(0)}_{1,2,3}\,$. Note that the construction in Section~\ref{sec:monodromies} can be formally extended to the case $\,T\rightarrow \infty\,$, \textit{i.e.} $\,\textrm{S}^{1} \rightarrow \mathbb{R}\,$, in order to study the patterns of symmetry breaking for the D3-brane background. However, the axions $\,\chi^{(0)}_{i}\,$ are no longer periodic in this case as $\,\frac{2 \pi}{T} \rightarrow
 0\,$.

Turning on the axions translated into the coordinates redefinition in (\ref{cob_angles}), thus inducing a non-trivial monodromy on $\,\textrm{S}^5 \times \textrm{S}^1\,$ by virtue of (\ref{G_etaeta_mod}). Using the embedding coordinates $\,\mathcal{Y}_{\underline{m}} \,$ (\ref{Y_coords_SO(2)^3}) on $\,\mathbb{R}^6\,$, the action of (\ref{cob_angles}) is encoded into a matrix element $\,h(\eta)\,$ that belongs to the $\,\textrm{SO}(2)^3\,$ Cartan subalgebra of $\,\textrm{SO}(6)\,$, namely,
\begin{equation}
\label{h_matrix}
h(\eta) = \begin{pmatrix}
h_{1} & & \\
 & h_{2}  & \\
 & & h_{3}
\end{pmatrix}
\hspace{6mm} \textrm{ with } \hspace{6mm}
h_{i} = \text{exp}\left(i \,\chi^{(0)}_i\, \sigma_2\, \eta\right) \in \textrm{SO}(2) \ .
\end{equation}
Observe that no constraint on the axions is needed in this case as $\,\textrm{G}=\textrm{SO}(6) \cong \textrm{SU}(4)\,$ has rank $\,3$. Looking at Section~\ref{sec:D3-brane_c=0} we identify the $\,\mathbb{T}^3\,$ with the angles $\,\theta_i\,$ parameterising independent rotations in the planes $\,(\mathcal{Y}_2,\mathcal{Y}_3)\,$, $\,(\mathcal{Y}_4,\mathcal{Y}_5)\,$ and $\,(\mathcal{Y}_6,\mathcal{Y}_7)\,$. In (\ref{cob_angles}) we provided the change of variables making the axions disappear locally. Up to a conventional sign for $\,\chi^{(0)}_{3}\,$, this agrees with the general result in (\ref{phi_theta}). Note also that the absence of an algebraic constraint for the axions $\,\chi_{i}\,$ when $\,\textrm{G}=\textrm{SO}(6) \,$ is precisely what allowed for
\begin{equation}
\sum_{i=1}^{3} \textrm{Re} z_{i} \approx c \, \sinh\Phi_{0}   \ ,
\end{equation}
in the deformed D3-brane solution controlling the UV behaviour of the RG flows in Section~\ref{sec:RG flows}. This generically causes an axionic breaking of the $\,\textrm{SO}(6)\,$ symmetry down to a subgroup of $\,\textrm{SU}(3)\,$. In addition, we have also verified that this construction reproduces the correct number of axions, coordinate redefinitions and patterns of symmetry breaking for the non-supersymmetric family of S-folds in \cite{Guarino:2020gfe}.

We will close this section by emphasising the importance of $\,h\,$ being an element of the symmetry group $\,\textrm{G}\,$. For example, if we do \textit{not} enforce that $\,h \in \textrm{SU}(3)\,$ for the $\mathcal{N}=1$ S-folds, the coordinate combination $\,3\beta + \psi\,$ maps to
\begin{equation}
\label{local_3beta+psi}
3\beta + \psi \,\, \rightarrow \,\, 3\beta + \psi  - \eta \, \sum_{i=1}^{3} \chi_i \ .
\end{equation}
Since the two-form potentials $\,\mathbb{B}^{\alpha}\,$ have an explicit dependence on $\,(3\beta + \psi )\,$ they will gain an explicit dependence on $\,\eta\,$ through the local coordinate transformation in (\ref{local_3beta+psi}), thus clashing with the equations of motion. Note that this is not a problem for the D3-brane as $\,\mathbb{B}^{\alpha}\,$ vanish identically and we can safely take $\,h \in \textrm{SO}(6)\,$.

\section{Conclusions}
\label{sec:conclus}

In this work we have investigated various aspects of type IIB S-fold backgrounds by using the $\,[\,\textrm{SO}(1,1) \times \textrm{SO}(6)\,] \ltimes \mathbb{R}^{12}\,$ gauged maximal supergravity as a four-dimensional effective description thereof and the $\,\textrm{E}_{7(7)}$-ExFT as a tool to uplift four-dimensional solutions of said supergravity to ten dimensions. The focus of the paper has been three-fold:

\begin{itemize}

\item[$\circ$] Firstly we have constructed new holographic RG flows on the D3-brane that connect an anisotropic deformation of $\,\textrm{SYM}_{4}\,$ in the UV to supersymmetric J-fold $\textrm{CFT}_{3}$'s in the IR. In addition, we have also provided examples of $\textrm{CFT}_{3}$ to $\textrm{CFT}_{3}$ flows between $\,\mathcal{N}=1\,$ J-fold CFT$_{3}$'s with either $\,\textrm{SU}(3)\,$ or $\,\textrm{SU}(2) \times \textrm{U}(1)\,$ flavour symmetry in the UV and the $\,\mathcal{N}=2\,$ J-fold CFT$_{3}$ with $\,\textrm{SU}(2)\,$ flavour symmetry in the IR. The role played by the axions as moduli fields dual to marginal deformations was analysed when characterising the UV behaviour of such RG flows.

\item[$\circ$] Secondly the ten-dimensional holographic description of such RG flows was investigated. The anisotropic deformation of $\,\textrm{SYM}_{4}\,$ was interpreted in a purely geometric manner and connected to the locally geometric $\,\textrm{SL}(2)_{\textrm{IIB}}\,$ twist matrix $\,A(\eta)\,$ generating the S-fold. In this manner we identified a set of five-dimensional one-forms $\,(\mathcal{F}_{(1)}{}^{\alpha}\,,\,\mathcal{F}_{(1)\alpha\beta}\,,\,\widetilde{\mathcal{F}}_{(1)})\,$ as the sources of anisotropy which came into $\,\textrm{SL}(2)_{\textrm{IIB}}\,$ representations. As mentioned in the main text, it would be interesting to characterise the operators triggering such a $\,\textrm{SYM}_{4}\,$ anisotropy in an $\,\textrm{SL}(2)_{\textrm{IIB}}\,$ covariant setup of the type investigated here.

\item[$\circ$] Thirdly we investigated the type IIB geometric origin of the four-dimensional axions dual to marginal deformations of the S-fold CFT$_3$'s. In all the cases analysed (which also include the D3-brane and the non-supersymmetric S-folds in \cite{Guarino:2020gfe}), and upon uplift to ten dimensions, the axions were shown to be locally reabsorbable in a reparameterisation of the angular coordinates on the five-sphere. However this reparameterisation induces a non-trivial $h$-monodromy on $\,\textrm{S}^5 \times \textrm{S}^1\,$ which is responsible for the various patterns of flavour symmetry breaking observed in the dual S-fold CFT$_{3}$'s. Building upon the recent results in \cite{1852633}, the possible patterns of flavour symmetry breaking induced by the $h$-monodromy were connected to the mapping torus $\,T(\textrm{S}^5)_h\,$.

\end{itemize}

Let us conclude by emphasising that, although a solid proof of a precise correspondence between 4D axions and monodromy-induced patterns of symmetry breaking in 10D is still lacking, the mechanism proposed in this work (see Figure~\ref{fig:monodromy_diagram}) correctly reproduces the known results (number of axions, patterns of flavour symmetry breaking, etc.) for all the S-folds captured by the $\,\mathbb{Z}_2^3$-invariant sector of the $\,\left[\textrm{SO}(6)\times \textrm{SO}(1,1)\right] \ltimes \mathbb{R}^{12}\,$ maximal supergravity, as well as for the D3-brane. It would also be interesting to assess whether the R-symmetry groups can be incorporated to the story, perhaps opening up new possibilities for RG flows possibly in different sectors of the theory. We hope to come back to this and related issues in the near future.

\section*{Acknowledgements}

We are grateful to Andr\'es Anabal\'on, Riccardo Argurio, Adrien Druart, Carlos Hoyos, Patrick Meessen and Mario Trigiante for conversations. The work of AG is supported by the Spanish government grant PGC2018-096894-B-100. The research of CS is supported by IISN-Belgium (convention 4.4503.15). CS is a Research Fellow of the F.R.S.-FNRS (Belgium).

\appendix

\section{5D \textit{vs} 4D}
\label{app:5Dvs4D}

This appendix collects various group theoretical considerations about the truncations that are of relevance for this work. We will also establish the connection with five-dimensional supergravity, which will help us to get a better understanding of our flows.

\subsection{5D supergravity: scalar fields and dual operators}

Let us consider the five-dimensional SO(6)-gauged supergravity. The scalar fields in the theory span the coset space
\begin{equation}
\mathcal{M}_{\textrm{scal}} = \frac{\textrm{E}_{6(6)}}{\textrm{USp}(8)} \ ,
\end{equation}
thus accounting for $\,42\,$ scalars. The group theoretical branching of such scalars under $\,\textrm{USp}(8) \supset \textrm{SU}(4) \times \textrm{U}(1)_{S}\,$ yields
\begin{equation}
\label{Branching_5D}
\begin{array}{ccl}
\textrm{USp}(8) &\supset& \textrm{SU}(4) \times \textrm{U}(1)_{S} \\[2mm]
\textbf{42} & \rightarrow & \textbf{1}_{4} \,\, + \,\, \textbf{1}_{-4} \,\, + \,\, \textbf{10}_{-2} \,\, + \,\, \overline{\textbf{10}}_{2} \,\,+ \,\, \textbf{20'}_{0}
\end{array}
\end{equation}
corresponding to the following holographic operators: gauge coupling $\,g \in \textbf{1}_{4}\,$, theta-angle $\,\theta \in \textbf{1}_{-4}\,$, fermion bilinears $\, \psi \, \psi \in \textbf{10}_{-2} \,$ and their conjugates $\, \bar{\psi} \, \bar{\psi} \in \overline{\textbf{10}}_{2}  \,$ and, lastly, scalar bilinears $\,\phi \, \phi \in \textbf{20'}_{0}\,$. As a result, the $\,\textrm{SO}(6) \cong \textrm{SU}(4)\,$ invariant sector of the 5D theory contains $\,2\,$ scalars associated with the two singlets $\,\textbf{1}_{\pm4}\,$. Lastly, the eight gravitini of the theory branch as
\begin{equation}
\label{Branching_5D_gravitini}
\begin{array}{ccl}
\textrm{USp}(8) &\supset& \textrm{SU}(4) \times \textrm{U}(1)_{S} \\[2mm]
\textbf{8} & \rightarrow & \textbf{4}_{1} \,\, + \,\, \bar{\textbf{4}}_{-1}
\end{array}
\end{equation}

\subsection{4D supergravity: scalar fields and 5D origin}

Let us now consider the four-dimensional supergravity with ${[\textrm{SO}(1,1) \times \textrm{SO}(6)] \ltimes \mathbb{R}^{12}}\,$ gauging. The scalar fields in the theory span the coset space
\begin{equation}
\mathcal{M}_{\textrm{scal}} = \frac{\textrm{E}_{7(7)}}{\textrm{SU}(8)} \ ,
\end{equation}
thus accounting for $\,70\,$ scalars. In order to establish a connection with the 5D theory, we must branch such scalars under $\,\textrm{SU}(8) \supset \textrm{USp}(8) \supset \textrm{SU}(4) \times \textrm{U}(1)_{S}\,$. The result of this procedure is
\begin{equation}
\label{Branching_4D}
\begin{array}{ccccl}
\textrm{SU}(8) &\supset& \textrm{USp}(8) &\supset& \textrm{SU}(4) \times \textrm{U}(1)_{S} \\[2mm]
\textbf{70} & \rightarrow&  \textbf{42} & \rightarrow & \textbf{1}_{4} \,\, + \,\, \textbf{1}_{-4} \,\, + \,\, \textbf{10}_{-2} \,\, + \,\, \overline{\textbf{10}}_{2} \,\,+ \,\, \textbf{20'}_{0} \\[2mm]
 &  &  \textbf{27} & \rightarrow & \textbf{15}_{0} \,\, + \,\, \textbf{6}_{2} \,\, + \,\, \textbf{6}_{-2} \\[2mm]
 &  &  \textbf{1} & \rightarrow & \textbf{1}_{0}
\end{array}
\end{equation}
The scalars descending from the $\,\textbf{42}\,$ are directly identified with the scalars (\ref{Branching_5D}) in the 5D theory. On the other hand, the scalars descending from the $\,\textbf{27}\,$ are connected with vectors $\,\textbf{15}_{0}\,$ and tensor fields (dual to vectors) $\,\textbf{6}_{\pm 2}\,$ in 5D. Lastly there is an additional scalar $\,\textbf{1}_{0}\,$ descending from the 5D metric (KK scalar).\footnote{The various entries in the right-most column of (\ref{Branching_4D}) are in one-to-one correspondence with the set of possible bosonic deformations of $\,\mathcal{N} = 4\,$ SYM$_{4}$. The latter are parameterised by the bosonic auxiliary fields of the off-shell $\,\mathcal{N} = 4\,$ conformal supergravity \cite{Maxfield:2016lok} (see also \cite{Arav:2020obl}), which transform in the displayed irreps of the (global) SU(4) R-symmetry group of undeformed $\,\mathcal{N} = 4\,$ SYM$_{4}$.} As a result, the $\,\textrm{SO}(6) \cong \textrm{SU}(4)\,$ invariant sector of the 4D theory contains $\,2+1\,$ scalars associated with the three singlets $\,\textbf{1}_{\pm4}\,$ and $\,\textbf{1}_{0}\,$. The eight gravitini of the theory decompose as
\begin{equation}
\label{Branching_4D_gravitini}
\begin{array}{ccl}
\textrm{SU}(8) &\supset& \textrm{SU}(4) \times \textrm{U}(1)_{S} \\[2mm]
\textbf{8} & \rightarrow & \textbf{4}_{1} \,\, + \,\, \bar{\textbf{4}}_{-1}
\end{array}
\end{equation}

\subsubsection*{SU(3) invariant sector}

Lets us look in more detail at the SU(3)-invariant sector that plays a central role in this work. The branching rules under the relevant embedding $\,\textrm{SU}(4) \times \textrm{U}(1)_{S} \supset \textrm{SU}(3) \times \textrm{U}(1) \times \textrm{U}(1)_{S}\,$ reads
\begin{equation}
\label{Branching_SU3}
\small{
\begin{array}{ccl}
 \textrm{SU}(4) \times \textrm{U}(1)_{S} &\supset& \textrm{SU}(3) \times \textrm{U}(1) \times \textrm{U}(1)_{S} \\[2mm]
\hline\\[-3mm]
 \textbf{1}_{4} \, + \, \textbf{1}_{-4}  & \rightarrow & \textbf{1}_{(0,4)} \, + \, \textbf{1}_{(0,-4)}  \\[2mm]
\textbf{10}_{-2} \, + \,  \overline{\textbf{10}}_{2}  & \rightarrow & \left( \,  \bar{\textbf{6}}_{(2,-2)}  \, + \,  \textbf{3}_{(-2,-2)}  \, + \,  \textbf{1}_{(-6,-2)} \, \right) \, + \, \left( \,  \textbf{6}_{(-2,2)} \, + \,  \bar{\textbf{3}}_{(2,2)}  \, + \,  \textbf{1}_{(6,2)} \, \right)  \\[2mm]
\textbf{20'}_{0} & \rightarrow &  \bar{\textbf{6}}_{(-4,0)}  \, + \, \textbf{8}_{(0,0)}  \, + \, \textbf{6}_{(4,0)} \\[2mm]
\hline\\[-3mm]
\textbf{15}_{0} & \rightarrow & \textbf{8}_{(0,0)}  \, + \, \textbf{3}_{(4,0)}  \, + \,  \bar{\textbf{3}}_{(-4,0)}  \, + \,   \textbf{1}_{(0,0)}  \\[2mm]
\textbf{6}_{2} \, + \,  \textbf{6}_{-2} & \rightarrow &  \left( \,  \textbf{3}_{(-2,2)}  \, + \,  \bar{\textbf{3}}_{(2,2)}  \, \right) \, + \, \left( \, \textbf{3}_{(-2,-2)}  \, + \,  \bar{\textbf{3}}_{(2,-2)}  \, \right)  \\[2mm]
\hline\\[-3mm]
\textbf{1}_{0}  & \rightarrow &  \textbf{1}_{(0,0)}   \\[1mm]
\hline\\[-3mm]
 \textbf{4}_{1} \, + \, \bar{\textbf{4}}_{-1}  & \rightarrow &  \textbf{3}_{(1,1)}   \, + \, \textbf{1}_{(-3,1)}  \, + \, \bar{\textbf{3}}_{(-1,-1)}   \, + \, \textbf{1}_{(3,-1)}  
\end{array}
}
\end{equation}
Then the $\,\textrm{SU}(3)\,$ invariant sector of the 4D theory preserves $\,\mathcal{N}=2\,$ supersymmetry and contains $\,2 \, [\, \textrm{denoted } (\phi,\sigma) \textrm{ in } \cite{Guarino:2019oct}  \,] \,+ \,2 \, [\,  \textrm{denoted }  (\zeta,\tilde{\zeta}) \textrm{ in } \cite{Guarino:2019oct}\,] \, + \, 1 \, [\,\textrm{Re}z_{1,2,3}\,] \,+\,  1\, [\,\textrm{Im}z_{1,2,3}\,]\,$ scalar fields respectively organised as: $\,\textbf{1}_{(0,\pm 4)} \subset \textbf{1}_{\pm 4}\,$ (dual to the gauge coupling $\,g\,$ and the theta-angle $\,\theta$) and $\,\textbf{1}_{(\mp 6,\mp 2)} \subset \textbf{10}_{-2} \textrm{ or } \overline{\textbf{10}}_{2}\,$ (dual to a fermion bilinear $\, \psi \,\psi \,$ and its conjugate $\, \bar{\psi} \, \bar{\psi} \,$) descending from 5D scalars, $\,\textbf{1}_{(0,0)} \subset \textbf{15}_{0}\,$ (dual to a one-form deformation $V_{\mu}$) descending from the unique SU(3)-invariant vector field in 5D, and $\,\textbf{1}_{(0,0)} \subset \textbf{1}_{0}\,$ (generically associated with placing $\,\mathcal{N} = 4\,$ SYM$_{4}$ on a curved manifold) descending from the 5D metric (KK scalar).

As a consequence of the above group theoretical decompositions, a configuration in an Einstein-scalar model in 4D that involves a non-trivial profile for the $\,\textbf{1}_{(0,0)} \in \textbf{15}_{0}\,$ scalar cannot be connected to a configuration in an Einstein-scalar model in 5D. This scalar is identified with the axion $\,\textrm{Re}z_{1,2,3}\,$ in the main text which is responsible for the global \textit{vs} local issues discussed in Section~\ref{sec:axions_SU3}. More precisely, when truncating the Einstein-scalar model from 5D to 4D one has
\begin{equation}
\textrm{5D:} \hspace{5mm} \mathcal{M}_{\textrm{scal}} = \dfrac{\textrm{SU}(2,1)}{\textrm{SU}(2)\times \textrm{U}(1)}  
\hspace{5mm} \longrightarrow \hspace{5mm}
\textrm{4D:} \hspace{5mm} \mathcal{M}_{\textrm{scal}} = \left.\mathbb{R}^{+}\right|_{\textrm{KK}} \times \dfrac{\textrm{SU}(2,1)}{\textrm{SU}(2)\times \textrm{U}(1)} \ .
\end{equation}
Finally, the $\,\mathbb{Z}_{2} \times \textrm{SU}(3)\,$ invariant sector investigated in this work preserves $\,\mathcal{N}=1\,$ and lies at the intersection between the $\,\textrm{SU}(3)\,$ and $\,\mathbb{Z}_{2}^3\,$ invariant sectors yielding
\begin{equation}
\textrm{4D:} \hspace{5mm} \left.\dfrac{\textrm{SL}(2)}{\textrm{SO}(2)}\right|_{z_{1,2,3}}    \times  \left.\dfrac{\textrm{SL}(2)}{\textrm{SO}(2)}\right|_{z_{4,5,6,7}} \,\, \subset \,\,  \left.\dfrac{\textrm{SL}(2)}{\textrm{SO}(2)}\right|_{z_{1,2,3}}  \times  \left.\dfrac{\textrm{SU}(2,1)}{\textrm{SU}(2)\times \textrm{U}(1)} \right|_{\textrm{Univ hyper}} \ ,
\end{equation}
with $\,z_{1,2,3} = \textrm{Re} z_{1,2,3} + i \, \textrm{Im}z_{1,2,3}\,$ and $z_{4,5,6,7} = - \frac{1}{2} \tilde{\zeta} + i e^{- \phi}$. This intersection corresponds to setting to zero the scalars $\,\sigma\,$ and $\,\zeta\,$ in the universal hypermultiplet parameterisation of \cite{Guarino:2019oct} which are not invariant under the action of $\,\mathbb{Z}_{2}^{*} \subset \mathbb{Z}_{2}^3\,$ as defined in \cite{Guarino:2020gfe}.

\subsubsection*{$\textrm{SU}(2) \times \textrm{U}(1)$ and $\,\textrm{SO}(4)\,$ invariant sectors}

These two sectors of the theory also play an important role in this work. The $\,\textrm{SU}(2) \times \textrm{U}(1)\,$ invariant sector is defined through the group theoretical embeddings and the corresponding gravitini decompositions
\begin{equation}
\label{Branching_4D_gravitini_SU2xU1}
\small{
\begin{array}{cclcl}
\textrm{SU}(8) &\supset& \textrm{SU}(4) \times \textrm{U}(1)_{S} &\supset& \textrm{SU}(2) \times \textrm{U}(1) \times \textrm{U}(1)' \times \textrm{U}(1)_{S} \\[2mm]
\textbf{8} & \rightarrow & \textbf{4}_{1} \,\, + \,\, \bar{\textbf{4}}_{-1} &\rightarrow   &  \textbf{2}_{(0,1,1)}   +  \textbf{1}_{(1,-1,1)}   +  \textbf{1}_{(-1,-1,1)}    +  \textbf{2}_{(0,-1,-1)}   + \textbf{1}_{(1,1,-1)}    +  \textbf{1}_{(-1,1,-1)} 
\end{array}
}
\end{equation}
Then this sector is non-supersymmetric as the four singlets in (\ref{Branching_4D_gravitini_SU2xU1}) are charged under the $\,\textrm{U}(1)\,$ factor. Note however that the $\,\textrm{SU}(2)\,$ invariant sector preserves $\,\mathcal{N}=4\,$ supersymmetry. On the other hand, the $\,\textrm{SO}(4) \cong \textrm{SU}(2) \times \textrm{SU}(2)^{(s)}\,$ invariant sector of the theory is defined through the group theoretical embeddings and the gravitini decompositions
\begin{equation}
\label{Branching_4D_gravitini_SO4}
\small{
\begin{array}{cclcl}
\textrm{SU}(8) &\supset& \textrm{SU}(4) \times \textrm{U}(1)_{S} &\supset& \textrm{SU}(2) \times\textrm{SU}(2)^{(s)} \times \textrm{U}(1)_{S} \\[2mm]
\textbf{8} & \rightarrow & \textbf{4}_{1} \,\, + \,\, \bar{\textbf{4}}_{-1} &\rightarrow   &  (\textbf{2},\textbf{2})_{1}  \, + \,  (\textbf{2},\textbf{2})_{-1} 
\end{array}
}
\end{equation}
thus being a non-supersymmetric sector.

\bibliographystyle{JHEP}
\bibliography{references}

\providecommand{\href}[2]{#2}\begingroup\raggedright\begin{thebibliography}{10}

\bibitem{Hull:1994ys}
C.M.~Hull and P.K.~Townsend, \emph{{Unity of superstring dualities}},
  \href{https://doi.org/10.1016/0550-3213(94)00559-W}{\emph{Nucl. Phys. B}
  {\bfseries 438} (1995) 109}
  [\href{https://arxiv.org/abs/hep-th/9410167}{{\ttfamily hep-th/9410167}}].

\bibitem{deWit:2007mt}
B.~de~Wit, H.~Samtleben and M.~Trigiante, \emph{{The Maximal D=4
  supergravities}},
  \href{https://doi.org/10.1088/1126-6708/2007/06/049}{\emph{JHEP} {\bfseries
  0706} (2007) 049} [\href{https://arxiv.org/abs/0705.2101}{{\ttfamily
  0705.2101}}].

\bibitem{deWit:1982ig}
B.~de~Wit and H.~Nicolai, \emph{{N=8 Supergravity}},
  \href{https://doi.org/10.1016/0550-3213(82)90120-1}{\emph{Nucl.Phys.}
  {\bfseries B208} (1982) 323}.

\bibitem{deWit:1986iy}
B.~de~Wit and H.~Nicolai, \emph{{The Consistency of the $S^7$ Truncation in
  $D=11$ Supergravity}},
  \href{https://doi.org/10.1016/0550-3213(87)90253-7}{\emph{Nucl.Phys.}
  {\bfseries B281} (1987) 211}.

\bibitem{Nicolai:2011cy}
H.~Nicolai and K.~Pilch, \emph{{Consistent Truncation of d = 11 Supergravity on
  AdS$_4 \times S^7$}},
  \href{https://doi.org/10.1007/JHEP03(2012)099}{\emph{JHEP} {\bfseries 1203}
  (2012) 099} [\href{https://arxiv.org/abs/1112.6131}{{\ttfamily 1112.6131}}].

\bibitem{Dall'Agata:2012bb}
G.~Dall'Agata, G.~Inverso and M.~Trigiante, \emph{{Evidence for a family of
  SO(8) gauged supergravity theories}},
  \href{https://doi.org/10.1103/PhysRevLett.109.201301}{\emph{Phys.Rev.Lett.}
  {\bfseries 109} (2012) 201301}
  [\href{https://arxiv.org/abs/1209.0760}{{\ttfamily 1209.0760}}].

\bibitem{Borghese:2012qm}
A.~Borghese, A.~Guarino and D.~Roest, \emph{{All $G_2$ invariant critical
  points of maximal supergravity}},
  \href{https://doi.org/10.1007/JHEP12(2012)108}{\emph{JHEP} {\bfseries 1212}
  (2012) 108} [\href{https://arxiv.org/abs/1209.3003}{{\ttfamily 1209.3003}}].

\bibitem{Borghese:2012zs}
A.~Borghese, G.~Dibitetto, A.~Guarino, D.~Roest and O.~Varela, \emph{{The
  SU(3)-invariant sector of new maximal supergravity}},
  \href{https://doi.org/10.1007/JHEP03(2013)082}{\emph{JHEP} {\bfseries 1303}
  (2013) 082} [\href{https://arxiv.org/abs/1211.5335}{{\ttfamily 1211.5335}}].

\bibitem{Borghese:2013dja}
A.~Borghese, A.~Guarino and D.~Roest, \emph{{Triality, Periodicity and
  Stability of SO(8) Gauged Supergravity}},
  \href{https://doi.org/10.1007/JHEP05(2013)107}{\emph{JHEP} {\bfseries 1305}
  (2013) 107} [\href{https://arxiv.org/abs/1302.6057}{{\ttfamily 1302.6057}}].

\bibitem{Anabalon:2013eaa}
A.~Anabalon and D.~Astefanesei, \emph{{Black holes in $\omega$-defomed gauged
  $N=8$ supergravity}},
  \href{https://doi.org/10.1016/j.physletb.2014.03.035}{\emph{Phys. Lett.}
  {\bfseries B732} (2014) 137}
  [\href{https://arxiv.org/abs/1311.7459}{{\ttfamily 1311.7459}}].

\bibitem{Lu:2014fpa}
H.~L{\"u}, Y.~Pang and C.N.~Pope, \emph{{An $\omega$ deformation of gauged STU
  supergravity}}, \href{https://doi.org/10.1007/JHEP04(2014)175}{\emph{JHEP}
  {\bfseries 04} (2014) 175} [\href{https://arxiv.org/abs/1402.1994}{{\ttfamily
  1402.1994}}].

\bibitem{Wu:2015ska}
S.-Q.~Wu and S.~Li, \emph{{Thermodynamics of Static Dyonic AdS Black Holes in
  the $\omega$-Deformed Kaluza-Klein Gauged Supergravity Theory}},
  \href{https://doi.org/10.1016/j.physletb.2015.05.013}{\emph{Phys. Lett.}
  {\bfseries B746} (2015) 276}
  [\href{https://arxiv.org/abs/1505.00117}{{\ttfamily 1505.00117}}].

\bibitem{Karndumri:2020bkc}
P.~Karndumri and C.~Maneerat, \emph{{Supersymmetric Janus solutions in
  $\omega$-deformed $N=8$ gauged supergravity}},
  \href{https://arxiv.org/abs/2012.15763}{{\ttfamily 2012.15763}}.

\bibitem{Guarino:2013gsa}
A.~Guarino, \emph{{On new maximal supergravity and its BPS domain-walls}},
  \href{https://doi.org/10.1007/JHEP02(2014)026}{\emph{JHEP} {\bfseries 1402}
  (2014) 026} [\href{https://arxiv.org/abs/1311.0785}{{\ttfamily 1311.0785}}].

\bibitem{Tarrio:2013qga}
J.~Tarrio and O.~Varela, \emph{{Electric/magnetic duality and RG flows in
  AdS$_4$/CFT$_3$}}, \href{https://doi.org/10.1007/JHEP01(2014)071}{\emph{JHEP}
  {\bfseries 1401} (2014) 071}
  [\href{https://arxiv.org/abs/1311.2933}{{\ttfamily 1311.2933}}].

\bibitem{deWit:2013ija}
B.~de~Wit and H.~Nicolai, \emph{{Deformations of gauged SO(8) supergravity and
  supergravity in eleven dimensions}},
  \href{https://arxiv.org/abs/1302.6219}{{\ttfamily 1302.6219}}.

\bibitem{Lee:2015xga}
K.~Lee, C.~Strickland-Constable and D.~Waldram, \emph{{New Gaugings and
  Non-Geometry}}, \href{https://doi.org/10.1002/prop.201700049}{\emph{Fortsch.
  Phys.} {\bfseries 65} (2017) 1700049}
  [\href{https://arxiv.org/abs/1506.03457}{{\ttfamily 1506.03457}}].

\bibitem{Dall'Agata:2014ita}
G.~Dall'Agata, G.~Inverso and A.~Marrani, \emph{{Symplectic Deformations of
  Gauged Maximal Supergravity}},
  \href{https://doi.org/10.1007/JHEP07(2014)133}{\emph{JHEP} {\bfseries 1407}
  (2014) 133} [\href{https://arxiv.org/abs/1405.2437}{{\ttfamily 1405.2437}}].

\bibitem{Inverso:2015viq}
G.~Inverso, \emph{{Electric-magnetic deformations of D = 4 gauged
  supergravities}}, \href{https://doi.org/10.1007/JHEP03(2016)138}{\emph{JHEP}
  {\bfseries 03} (2016) 138}
  [\href{https://arxiv.org/abs/1512.04500}{{\ttfamily 1512.04500}}].

\bibitem{Guarino:2015jca}
A.~Guarino, D.L.~Jafferis and O.~Varela, \emph{{The string origin of dyonic N=8
  supergravity and its simple Chern-Simons duals}},
  \href{https://doi.org/10.1103/PhysRevLett.115.091601}{\emph{Phys. Rev. Lett.}
  {\bfseries 115} (2015) 091601}
  [\href{https://arxiv.org/abs/1504.08009}{{\ttfamily 1504.08009}}].

\bibitem{Inverso:2016eet}
G.~Inverso, H.~Samtleben and M.~Trigiante, \emph{{Type II supergravity origin
  of dyonic gaugings}},
  \href{https://doi.org/10.1103/PhysRevD.95.066020}{\emph{Phys. Rev.}
  {\bfseries D95} (2017) 066020}
  [\href{https://arxiv.org/abs/1612.05123}{{\ttfamily 1612.05123}}].

\bibitem{Guarino:2015qaa}
A.~Guarino and O.~Varela, \emph{{Dyonic ISO(7) supergravity and the duality
  hierarchy}}, \href{https://doi.org/10.1007/JHEP02(2016)079}{\emph{JHEP}
  {\bfseries 02} (2016) 079}
  [\href{https://arxiv.org/abs/1508.04432}{{\ttfamily 1508.04432}}].

\bibitem{Guarino:2019jef}
A.~Guarino, J.~Tarrio and O.~Varela, \emph{{Halving ISO(7) supergravity}},
  \href{https://doi.org/10.1007/JHEP11(2019)143}{\emph{JHEP} {\bfseries 11}
  (2019) 143} [\href{https://arxiv.org/abs/1907.11681}{{\ttfamily
  1907.11681}}].

\bibitem{Guarino:2020jwv}
A.~Guarino, J.~Tarrio and O.~Varela, \emph{{Brane-jet stability of
  non-supersymmetric AdS vacua}},
  \href{https://doi.org/10.1007/JHEP09(2020)110}{\emph{JHEP} {\bfseries 09}
  (2020) 110} [\href{https://arxiv.org/abs/2005.07072}{{\ttfamily
  2005.07072}}].

\bibitem{Bobev:2020qev}
N.~Bobev, T.~Fischbacher, F.F.~Gautason and K.~Pilch, \emph{{New AdS$_4$ Vacua
  in Dyonic ISO(7) Gauged Supergravity}},
  \href{https://arxiv.org/abs/2011.08542}{{\ttfamily 2011.08542}}.

\bibitem{Guarino:2017eag}
A.~Guarino and J.~Tarrio, \emph{{BPS black holes from massive IIA on S$^6$}},
  \href{https://arxiv.org/abs/1703.10833}{{\ttfamily 1703.10833}}.

\bibitem{Guarino:2017pkw}
A.~Guarino, \emph{{BPS black hole horizons from massive IIA}},
  \href{https://doi.org/10.1007/JHEP08(2017)100}{\emph{JHEP} {\bfseries 08}
  (2017) 100} [\href{https://arxiv.org/abs/1706.01823}{{\ttfamily
  1706.01823}}].

\bibitem{Suh:2018nmp}
M.~Suh, \emph{{Supersymmetric Janus solutions of dyonic $ISO(7)$-gauged
  $\mathcal{N}\,=\,8$ supergravity}},
  \href{https://doi.org/10.1007/JHEP04(2018)109}{\emph{JHEP} {\bfseries 04}
  (2018) 109} [\href{https://arxiv.org/abs/1803.00041}{{\ttfamily
  1803.00041}}].

\bibitem{Guarino:2016ynd}
A.~Guarino, J.~Tarrio and O.~Varela, \emph{{Romans-mass-driven flows on the
  D2-brane}}, \href{https://doi.org/10.1007/JHEP08(2016)168}{\emph{JHEP}
  {\bfseries 08} (2016) 168}
  [\href{https://arxiv.org/abs/1605.09254}{{\ttfamily 1605.09254}}].

\bibitem{Guarino:2019snw}
A.~Guarino, J.~Tarrio and O.~Varela, \emph{{Flowing to $\mathcal{N}=3$
  Chern--Simons-matter theory}},
  \href{https://arxiv.org/abs/1910.06866}{{\ttfamily 1910.06866}}.

\bibitem{Guarino:2015vca}
A.~Guarino and O.~Varela, \emph{{Consistent $ \mathcal{N}=8 $ truncation of
  massive IIA on S$^{6}$}},
  \href{https://doi.org/10.1007/JHEP12(2015)020}{\emph{JHEP} {\bfseries 12}
  (2015) 020} [\href{https://arxiv.org/abs/1509.02526}{{\ttfamily
  1509.02526}}].

\bibitem{Varela:2015uca}
O.~Varela, \emph{{AdS$_{4}$ solutions of massive IIA from dyonic ISO(7)
  supergravity}}, \href{https://doi.org/10.1007/JHEP03(2016)071}{\emph{JHEP}
  {\bfseries 03} (2016) 071}
  [\href{https://arxiv.org/abs/1509.07117}{{\ttfamily 1509.07117}}].

\bibitem{Romans:1985tz}
L.~Romans, \emph{{Massive N=2a Supergravity in Ten-Dimensions}},
  \href{https://doi.org/10.1016/0370-2693(86)90375-8}{\emph{Phys.Lett.}
  {\bfseries B169} (1986) 374}.

\bibitem{Hosseini:2017fjo}
S.M.~Hosseini, K.~Hristov and A.~Passias, \emph{{Holographic microstate
  counting for AdS$_{4}$ black holes in massive IIA supergravity}},
  \href{https://doi.org/10.1007/JHEP10(2017)190}{\emph{JHEP} {\bfseries 10}
  (2017) 190} [\href{https://arxiv.org/abs/1707.06884}{{\ttfamily
  1707.06884}}].

\bibitem{Benini:2017oxt}
F.~Benini, H.~Khachatryan and P.~Milan, \emph{{Black hole entropy in massive
  Type IIA}}, \href{https://doi.org/10.1088/1361-6382/aa9f5b}{\emph{Class.
  Quant. Grav.} {\bfseries 35} (2018) 035004}
  [\href{https://arxiv.org/abs/1707.06886}{{\ttfamily 1707.06886}}].

\bibitem{Guarino:2019oct}
A.~Guarino and C.~Sterckx, \emph{{S-folds and (non-)supersymmetric Janus
  solutions}}, \href{https://doi.org/10.1007/JHEP12(2019)113}{\emph{JHEP}
  {\bfseries 12} (2019) 113}
  [\href{https://arxiv.org/abs/1907.04177}{{\ttfamily 1907.04177}}].

\bibitem{Guarino:2020gfe}
A.~Guarino, C.~Sterckx and M.~Trigiante, \emph{{$\mathcal{N}=2$ supersymmetric
  S-folds}}, \href{https://doi.org/10.1007/JHEP04(2020)050}{\emph{JHEP}
  {\bfseries 04} (2020) 050}
  [\href{https://arxiv.org/abs/2002.03692}{{\ttfamily 2002.03692}}].

\bibitem{Gallerati:2014xra}
A.~Gallerati, H.~Samtleben and M.~Trigiante, \emph{{The N$>$2 supersymmetric
  AdS vacua in maximal supergravity}},
  \href{https://arxiv.org/abs/1410.0711}{{\ttfamily 1410.0711}}.

\bibitem{Assel:2018vtq}
B.~Assel and A.~Tomasiello, \emph{{Holographic duals of 3d S-fold CFTs}},
  \href{https://doi.org/10.1007/JHEP06(2018)019}{\emph{JHEP} {\bfseries 06}
  (2018) 019} [\href{https://arxiv.org/abs/1804.06419}{{\ttfamily
  1804.06419}}].

\bibitem{Bobev:2019jbi}
N.~Bobev, F.F.~Gautason, K.~Pilch, M.~Suh and J.~Van~Muiden, \emph{{Janus and
  J-fold Solutions from Sasaki-Einstein Manifolds}},
  \href{https://doi.org/10.1103/PhysRevD.100.081901}{\emph{Phys. Rev.}
  {\bfseries D100} (2019) 081901}
  [\href{https://arxiv.org/abs/1907.11132}{{\ttfamily 1907.11132}}].

\bibitem{Bobev:2020fon}
N.~Bobev, F.F.~Gautason, K.~Pilch, M.~Suh and J.~van Muiden, \emph{{Holographic
  interfaces in $ \mathcal{N} $ = 4 SYM: Janus and J-folds}},
  \href{https://doi.org/10.1007/JHEP05(2020)134}{\emph{JHEP} {\bfseries 05}
  (2020) 134} [\href{https://arxiv.org/abs/2003.09154}{{\ttfamily
  2003.09154}}].

\bibitem{Arav:2021tpk}
I.~Arav, K.C.M.~Cheung, J.P.~Gauntlett, M.M.~Roberts and C.~Rosen, \emph{{A new
  family of $AdS_4$ S-folds in type IIB string theory}},
  \href{https://arxiv.org/abs/2101.07264}{{\ttfamily 2101.07264}}.

\bibitem{Mateos:2011ix}
D.~Mateos and D.~Trancanelli, \emph{{The anisotropic N=4 super Yang-Mills
  plasma and its instabilities}},
  \href{https://doi.org/10.1103/PhysRevLett.107.101601}{\emph{Phys. Rev. Lett.}
  {\bfseries 107} (2011) 101601}
  [\href{https://arxiv.org/abs/1105.3472}{{\ttfamily 1105.3472}}].

\bibitem{Conde:2016hbg}
E.~Conde, H.~Lin, J.M.~Penin, A.V.~Ramallo and D.~Zoakos,
  \emph{{D3\textendash{}D5 theories with unquenched flavors}},
  \href{https://doi.org/10.1016/j.nuclphysb.2016.11.016}{\emph{Nucl. Phys. B}
  {\bfseries 914} (2017) 599}
  [\href{https://arxiv.org/abs/1607.04998}{{\ttfamily 1607.04998}}].

\bibitem{Hoyos:2020zeg}
C.~Hoyos, N.~Jokela, J.M.~Pen\'\i{}n and A.V.~Ramallo, \emph{{Holographic
  spontaneous anisotropy}},
  \href{https://doi.org/10.1007/JHEP04(2020)062}{\emph{JHEP} {\bfseries 04}
  (2020) 062} [\href{https://arxiv.org/abs/2001.08218}{{\ttfamily
  2001.08218}}].

\bibitem{1852633}
A.~Giambrone, E.~Malek, H.~Samtleben and M.~Trigiante, \emph{{Global Properties
  of the Conformal Manifold for S-Fold Backgrounds}},
  \href{https://arxiv.org/abs/2103.10797}{{\ttfamily 2103.10797}}.

\bibitem{Klebanov:1999tb}
I.R.~Klebanov and E.~Witten, \emph{{AdS / CFT correspondence and symmetry
  breaking}}, \href{https://doi.org/10.1016/S0550-3213(99)00387-9}{\emph{Nucl.
  Phys.} {\bfseries B556} (1999) 89}
  [\href{https://arxiv.org/abs/hep-th/9905104}{{\ttfamily hep-th/9905104}}].

\bibitem{Hohm:2013uia}
O.~Hohm and H.~Samtleben, \emph{{Exceptional Field Theory II: E$_{7(7)}$}},
  \href{https://doi.org/10.1103/PhysRevD.89.066017}{\emph{Phys.Rev.} {\bfseries
  D89} (2014) 066017} [\href{https://arxiv.org/abs/1312.4542}{{\ttfamily
  1312.4542}}].

\bibitem{Hohm:2014qga}
O.~Hohm and H.~Samtleben, \emph{{Consistent Kaluza-Klein Truncations via
  Exceptional Field Theory}},
  \href{https://doi.org/10.1007/JHEP01(2015)131}{\emph{JHEP} {\bfseries 1501}
  (2015) 131} [\href{https://arxiv.org/abs/1410.8145}{{\ttfamily 1410.8145}}].

\bibitem{Jain:2014vka}
S.~Jain, N.~Kundu, K.~Sen, A.~Sinha and S.P.~Trivedi, \emph{{A Strongly Coupled
  Anisotropic Fluid From Dilaton Driven Holography}},
  \href{https://doi.org/10.1007/JHEP01(2015)005}{\emph{JHEP} {\bfseries 01}
  (2015) 005} [\href{https://arxiv.org/abs/1406.4874}{{\ttfamily 1406.4874}}].

\bibitem{Maxfield:2016lok}
T.~Maxfield, \emph{{Supergravity Backgrounds for Four-Dimensional Maximally
  Supersymmetric Yang-Mills}},
  \href{https://doi.org/10.1007/JHEP02(2017)065}{\emph{JHEP} {\bfseries 02}
  (2017) 065} [\href{https://arxiv.org/abs/1609.05905}{{\ttfamily
  1609.05905}}].

\bibitem{Arav:2020obl}
I.~Arav, K.C.M.~Cheung, J.P.~Gauntlett, M.M.~Roberts and C.~Rosen,
  \emph{{Spatially modulated and supersymmetric mass deformations of $
  \mathcal{N} $ = 4 SYM}},
  \href{https://doi.org/10.1007/JHEP11(2020)156}{\emph{JHEP} {\bfseries 11}
  (2020) 156} [\href{https://arxiv.org/abs/2007.15095}{{\ttfamily
  2007.15095}}].

\end{thebibliography}\endgroup

\end{document}